\documentclass[twocolumn]{aastex63}
\usepackage{amsmath}
\usepackage{graphicx}
\usepackage{natbib}
\usepackage[scaled]{helvet}
\usepackage{epsfig}
\usepackage{url}
\bibpunct{(}{)}{;}{a}{}{,}
\usepackage{textcomp}
\usepackage{mathpazo}
\usepackage{xspace}
\usepackage{hyperref}
\usepackage{apjfonts}
\usepackage{mathrsfs}
\usepackage{verbatim}
\usepackage{threeparttable}

\usepackage{color}
\usepackage[normalem]{ulem}
\usepackage{multirow}

\newcommand{\bjdtdb}{\ensuremath{\rm {BJD_{TDB}}}}
\newcommand{\feh}{\ensuremath{\left[{\rm Fe}/{\rm H}\right]}}

\newcommand{\teff}{\ensuremath{T_{\rm eff}}\xspace}

\newcommand{\ecosw}{\ensuremath{e\cos{\omega_*}}}
\newcommand{\esinw}{\ensuremath{e\sin{\omega_*}}}
\newcommand{\msun}{\ensuremath{\,M_\Sun}}
\newcommand{\rsun}{\ensuremath{\,R_\Sun}}
\newcommand{\lsun}{\ensuremath{\,L_\Sun}}
\newcommand{\mj}{\ensuremath{\,M_{\rm J}}}
\newcommand{\rj}{\ensuremath{\,R_{\rm J}}}

\newcommand{\re}{\ensuremath{\,R_{\rm \Earth}}\xspace}
\newcommand{\me}{\ensuremath{\,M_{\rm \Earth}}\xspace}
\newcommand{\fave}{\langle F \rangle}
\newcommand{\fluxcgs}{10$^9$ erg s$^{-1}$ cm$^{-2}$}

\newcommand{\Kepler}{{\it Kepler}\xspace}
\newcommand{\Ktwo}{{\it K2}\xspace}
\newcommand{\ktwo}{{\it K2}\xspace}
\newcommand{\tess}{{\it TESS}\xspace}

\newcommand{\be}{\begin{equation}}
\newcommand{\ee}{\end{equation}}

\newcommand{\TESS}{{\it TESS}}

\begin{document}

\title{The \textit{K2} \& \textit{TESS} Synergy II: Revisiting 26 systems in the \textit{TESS} Primary Mission}

\newcommand{\cfa}{Center for Astrophysics \textbar \ Harvard \& Smithsonian, 60 Garden St, Cambridge, MA 02138, USA}
\newcommand{\msu}{Center for Data Intensive and Time Domain Astronomy, Department of Physics and Astronomy, Michigan State University, East Lansing, MI 48824, USA}
\newcommand{\umich}{Astronomy Department, University of Michigan, 1085 S University Avenue, Ann Arbor, MI 48109, USA}
\newcommand{\utaustin}{Department of Astronomy, The University of Texas at Austin, Austin, TX 78712, USA}
\newcommand{\MIT}{Department of Physics and Kavli Institute for Astrophysics and Space Research, Massachusetts Institute of Technology, Cambridge, MA 02139, USA}
\newcommand{\MITEPS}{Department of Earth, Atmospheric and Planetary Sciences, Massachusetts Institute of Technology,  Cambridge,  MA 02139, USA}
\newcommand{\uflorida}{Department of Astronomy, University of Florida, 211 Bryant Space Science Center, Gainesville, FL, 32611, USA}
\newcommand{\riverside}{Department of Earth and Planetary Sciences, University of California, Riverside, CA 92521, USA}
\newcommand{\usq}{Centre for Astrophysics, University of Southern Queensland, West Street, Toowoomba, QLD 4350, Australia}
\newcommand{\ames}{NASA Ames Research Center, Moffett Field, CA, 94035, USA}
\newcommand{\geneva}{Geneva Observatory, University of Geneva, Chemin des Mailettes 51, 1290 Versoix, Switzerland}
\newcommand{\uw}{Astronomy Department, University of Washington, Seattle, WA 98195 USA}
\newcommand{\warwick}{Department of Physics, University of Warwick, Gibbet Hill Road, Coventry CV4 7AL, UK}
\newcommand{\warwickceh}{Centre for Exoplanets and Habitability, University of Warwick, Gibbet Hill Road, Coventry CV4 7AL, UK}
\newcommand{\princeton}{Department of Astrophysical Sciences, Princeton University, 4 Ivy Lane, Princeton, NJ, 08544, USA}
\newcommand{\liege}{Space Sciences, Technologies and Astrophysics Research (STAR) Institute, Universit\'e de Li\`ege, 19C All\'ee du 6 Ao\^ut, 4000 Li\`ege, Belgium}
\newcommand{\vanderbilt}{Department of Physics and Astronomy, Vanderbilt University, Nashville, TN 37235, USA}
\newcommand{\fisk}{Department of Physics, Fisk University, 1000 17th Avenue North, Nashville, TN 37208, USA}
\newcommand{\columbia}{Department of Astronomy, Columbia University, 550 West 120th Street, New York, NY 10027, USA}
\newcommand{\toronto}{Dunlap Institute for Astronomy and Astrophysics, University of Toronto, Ontario M5S 3H4, Canada}
\newcommand{\unc}{Department of Physics and Astronomy, University of North Carolina at Chapel Hill, Chapel Hill, NC 27599, USA}
\newcommand{\iac}{Instituto de Astrof\'isica de Canarias (IAC), E-38205 La Laguna, Tenerife, Spain}
\newcommand{\lalaguna}{Departamento de Astrof\'isica, Universidad de La Laguna (ULL), E-38206 La Laguna, Tenerife, Spain}
\newcommand{\louisville}{Department of Physics and Astronomy, University of Louisville, Louisville, KY 40292, USA}
\newcommand{\aavso}{American Association of Variable Star Observers, 49 Bay State Road, Cambridge, MA 02138, USA}
\newcommand{\utokyo}{The University of Tokyo, 7-3-1 Hongo, Bunky\={o}, Tokyo 113-8654, Japan}
\newcommand{\naoj}{National Astronomical Observatory of Japan, 2-21-1 Osawa, Mitaka, Tokyo 181-8588, Japan}
\newcommand{\jstpresto}{JST, PRESTO, 7-3-1 Hongo, Bunkyo-ku, Tokyo 113-0033, Japan}
\newcommand{\astrobiojapan}{Astrobiology Center, 2-21-1 Osawa, Mitaka, Tokyo 181-8588, Japan}
\newcommand{\ctio}{Cerro Tololo Inter-American Observatory, Casilla 603, La Serena, Chile}
\newcommand{\noirlab}{NOIRLab/Cerro Tololo Inter-American Observatory, Casilla 603, La Serena, Chile}
\newcommand{\nexsci}{Caltech IPAC -- NASA Exoplanet Science Institute 1200 E. California Ave, Pasadena, CA 91125, USA}
\newcommand{\ucsc}{Department of Astronomy and Astrophysics, University of
California, Santa Cruz, CA 95064, USA}
\newcommand{\gsfc}{Exoplanets and Stellar Astrophysics Laboratory, Code 667, NASA Goddard Space Flight Center, Greenbelt, MD 20771, USA}
\newcommand{\sgtinc}{SGT, Inc./NASA AMES Research Center, Mailstop 269-3, Bldg T35C, P.O. Box 1, Moffett Field, CA 94035, USA}
\newcommand{\chile}{Center of Astro-Engineering UC, Pontificia Universidad Cat\'olica de Chile, Av. Vicu\~{n}a Mackenna 4860, 7820436 Macul, Santiago, Chile}
\newcommand{\Pontificia}{Facultad de Ingeniería y Ciencias, Universidad Adolfo Ib\'a\~nez, Av. Diagonal las Torres 2640, Pe\~nalol\'en, Santiago, Chile}
\newcommand{\Millennium}{Millennium Institute for Astrophysics, Chile}
\newcommand{\maxplank}{Max-Planck-Institut f\"ur Astronomie, K\"onigstuhl 17, Heidelberg 69117, Germany}
\newcommand{\utdallas}{Department of Physics, The University of Texas at Dallas, 800 West
Campbell Road, Richardson, TX 75080-3021 USA}
\newcommand{\MauryLewin}{Maury Lewin Astronomical Observatory, Glendora, CA 91741, USA}
\newcommand{\umbc}{University of Maryland, Baltimore County, 1000 Hilltop Circle, Baltimore, MD 21250, USA}
\newcommand{\osu}{Department of Astronomy, The Ohio State University, 140 West 18th Avenue, Columbus, OH 43210, USA}
\newcommand{\MITAA}{Department of Aeronautics and Astronautics, MIT, 77 Massachusetts Avenue, Cambridge, MA 02139, USA}
\newcommand{\openu}{School of Physical Sciences, The Open University, Milton Keynes MK7 6AA, UK}
\newcommand{\swarthmore}{Department of Physics and Astronomy, Swarthmore College, Swarthmore, PA 19081, USA}
\newcommand{\seti}{SETI Institute, Mountain View, CA 94043, USA}
\newcommand{\lehigh}{Department of Physics, Lehigh University, 16 Memorial Drive East, Bethlehem, PA 18015, USA}
\newcommand{\utah}{Department of Physics and Astronomy, University of Utah, 115 South 1400 East, Salt Lake City, UT 84112, USA}
\newcommand{\USNA}{Department of Physics, United States Naval Academy, 572C Holloway Rd., Annapolis, MD 21402, USA}
\newcommand{\DTM}{Department of Terrestrial Magnetism, Carnegie Institution for Science, 5241 Broad Branch Road, NW, Washington, DC 20015, USA}
\newcommand{\UPenn}{The University of Pennsylvania, Department of Physics and Astronomy, Philadelphia, PA, 19104, USA}
\newcommand{\montana}{Department of Physics and Astronomy, University of Montana, 32 Campus Drive, No. 1080, Missoula, MT 59812 USA}
\newcommand{\psu}{Department of Astronomy \& Astrophysics, The Pennsylvania State University, 525 Davey Lab, University Park, PA 16802, USA}
\newcommand{\psust}{Center for Exoplanets and Habitable Worlds, The Pennsylvania State University, 525 Davey Lab, University Park, PA 16802, USA}
\newcommand{\Kutztown}{Department of Physical Sciences, Kutztown University, Kutztown, PA 19530, USA}
\newcommand{\udel}{Department of Physics \& Astronomy, University of Delaware, Newark, DE 19716, USA}
\newcommand{\Westminster}{Department of Physics, Westminster College, New Wilmington, PA 16172}
\newcommand{\steward}{Department of Astronomy and Steward Observatory, University of Arizona, Tucson, AZ 85721, USA}
\newcommand{\saao}{South African Astronomical Observatory, PO Box 9, Observatory, 7935, Cape Town, South Africa}
\newcommand{\salt}{Southern African Large Telescope, PO Box 9, Observatory, 7935, Cape Town, South Africa}
\newcommand{\ssl}{Societ\`{a} Astronomica Lunae, Italy}
\newcommand{\spot}{Spot Observatory, Nashville, TN 37206, USA}
\newcommand{\txamGP}{George P.\ and Cynthia Woods Mitchell Institute for Fundamental Physics and Astronomy, Texas A\&M University, College Station, TX77843 USA}
\newcommand{\txam}{Department of Physics and Astronomy, Texas A\&M university, College Station, TX 77843 USA}
\newcommand{\txammul}{Munnerlyn Astronomical Instrumentation Laboratory, Department of Physics \& Astronomy, Texas A\&M university, College Station, TX 77843 USA}
\newcommand{\wellesley}{Department of Astronomy, Wellesley College, Wellesley, MA 02481, USA}
\newcommand{\Wesleyan}{Astronomy Department and Van Vleck Observatory, Wesleyan University, Middletown, CT 06459, USA}
\newcommand{\inaf}{INAF -- Osservatorio Astronomico di Padova, Vicolo dell'Osservatorio 5, I-35122, Padova, Italy}
\newcommand{\byu}{Department of Physics and Astronomy, Brigham Young University, Provo, UT 84602, USA}
\newcommand{\Hazelwood}{Hazelwood Observatory, Churchill, Victoria, Australia}
\newcommand{\pest}{Perth Exoplanet Survey Telescope}
\newcommand{\Winer}{Winer Observatory, PO Box 797, Sonoita, AZ 85637, USA}
\newcommand{\icpo}{Ivan Curtis Private Observatory}
\newcommand{\elsauce}{El Sauce Observatory, Chile}
\newcommand{\crow}{Atalaia Group \& CROW Observatory, Portalegre, Portugal}
\newcommand{\dfus}{Dipartimento di Fisica ``E.R.Caianiello'', Universit\`a di Salerno, Via Giovanni Paolo II 132, Fisciano 84084, Italy}
\newcommand{\indfn}{Istituto Nazionale di Fisica Nucleare, Napoli, Italy}
\newcommand{\sotes}{Gabriel Murawski Private Observatory (SOTES)}
\newcommand{\lco}{Las Cumbres Observatory Global Telescope, 6740 Cortona Dr., Suite 102, Goleta, CA 93111, USA}
\newcommand{\ucsb}{Department of Physics, University of California, Santa Barbara, CA 93106-9530, USA}
\newcommand{\yale}{Department of Astronomy, Yale University, 52 Hillhouse Avenue, New Haven, CT 06511, USA}
\newcommand{\eso}{European Southern Observatory, Alonso de C\'ordova 3107, Vitacura, Casilla 19001, Santiago, Chile}
\newcommand{\stsci}{Space Telescope Science Institute, Baltimore, MD 21218, USA}
\newcommand{\keele}{Astrophysics Group, Keele University, Staffordshire ST5 5BG, UK}
\newcommand{\gsfcsellers}{GSFC Sellers Exoplanet Environments Collaboration, NASA Goddard Space Flight Center, Greenbelt, MD 20771 }
\newcommand{\usno}{U.S. Naval Observatory, Washington, DC 20392, USA}
\newcommand{\kansas}{Department of Physics and Astronomy, University of Kansas, 1251 Wescoe Hall Dr., Lawrence, KS 66045, USA}
\newcommand{\gmu}{George Mason University, 4400 University Drive MS 3F3, Fairfax, VA 22030, USA}
\newcommand{\unsw}{Exoplanetary Science at UNSW, School of Physics, UNSW Sydney, NSW 2052, Australia}
\newcommand{\sifa}{School of Physics, Sydney Institute for Astronomy (SIfA), The University of Sydney, NSW 2006, Australia}
\newcommand{\nanjing}{School of Astronomy and Space Science, Key Laboratory of Modern Astronomy and Astrophysics in Ministry of Education, Nanjing University, Nanjing 210046, Jiangsu, China}
\newcommand{\berkely}{Department of Astronomy, University of California Berkeley, Berkeley, CA 94720-3411, USA}
\newcommand{\bhicfa}{Black Hole Initiative at Harvard University, 20 Garden Street, Cambridge, MA 02138, USA}
\newcommand{\Silesian}{Department of Electronics, Electronical Engineering and Microelectronics, Silesian University of Techhnology Akademicka 16, 44-100 Gliwice, Poland}
\newcommand{\Patashnick}{Patashnick Voorheesville Observatory, Voorheesville, NY 12186, USA}
\newcommand{\austincollege}{Physics Department, Austin College, 900 North Grand Avenue, Sherman TX 75090, USA}
\newcommand{\Tsinghua}{Department of Astronomy, Tsinghua University, Beijing 100084, China}
\newcommand{\Tsinghuaschool}{Tsinghua International School, Beijing 100084, China}
\newcommand{\chinaNAO}{National Astronomical Observatories, Chinese Academy of Sciences, 20A Datun Road, Chaoyang District, Beijing 100012, China}
\newcommand{\Tautenburg}{Th{\"u}ringer Landessternwarte Tautenburg, Sternwarte 5, 07778 Tautenburg, Germany}
\newcommand{\brierfield}{Brierfield Observatory, New South Wales, Australia}
\newcommand{\Indiana}{Indiana University Department of Astronomy, SW319, 727 E 3rd Street, Bloomington, IN 47405 USA}
\newcommand{\wisconsin}{Department of Astronomy, University of Wisconsin-Madison, Madison, WI 53706, USA}
\newcommand{\protologic}{Proto-Logic Consulting LLC, Washington, DC 20009, USA}
\newcommand{\ASTRAVEO}{ASTRAVEO LLC, PO Box 1668, MA 01931}
\newcommand{\TJHS}{Thomas Jefferson High School, 6560 Braddock Rd, Alexandria, VA 22312 USA}
\newcommand{\ucatchile}{Instituto de Astrof\'isica, Facultad de F\'isica, Pontificia Universidad Cat\'olica de Chile}
\newcommand{\lasa}{Liberal Arts and Science Academy, Austin, Texas 78724, USA}
\newcommand{\gemini}{Gemini Observatory/NSF’s NOIRLab, 670 N. A’ohoku Place, Hilo, HI, 96720, USA}
\newcommand{\umd}{Department of Astronomy, University of Maryland, College Park, College Park, MD}
\newcommand{\ucscchile}{Departamento de Matem\'atica y F\i'sica Aplicadas, Universidad Cat\'olica de la Sant\'isima Concepci\'on, Alonso de Rivera 2850, Concepci\'on, Chile}

\newcommand{\eberly}{\altaffiliation{Eberly Research Fellow}}
\newcommand{\torres}{\altaffiliation{Juan Carlos Torres Fellow}}
\newcommand{\sagan}{\altaffiliation{NASA Sagan Fellow}}
\newcommand{\bernoulli}{\altaffiliation{Bernoulli fellow}}
\newcommand{\gruber}{\altaffiliation{Gruber fellow}}
\newcommand{\kavli}{\altaffiliation{Kavli Fellow}}
\newcommand{\peg}{\altaffiliation{51 Pegasi b Fellow}}
\newcommand{\pappalardo}{\altaffiliation{Pappalardo Fellow}}
\newcommand{\hubble}{\altaffiliation{NASA Hubble Fellow}}
\newcommand{\nsf}{\altaffiliation{National Science Foundation Graduate Research Fellow}}

\correspondingauthor{Erica Thygesen} 
\email{thygesen@msu.edu}

\author[0000-0002-9165-6245]{Erica Thygesen} 
\affiliation{\msu}

\author[0000-0003-0941-4181]{Jessica A. Ranshaw} 
\affiliation{\Indiana}

\author[0000-0001-8812-0565]{Joseph E. Rodriguez} 
\affiliation{\msu}

\author[0000-0001-7246-5438]{Andrew Vanderburg} 
\affiliation{\MIT}

\author[0000-0002-8964-8377]{Samuel N. Quinn} 
\affiliation{\cfa}

\author[0000-0003-3773-5142]{Jason D. Eastman}  
\affiliation{\cfa}

\author[0000-0001-6637-5401]{Allyson Bieryla} 
\affiliation{\cfa}

\author[0000-0001-9911-7388]{David W. Latham} 
\affiliation{\cfa}

\author{Roland K. Vanderspek}
\affiliation{\MIT}

\author[0000-0002-4715-9460]{Jon M. Jenkins}
\affiliation{\ames}

\author[0000-0003-1963-9616]{Douglas A. Caldwell}
\affiliation{\seti}
\affiliation{\ames}

\author[0000-0002-4404-5505]{Mma Ikwut-Ukwa} 
\affiliation{\cfa}

\author[0000-0001-8020-7121]{Knicole D.\ Col\'on} 
\affiliation{\gsfc}

\author{Jessie Dotson} 
\affiliation{\ames}

\author[0000-0002-3385-8391]{Christina Hedges} 
\affiliation{\ames}

\author[0000-0001-6588-9574]{Karen A.\ Collins}
\affiliation{\cfa}

\author[0000-0002-2830-5661]{Michael L. Calkins} 
\affiliation{\cfa}

\author{Perry Berlind} 
\affiliation{\cfa}

\author[0000-0002-9789-5474]{Gilbert A. Esquerdo} 
\affiliation{\cfa}








\shorttitle{The \ktwo\ \& \TESS\ Synergy II}
\shortauthors{Thygesen et al.}

\begin{abstract}

The legacy of NASA's \ktwo mission has provided hundreds of transiting exoplanets that can be revisited by new and future facilities for further characterization, with a particular focus on studying the atmospheres of these systems. However, the majority of \ktwo-discovered exoplanets have typical uncertainties on future times of transit within the next decade of greater than four hours, making observations less practical for many upcoming facilities. Fortunately, NASA's Transiting Exoplanet Survey Satellite (\tess) mission is reobserving most of the sky, providing the opportunity to update the ephemerides for $\sim$300 \ktwo systems. In the second paper of this series, we reanalyze 26 single-planet, \ktwo-discovered systems that were observed in the \tess primary mission by globally fitting their \ktwo and \tess lightcurves (including extended mission data where available), along with any archival radial velocity measurements. As a result of the faintness of the \ktwo\ sample, 13 systems studied here do not have transits detectable by \tess. In those cases, we re-fit the \ktwo lightcurve and provide updated system parameters. For the 23 systems with $M_* \gtrsim 0.6~ M_\odot$, we determine the host star parameters using a combination of {\it Gaia} parallaxes, Spectral Energy Distribution (SED) fits, and MESA Isochrones and Stellar Tracks (MIST) stellar evolution models. Given the expectation of future \tess\ extended missions, efforts like the \ktwo\ \& \tess\ Synergy project will ensure the accessibility of transiting planets for future characterization while leading to a self-consistent catalog of stellar and planetary parameters for future population efforts.

\end{abstract}

\section{Introduction}
\label{sec:intro}

The past two decades have been fruitful for exoplanet discovery, with over 5000 exoplanets confirmed to date\footnote{\url{https://exoplanetarchive.ipac.caltech.edu/}}. While new discoveries are still being made, we are simultaneously venturing into an era of exploring known systems in further detail, with a variety of dedicated efforts for exoplanet characterization. Facilities that are operational or expected to be online in the next decade such as JWST \citep{Gardner:2006,Beichman:2020}, 39 m European Southern Observatory Extremely Large Telescope (ELT; \citealt{Udry-ELT:2014}), Nancy Grace Roman Space Telescope (e.g. \citealt{Carrion-Gonzalez:2021}), Giant Magellan Telescope \citep{Johns:2012GMT} and Atmospheric Remote-sensing Infrared Exoplanet Large-survey (ARIEL; \citealt{Tinetti-ARIEL:2018,Tinetti-ARIEL:2021}) will provide key information about the atmospheres of exoplanets, and insight into their formation and evolutionary processes. However, these ongoing and future endeavors to reobserve known transiting exoplanets heavily rely on precisely knowing the transit time, which is challenged by the degradation of the ephemeris over time.

\begin{figure*}[th]
    \centering
    \includegraphics[width=\linewidth]{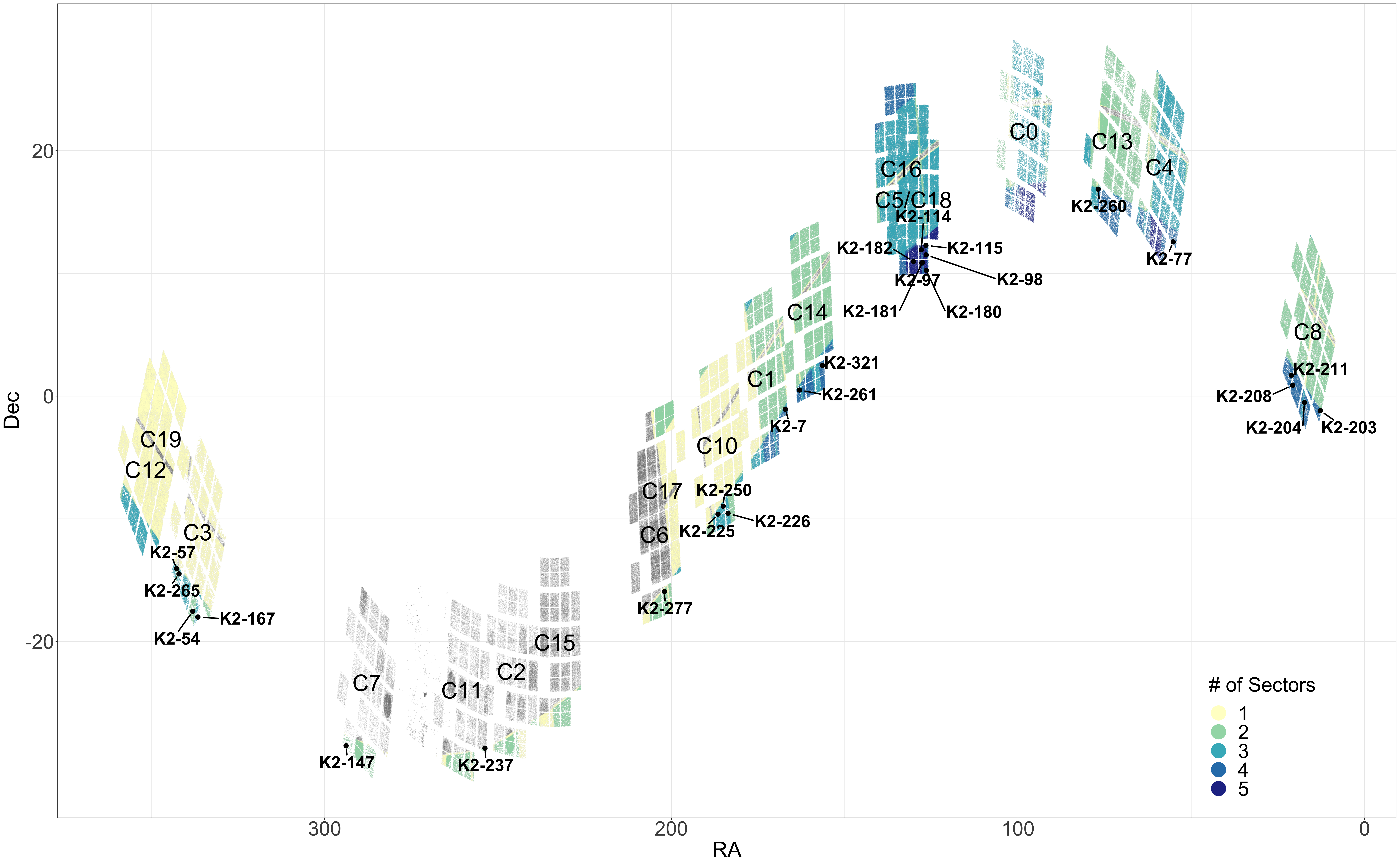}
    \caption{Overlap between \ktwo campaigns and \tess sectors. The number of times each \ktwo target was observed in \tess sectors is indicated by the color, with gray indicating no \tess overlap as of Sector 46. The systems analyzed in this study are labeled.}
    \label{fig:K2_TESS_overlap}
\end{figure*}

Most exoplanets and candidates found to date were originally discovered by the \Kepler mission \citep{Borucki:2010}. \Kepler was launched in 2009 with the goal of understanding the demographics of transiting exoplanets. This mission was a success, having discovered $\sim$2700 confirmed planets with a further $\sim$2000 candidates\footnote{\url{https://exoplanetarchive.ipac.caltech.edu/}}, in addition to advancing our understanding of the host stars they orbit (e.g. \citealt{Bastien:2013,Berger:2020a,Berger:2020b}). However, by May of 2013 two of the four reaction wheels on the spacecraft had failed, severely limiting the pointing of \Kepler, threatening to end the mission. A solution was conceived to point the spacecraft at the ecliptic to reduce torque from Solar radiation pressure, so that the remaining two reaction wheels, along with the thrusters, could maintain sufficient stability. This saw \Kepler successfully reborn as the \Ktwo mission \citep{Howell:2014}. While \Kepler continuously pointed at one region of sky, the necessity of \Ktwo being aimed along the ecliptic opened up an opportunity to study different populations of stars. \Ktwo continued on the path of exoplanet discovery, with currently $\sim$500 confirmed planets and another $\sim$1000 candidates found by the time the spacecraft retired in 2018 when fuel for the thrusters ran out \citep{Vanderburg:2016b,Zink:2021,Kruse:2019,Pope:2016,Livingston:2018-K2C5-8,Crossfield:2016,Dattilo:2019}.

Unfortunately, many of the known planets discovered by the \Ktwo mission have not been reobserved since their discovery, leading to future transit time uncertainties of many hours \citep{Ikwut-Ukwa-SynergyI:2020}. This has recently changed with the launch of NASA's \textit{Transiting Exoplanet Survey Satellite} (\tess) mission in 2018 \citep{Ricker:2015}, the successor to the \Kepler and \Ktwo missions. The two-year primary mission of \tess aimed to observe more than 200,000 stars at two-minute cadence across $\sim$75\% of the sky. To date, \tess has found $\sim$280 confirmed planets and another $\sim$6100 candidates\footnote{\url{https://nexsci.caltech.edu/}}. Even though \ktwo targeted the ecliptic plane and the \tess primary mission only skimmed the edges of some \ktwo fields, there are $\sim$30 systems that were observed by both (single- and multi-planet systems). This provides an opportunity to begin updating the ephemerides and parameters of \ktwo systems that have been reobserved by \tess. The first extended mission of \tess began during 2020, and includes sectors dedicated to the ecliptic plane, providing more substantial overlap of a further $\sim$300 systems with the \ktwo fields\footnote{\url{https://heasarc.gsfc.nasa.gov/docs/tess/the-tess-extended-mission.html}} (Figure \ref{fig:K2_TESS_overlap}). With \tess scheduled to reobserve nearly the entire sky during its extended missions, it will be a useful tool for refreshing the ephemerides of thousands of transiting exoplanets.

Currently, many known exoplanets do not have sufficiently accurate projected transit times to plan observations with future missions. Even \tess ephemerides will need to be updated as most \tess planets will have transit time uncertainties exceeding 30 minutes in the era of JWST \citep{Dragomir:2020}. With the wealth of data coming from ongoing surveys like \tess and the ability to follow up many planets with small aperture ($<$1 m) telescopes \citep{Collins:2018}, many efforts have begun to keep the ephemerides of transiting planets from going stale, like the ExoClock Project \citep{Kokori-ExoclockI:2021,Kokori-ExoclockII:2022} for future ARIEL targets and the \ktwo \& \tess Synergy \citep{Ikwut-Ukwa-SynergyI:2020}. Ephemeris refinement programs focused on citizen science \citep{Zellem:2019,Zellem:2020} and high-school students (e.g. ORBYTS; \citealt{Edwards:2019,Edwards:2020,Edwards:2021}) also provide opportunities to actively engage the public while contributing to an essential aspect of future exoplanet characterization. These efforts will be key to making a large number of systems accessible for future facilities.

A continual renewal of ephemerides also presents an opportunity to create self-consistent catalogs of exoplanets and their parameters, which not only helps to plan for future missions, but also allows for appropriate population studies using data that have been uniformly prepared. While the vast amount of data available per system makes this a challenge, the advent of new exoplanet fitting suites to globally analyze large quantities of data, like {\tt Juliet} \citep{Espinoza-Juliet:2019}, {\tt EXOFASTv2} \citep{Eastman:2013,Eastman:2017,Eastman:2019}, {\tt Allesfitter} \citep{allesfitter-paper} and {\tt exoplanet} \citep{Foreman-Mackey-exoplanet:2021}, has made it possible to individually model the available observations for a large sample of exoplanetary systems. These types of studies are necessary to uncover large-scale trends or mechanisms that may play important roles in planet formation and evolution. A renowned example is the radius valley of small planets \citep{Fulton:2017}, which was achieved through more accurate and consistent handling of host star parameters for over 2000 planets from the California-\textit{Kepler} Survey. 

A case study for updating \ktwo ephemerides and system parameters with new \tess data was presented in the first paper of this series \citep{Ikwut-Ukwa-SynergyI:2020}, where four \Ktwo-discovered systems (K2-114, K2-167, K2-237 and K2-261) were reanalyzed by performing global fits using \Ktwo and \tess lightcurves. This resulted in the uncertainties for the transit times of all four planets being reduced from multiple hours to between 3-26 minutes (at a one sigma level) throughout the expected span of the JWST primary mission, showcasing the value of combining the \Ktwo and \tess data. We continue this work by reanalyzing a sample of 26 single-planet systems observed with \Ktwo and the primary \tess mission (including refitting the original four systems for consistency), while also making use of archival radial velocities, \textit{Gaia} parallaxes and any currently available lightcurves from the \tess extended mission. We focus on previously-confirmed single-planet systems, but future papers in this series are expected to reanalyze all \ktwo systems (including multi-planet systems) as part of an ongoing \tess guest investigator program (G04205, PI Rodriguez). Updated transit times will be made available to the community throughout this series through the Exoplanet Follow-up Observing Program for TESS (ExoFOP)\footnote{\url{https://exofop.ipac.caltech.edu/tess/}}.

In \S \ref{sec:obs} we describe how we obtained and prepared the data used in our global fits. \S \ref{sec:GlobalModel} outlines how we ran the {\tt EXOFASTv2} analysis, and \S \ref{sec:discussion} presents our results along with any peculiarities for specific systems. Our conclusions are summarized in \S \ref{sec:conclusion}.

\section{Observations and Archival Data} \label{sec:obs}

\begin{figure}
    \centering
    \includegraphics[width=\linewidth]{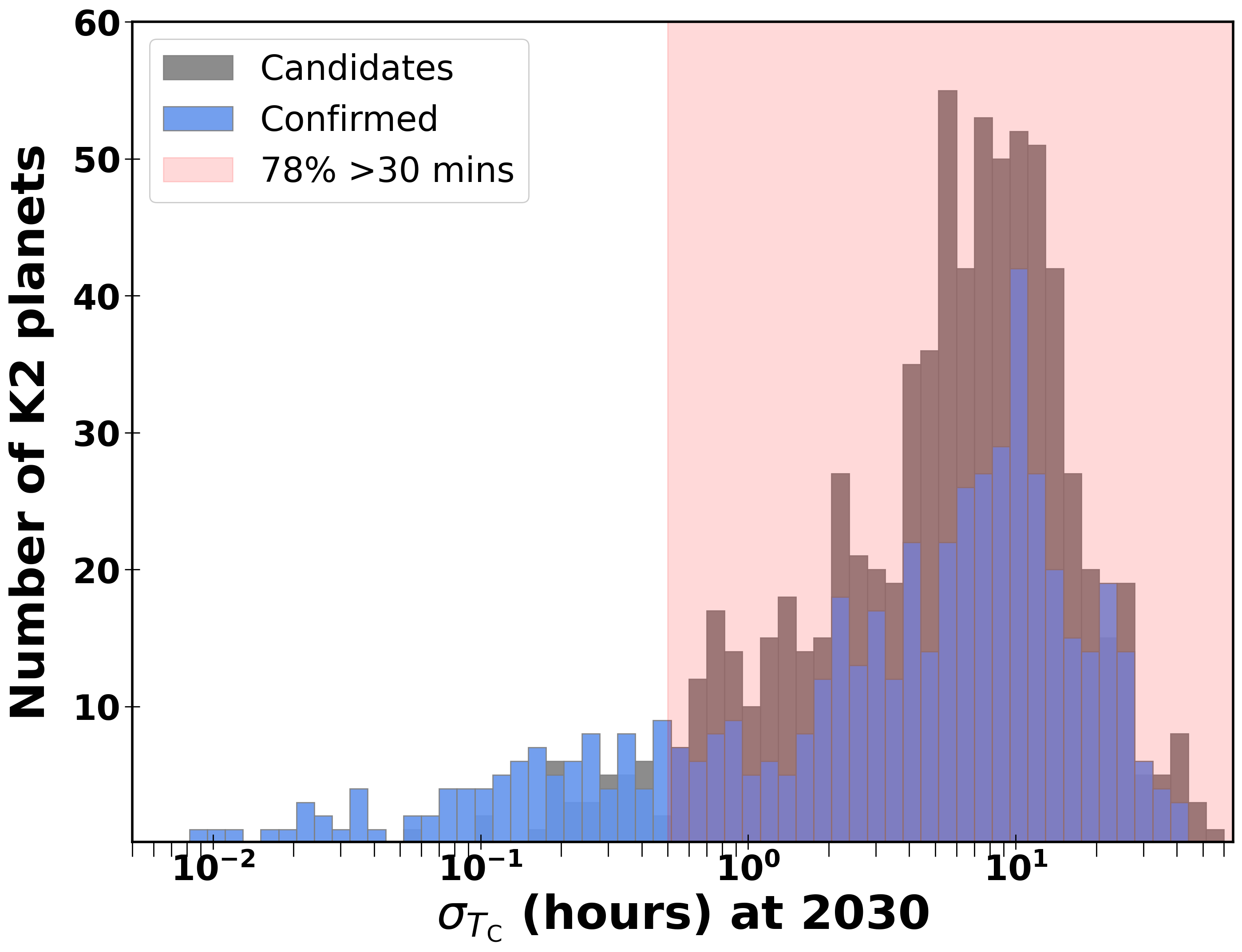}
    \caption{Uncertainty of the transit time ($\sigma_{T_c}$) for \ktwo candidate and confirmed planets at the year 2030, based on the discovery ephemeris. The majority of planets have uncertainties greater than 30 minutes (indicated by the red region) in the era of JWST, making these challenging to reobserve. Values taken from the NASA Exoplanet Archive (NEA) default parameter sets.}
    \label{fig:k210year}
\end{figure}

Given that most known \ktwo-discovered exoplanet systems will have uncertainties larger than 30 minutes (see Figure \ref{fig:k210year}), we take advantage of the high-quality data obtained with \Ktwo and \tess, simultaneously fitting the photometry and archival spectroscopy to update system parameters for 26 \ktwo systems. Here we describe the techniques used to obtain and process \ktwo and \tess lightcurves, as well as radial velocities from existing literature.

\subsection{{\it K2} Photometry}
\label{sec:k2}
Each of these stars was observed by the \Kepler\ spacecraft during its \ktwo extended mission \citep{Howell:2014}. During \ktwo, the spacecraft's roll angle drifted significantly due to the failure of two reaction wheels, which introduced significant systematic errors into its lightcurves\footnote{The two remaining reaction wheels onboard \ktwo could control the position of the telescope’s boresight, but the roll angle could only be controlled by occasional firing of the thrusters about every 6 hours as radiation pressure caused the telescope to slowly roll about its long axis.}. Over the course of the mission, a number of different techniques and methods were developed to mitigate these errors (e.g. \citealt{Aigrain:2016,Barros:2016,Luger:2016,Lund:2015,Pope:2019}). In this work, we used the methods of \citet{Vanderburg:2014} and \citet{Vanderburg:2016b} to derive a rough systematics correction. In brief, these methods involve extracting raw lightcurves from a series of 20 different photometric apertures, correlating short timescale variations in the raw lightcurves with the spacecraft's roll angle (which changes rapidly due to K2's unstable pointing), and subtracting variability correlated with the spacecraft's roll angle. The process of correlating and subtracting variability correlated with the roll angle is performed iteratively until the only remaining variations in the lightcurve are unrelated to the spacecraft's roll. Finally, we select the aperture that produces the most precise lightcurve among the 20 originally extracted. Then, we refined the systematics correction by simultaneously fitting the transits for each planet along with the systematics correction and low-frequency stellar variability, prior to the final global fit. Most of the data we analyzed were collected in 30-minute long-cadence data, but when available, we analyzed 1-minute short-cadence exposures for better time sampling. For all systems, we only included out-of-transit data from one full transit duration before and after each transit. This is to optimize the balance between having enough data points to establish the baseline flux of the star and lengthening the runtime of the fits due to having more data.

\subsection{{\it TESS} Photometry}
\label{sec:TESS}

While all 26 systems were initially observed by \tess in the primary mission, each was reobserved in at least one sector of the first extended mission. We therefore included \tess lightcurves from the primary and extended missions up to and including Sector 46 (as of February 1, 2022). This was the final sector dedicated to the ecliptic plane for the first extended mission. Future efforts in this series will analyze systems that were first observed by \tess during the first extended mission and beyond.

We used the Python package \textit{Lightkurve} \citep{Lightkurve:2018} to retrieve \tess lightcurves from the Mikulski Archive for Space Telescopes (MAST). Three systems within the footprint of the \tess primary mission (K2-42, K2-132/TOI 2643 and K2-267/TOI 2461) did not have corresponding retrievable lightcurves, which is likely due to being too close to the edge of the detector, so we excluded these from the current analysis. For the \tess lightcurves, we used the Pre-search Data Conditioned Simple Aperture Photometry (PDCSAP) flux, which is the target flux within the optimal \tess aperture that has been corrected for systematics with the PDC module \citep{Stumpe:2012,Stumpe:2014,Smith:2012}. Typically, observations for each sector are processed through the Science Processing Operations Center (SPOC) pipeline at the NASA Ames Research Center \citep{Jenkins:2016}. The SPOC pipeline takes in the raw data and applies corrections for systematics, runs diagnostic tests and identifies transits, resulting in a calibrated lightcurve that can be used for analysis.

\tess science observations are taken at 20-second and 2-minute cadences (the former only becoming available from the first extended mission), while the Full Frame Images (FFIs) are created every 30 minutes during the primary mission, and every 10 minutes since the first extended mission. For our global analysis, (see \S \ref{sec:GlobalModel}) we used the shortest cadence available, preferentially using data processed through SPOC \citep{Jenkins:2016, Caldwell:2020}. The increased timing precision of short cadence observations is only valuable if there is a significant detection of the transit. For this reason, and since \tess is optimized for targets with brighter magnitudes than those of \ktwo, we binned lightcurves observed at 20-second cadence to two minutes to increase signal-to-noise. 

If a TESS-SPOC FFI lightcurve was not available for a particular sector, we extracted the lightcurve using a custom pipeline as described in \cite{Vanderburg:2019}. The pipeline uses a series of 20 apertures from which lightcurves are extracted and corrected for systematic errors from the spacecraft by decorrelating the flux with the mean and standard deviation of the quaternion time series. Dilution from neighbouring stars within the TIC is corrected for within each aperture, which takes into account the TESS pixel response function. The final aperture used for the lightcurve extraction is selected as the one that minimized the scatter in the photometry. Recent efforts have compared this custom pipeline with other FFI pipelines \citep{Rodriguez:2022-Cargo2}, supporting our adoption of this pipeline. The list of available lightcurves (as of February 1, 2022) is shown in Table \ref{tab:obs}.

\begin{table*}[t]
\centering
\scriptsize
\caption{Target list and data used in this analysis.}
\begin{tabular}{lrlrcccccc}

\hline
TIC ID & TOI & KID & EPIC ID & \ktwo Campaign & \multicolumn{2}{c}{\tess Sector} & RV instrument & \ktwo Reference & \tess SNR\\ 
   &   &   &   &   &  (2 min) & (FFI) & & & \\
\hline

53210555 & --- & K2-7 & 201393098 & C1 & {\color{red}9, 36, 45, 46} & --- & --- & 1 & 5.62 \\
12822545 & --- & K2-54$^\dagger$ & 205916793 & C3 & {\color{red}2, 42} & --- & --- & 2 & 1.76\\
146799150 & --- & K2-57 & 206026136 & C3 & {\color{red}2, 29} & --- & --- & 2 & 1.99\\
435339847 & 4544.01 & K2-77 & 210363145 & C4 & 5$^\prime$, 42$^\prime$, 43$^\prime$, 44$^\prime$ & --- & --- & 3 & 13.52 \\
366568760 & 5121.01 & K2-97 & 211351816 & C5, C18 & 7$^\prime$, 44$^\prime$, 45$^\prime$, 46$^\prime$ & --- & LEVY$^1$ (6), HIRES$^2$ (18) & 4 & 16.76 \\
366410512 & 5101.01 & K2-98 & 211391664 & C5, C18 & 7, 34, 44, 45, 46 & --- & FIES$^3$ (4), HARPS$^3$ (4), HARPSN$^3$ (4)& 5 & 20.93\\
366576758 & 514.01 & K2-114 & 211418729 & C5, C18 & 7, 44, 45, 46 & --- & HIRES$^4$ (5) & 6 & 134.03 \\
7020254 & 4316.01 & K2-115 & 211442297 & C5, C18 & 7, 34, 45, 46 & --- & HIRES$^4$ (7) & 6 & 88.19 \\
398275886 & --- & K2-147$^\dagger$ & 213715787 & C7 & {\color{red}27} & {\color{red}13} & --- & 7 & 2.91 \\
69747919 & 1407.01 & K2-167 & 205904628 & C3 & 2, 28, 42, & --- & --- & 3 & 13.82\\
366411016 & 5529.01 & K2-180 & 211319617 & C5, C18 & 34, 44, 45, 46 & 7$^*$ & HARPSN$^5$ (12) & 8 & 12.03 \\
366528389 & --- & K2-181 & 211355342 & C5, C18& {\color{red}7, 44, 45, 46} & --- & --- & 3 & 5.74 \\
366631954 & 5068.01 & K2-182 & 211359660 & C5, C18 & 34, 44, 45, 46 & 7 & HIRES$^6$ (12) & 9 & 32.39 \\
333605244 & --- & K2-203 & 220170303 & C8 & {\color{red}30, 42, 43} & {\color{red}3} & --- & 3 & 3.17 \\
248351386 & --- & K2-204 & 220186645 & C8 & {\color{red}30, 42, 43} & {\color{red}3} & --- & 3 & 5.44 \\
399722652 & --- & K2-208 & 220225178 & C8 & {\color{red}30, 42, 43} & {\color{red}3} & --- & 3 & 4.76 \\
399731211 & --- & K2-211 & 220256496 & C8 & {\color{red}30, 42, 43} & {\color{red}3} & --- & 7 & 2.90 \\
98677125 & --- & K2-225 & 228734900 & C10 & {\color{red}36, 46} & {\color{red}10} & --- & 3 & 2.93 \\
176938958 & --- & K2-226 & 228736155 & C10 & {\color{red}36, 46} & {\color{red}10} & --- & 3 & 3.82 \\
16288184 & 1049.01 & K2-237 & 229426032 & C11 & 12, 39 & --- & CORALIE$^7$ (9), HARPS$^{7,8}$ (4,7), FIES$^8$ (9)& 10 & 129.83 \\
98591691 & --- & K2-250 & 228748826 & C10 & {\color{red}36, 46} & {\color{red}10}  & --- & 11 & 3.79 \\
293612446 & 2466.01 & K2-260 & 246911830 & C13 & 32, 43 & 5$^*$ & FIES$^9$ (18)& 12 & 98.44 \\
281731203 & 685.01 & K2-261 & 201498078 & C14 & 9, 35, 45, 46 & --- & FIES$^9$ (12), HARPS$^9$ (10), HARPSN$^9$ (8)& 12 & 83.58 \\
146364192 & --- & K2-265 & 206011496 & C3 & {\color{red}29, 42} & {\color{red}2} & HARPS$^{10}$ (138)& 13 & 6.01 \\
404421005 & 4628.01 & K2-277 & 212357477 & C6 & 10, 37$^\prime$ & --- & --- & 4 & 8.75 \\
277833995 & 5524.01 & K2-321$^\dagger$ & 248480671 & C14 & 8$^\prime$, 45$^\prime$, 46$^\prime$ & 35$^\prime$ (10min) & --- & 14 & 9.21 \\ \hline
\end{tabular}
\begin{flushleft}
 \footnotesize{ \textbf{\textsc{Notes:}}
 \tess lightcurves taken at 20 second cadence were prioritised, and binned to two minutes. Where short cadence observations were not available, FFIs were used. \tess sectors in which transits had SNR$\leq$7 and thus were too shallow to be recovered are colored red. We incorporated previous RV measurements that were taken from the previous studies listed here. The number in parentheses following the RV instrument indicates the number of measurements. \ktwo references are previous analyses with which we compare our updated ephemerides in \S\ref{sec:discussion}. \\
 $^\dagger$ The host stars in these systems were classed as low mass ($\lesssim 0.6~ M_\odot$), so we did not include the SEDs in the global fits. See \S \ref{sec:GlobalModel} for details. \\
 $^\prime$ The full lightcurves for these were used to ensure the transit was able to be detected. All other lightcurves were sliced as discussed in \S \ref{sec:obs}. \\ 
 $^*$ A custom pipeline was used to extract lightcurves for sectors without TESS-SPOC FFIs as discussed in \S \ref{sec:TESS}. \\
 References for RV measurements: $^1$\cite{Grunblatt:2016}, $^2$\cite{Grunblatt:2018}, $^3$\cite{Barragan:2016}, $^4$\cite{Shporer:2017}, $^5$\cite{Korth:2019}, $^6$\cite{AkanaMurphy:2021}, $^7$\cite{Soto:2018}, $^8$\cite{Smith:2019}, $^9$\cite{Johnson2018}, $^{10}$\cite{Lam:2018} \\
 \ktwo references: 1 - \cite{Montet:2015}, 2 - \cite{Crossfield:2016}, 3 - \cite{Mayo:2018}, 4 - \cite{Livingston:2018-K2C5-8}, 5 - \cite{Barragan:2016}, 6 - \cite{Shporer:2017}, 7 - \cite{Adams:2021}, 8 - \cite{Korth:2019}, 9 - \cite{AkanaMurphy:2021}, 10 - \cite{Soto:2018}, 11 - \cite{Livingston:2018A-K2C10}, 12 - \cite{Johnson:2018}, 13 - \cite{Lam:2018}, \cite{CastroGonzalez:2020}
}
\end{flushleft}
\label{tab:obs}
\end{table*}

After retrieving the \tess lightcurves for our targets, we processed them further for our own analysis, assuming values for transit duration, time of conjunction ($T_c$) and period from the NASA Exoplanet Archive (NEA). To flatten the out-of-transit lightcurve for fitting, we used {\tt keplerspline}\footnote{\url{https://github.com/avanderburg/keplerspline}}, a spline-fitting routine to model and remove any variability from the star or remaining systematics \citep{Vanderburg:2014}. Within {\tt keplerspline}, the spacing between breaks in the spline to handle discontinuities is optimized by minimizing the Bayesian Information Criterion (BIC) for different break points (see \citealt{Shallue:2018} for further methodology). We applied a constant per-point error for the photometry, calculated as the median absolute deviation of the out-of-transit flattened lightcurve, although this error is optimized within our analysis since {\tt EXOFASTv2} fits a jitter term. If any lightcurve had large outliers or features that may influence our transit fit, we used only the data that had no bad quality flags within \textit{Lightkurve} (this was only the case for K2-250 and K2-260). To reduce the individual runtime for each system, we excluded the out-of-transit baseline of the \tess lightcurves from the {\tt EXOFASTv2} fit other than one full transit duration before and after each transit (as with the \ktwo lightcurves). However, for systems whose transits were not readily visually identified in the \tess data (K2-77, K2-97, K2-277 and K2-321; see Table \ref{tab:obs}), we included all out-of-transit photometry to account for any large uncertainties in the time of transits during the \tess epochs.

\subsection{Archival Spectroscopy}
\label{sec:RVs}

\begin{figure*}[h]
	\centering\vspace{.0in}
	\includegraphics[width=0.33\linewidth]{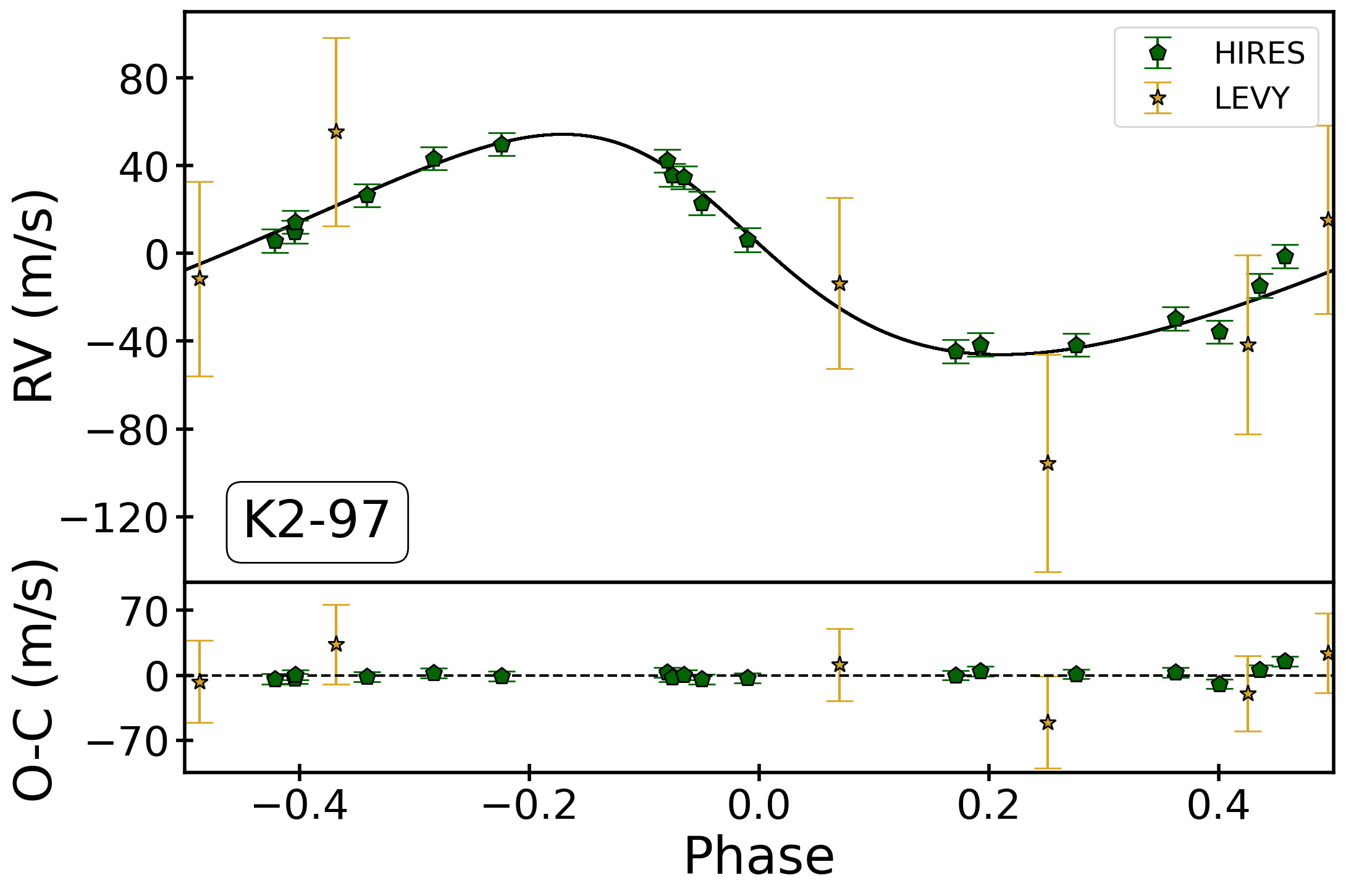}
	\includegraphics[width=0.33\linewidth]{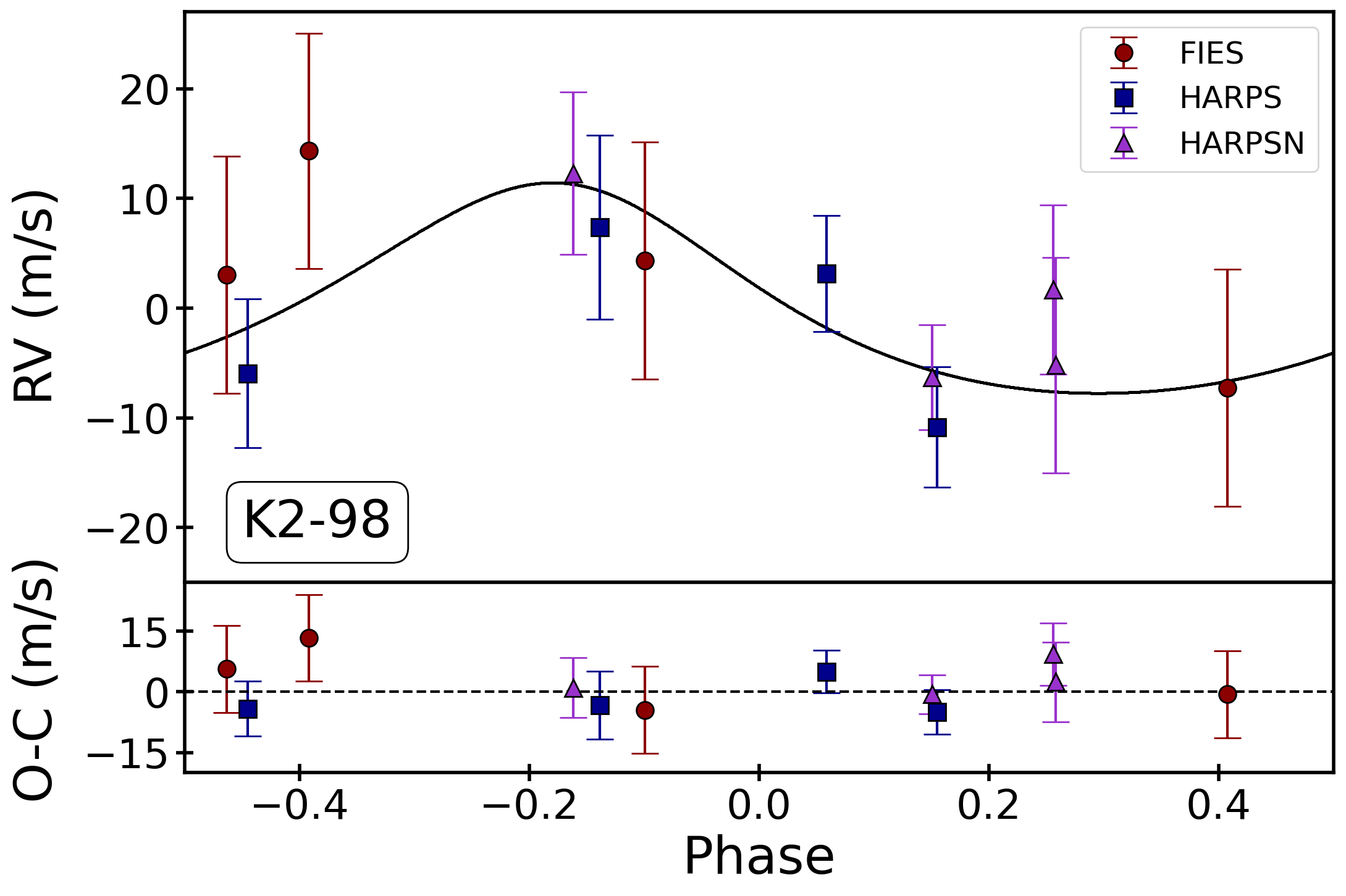}
	\includegraphics[width=0.33\linewidth]{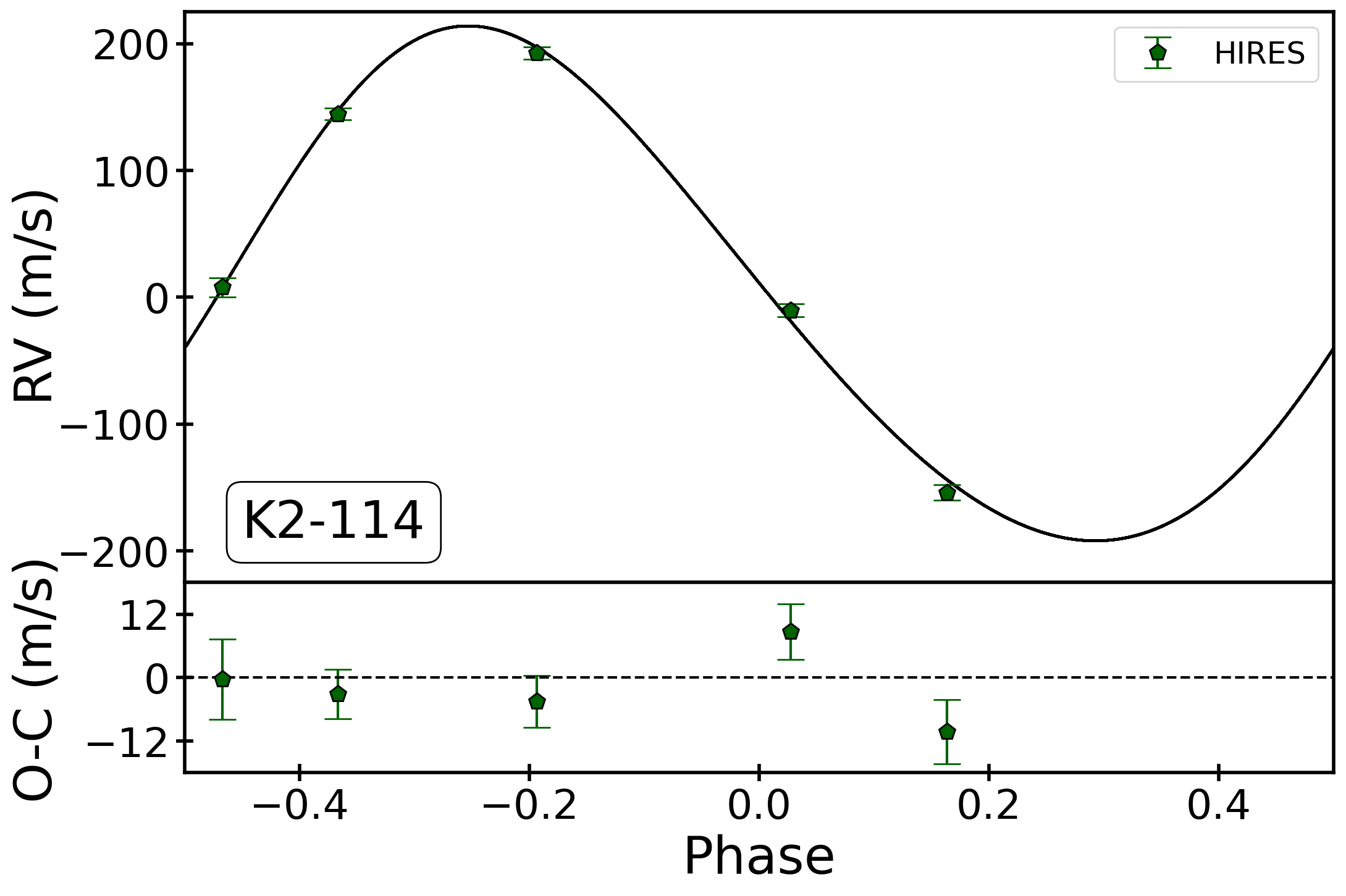}
	\includegraphics[width=0.33\linewidth]{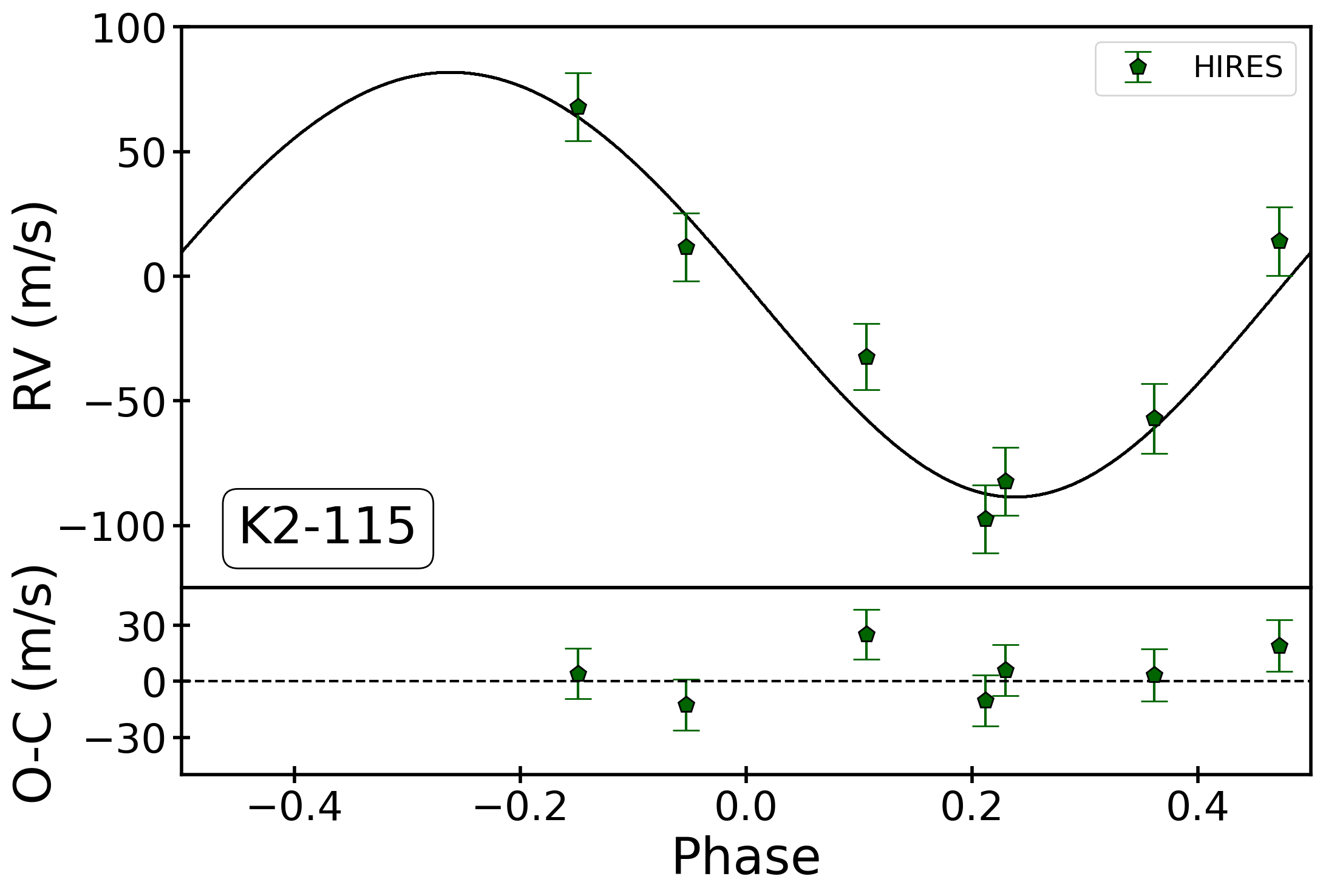}
	\includegraphics[width=0.33\linewidth]{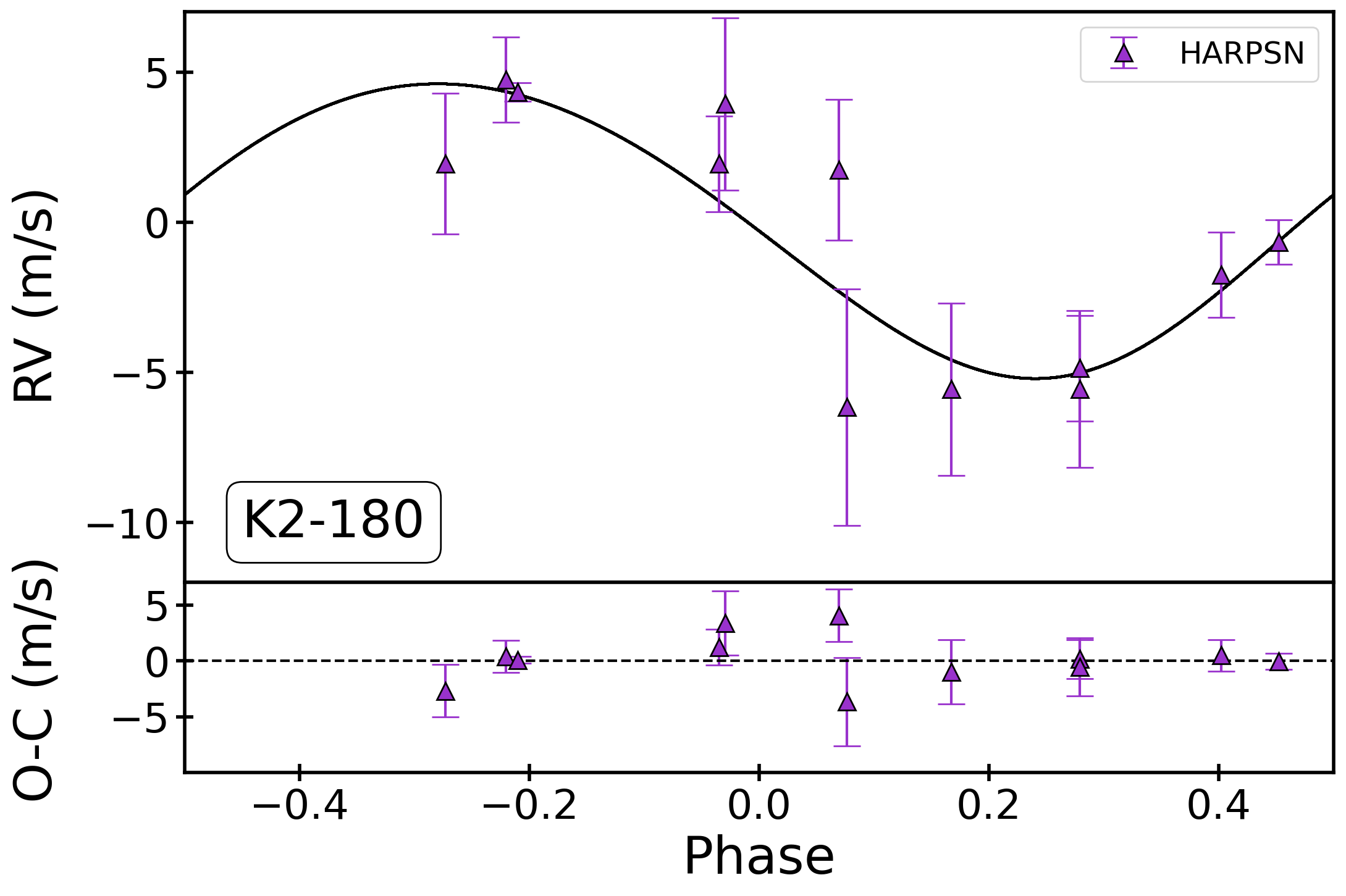}
	\includegraphics[width=0.33\linewidth]{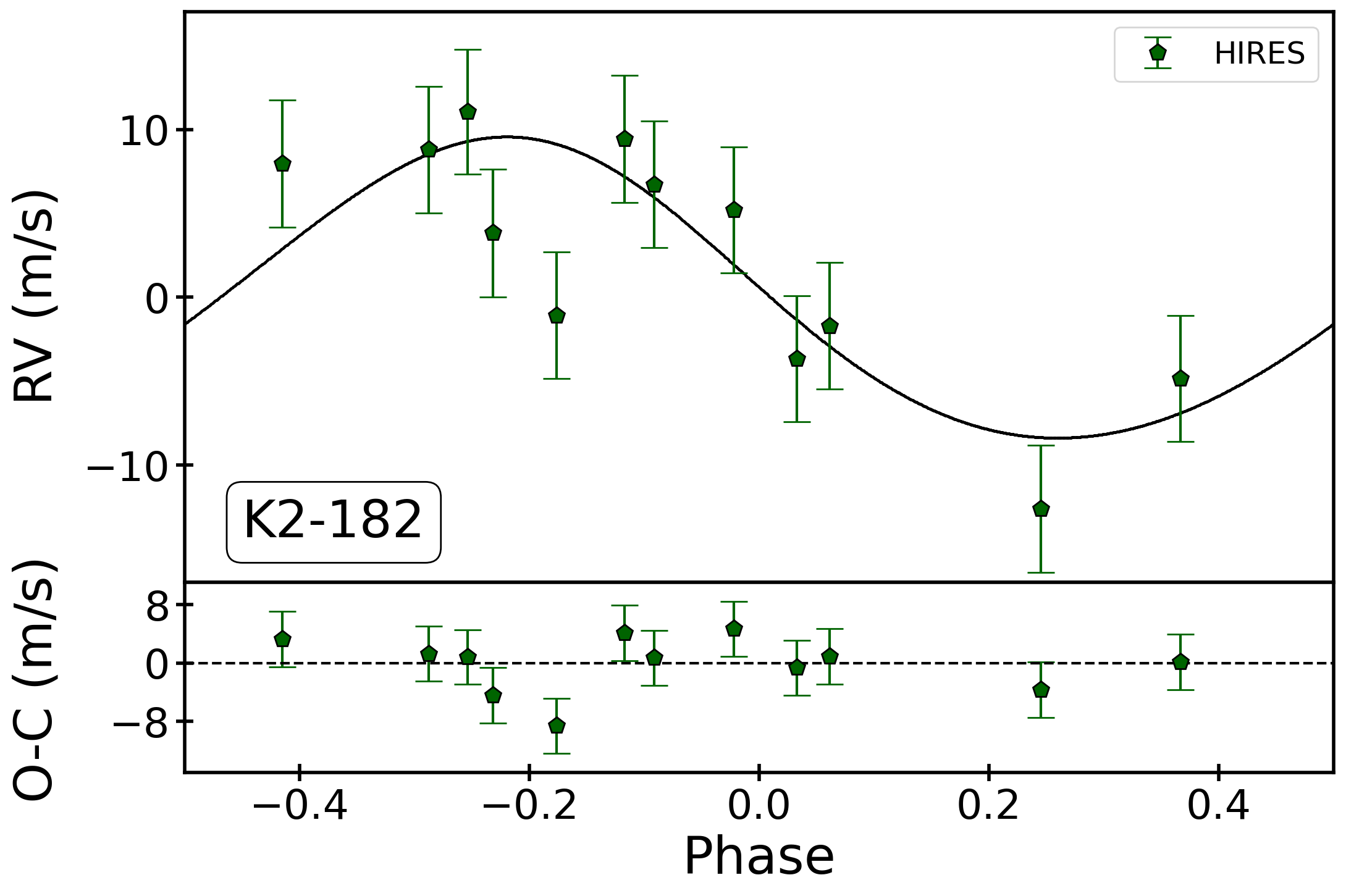}
	\includegraphics[width=0.33\linewidth]{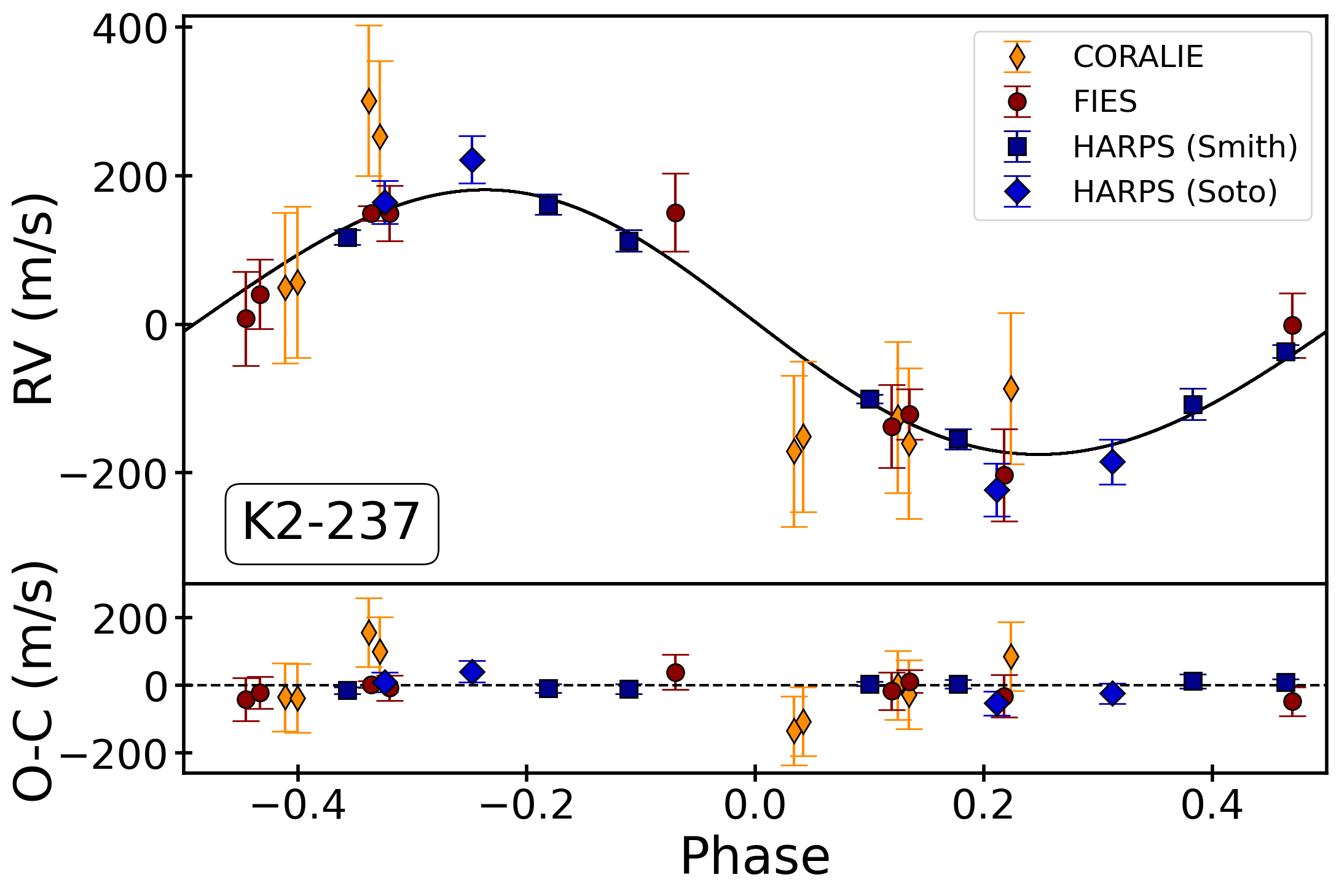}
	\includegraphics[width=0.33\linewidth]{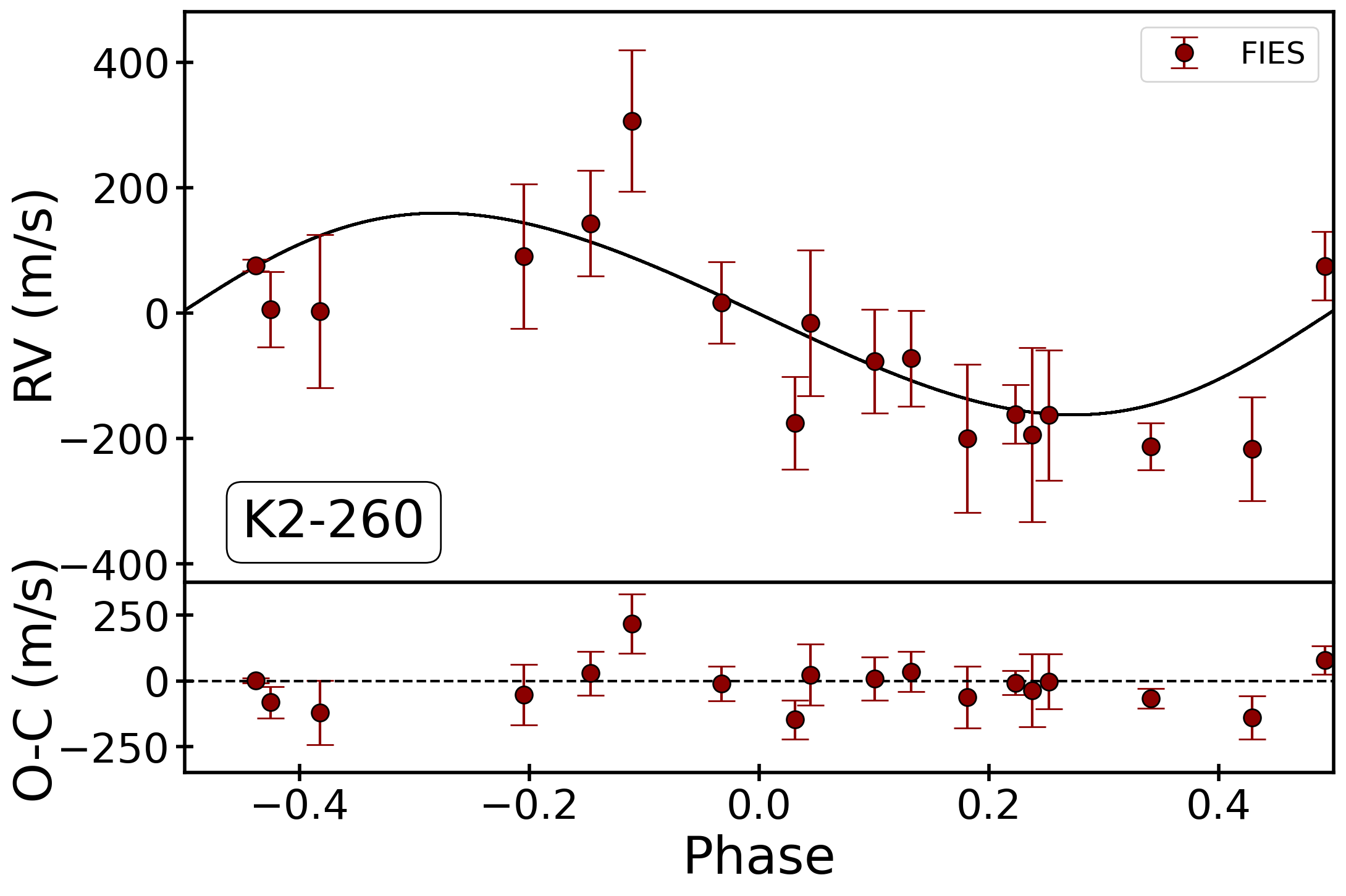}
	\includegraphics[width=0.33\linewidth]{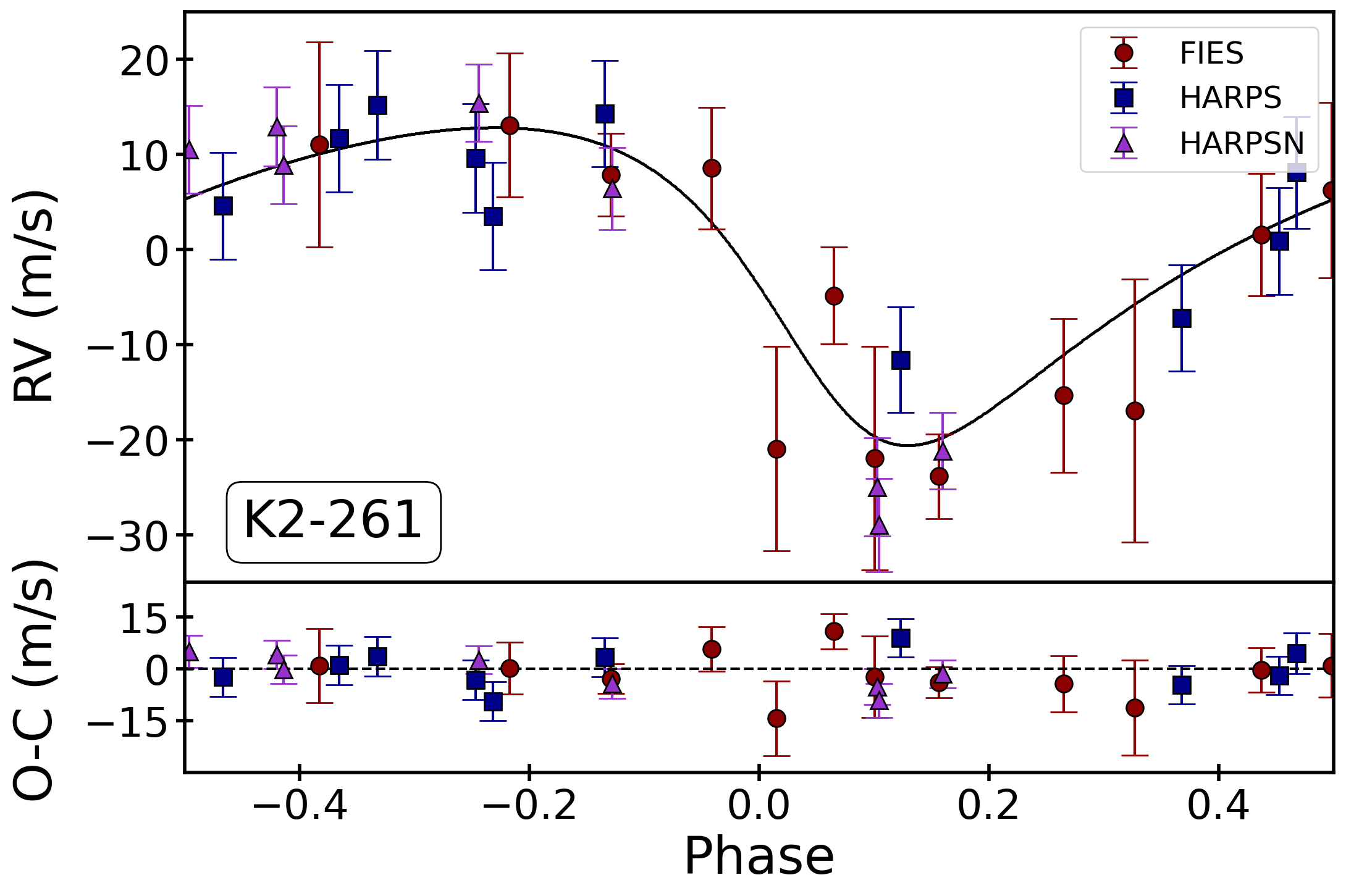}
	\includegraphics[width=0.33\linewidth]{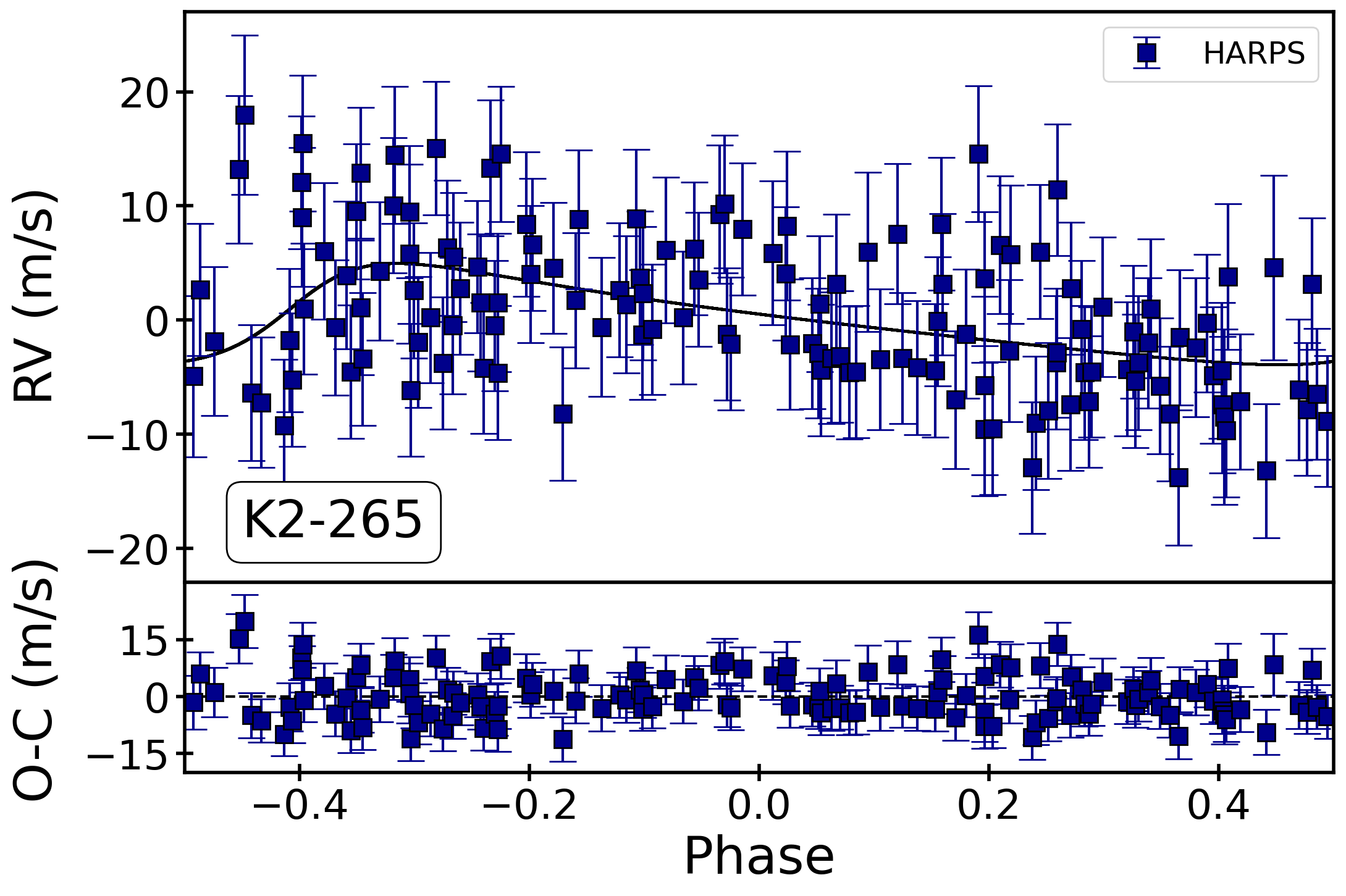}\\
	\caption{Radial velocities for the 10 systems with archival spectroscopic measurements. The best fit model from {\tt EXOFASTv2} is shown in each subplot. Each set of RVs are phased using the best fit period and $T_c$ determined in the fit, and the residuals are shown below each dataset. The references for each set of RVs are listed in Table \ref{tab:obs}.}
	\label{fig:RVs} 
\end{figure*}

We identified spectroscopic observations from the literature for 10 of the 26 total targets (Figure \ref{fig:RVs}; K2-97, K2-98, K2-114, K2-115, K2-180, K2-182, K2-237, K2-260, K2-261 and K2-265; \citealt{Grunblatt:2016,Grunblatt:2018,Barragan:2016,Shporer:2017,Korth:2019,AkanaMurphy:2021,Soto:2018,Smith:2019,Johnson2018,Lam:2018}). We selected data sets with four or more RV measurements to ensure more degrees of freedom in the global fit, thus avoiding overfitting the data. For this reason we do not include RVs for K2-77 \citep{Gaidos:2017} and K2-147 \citep{Hirano:2018a}. Table \ref{tab:obs} lists the analyses from which we obtained each set of RVs that we incorporated in the global analysis (see \S \ref{sec:GlobalModel}). All but one of the systems that have RVs also have significant \tess transits (see \S \ref{sec:GlobalModel}), which is an outcome of spectroscopic measurements preferentially targeting brighter stars. The archival RVs were obtained from the following instruments: the Levy spectrometer on the 2.4m Automated Planet Finder at Lick Observatory, the High Resolution Echelle Spectrometer (HIRES) on the Keck-I Telescope \citep{Vogt:1994}, the FIbre-fed Echelle Spectrograph (FIES) on the 2.56m Nordic Optical Telescope at Roque de los Muchachos Observatory \cite{Frandsen:1999}, the High Accuracy Radial velocity Planet Searcher (HARPS) spectrograph on the 3.6m telescope at La Silla Observatory \citep{Mayor:2003}, HARPS-N on the 3.58m Telescopio Nazionale Galileo at the Roque de los Muchachos Observatory \citep{Consentino:2012}, and the CORALIE spectrograph on the Swiss 1.2m Leonhard Euler Telescope at La Silla Observatory \citep{Queloz:2000}.

If any determination for the host star's metallicity ([Fe/H]) was available, we included it as a prior in the fit to better constrain the host star parameters. For consistency, we used metallicity priors for most of the systems from spectra obtained using the Tillinghast Reflector Echelle Spectrograph (TRES; \citealt{furesz:2008}) on the 1.5m Tillinghast Reflector at the Fred L. Whipple Observatory (FLWO). Starting points were used for other stellar parameters where available, but no prior constraints were placed on any other values. We assumed the RV extraction and metallicity determination was done correctly in the discovery data. An RV jitter term is fit within the {\tt EXOFASTv2} analysis to ensure the uncertainties are properly estimated. In the cases of five or fewer RVs, we placed conservative uniform bounds on the variance of the jitter. The jitter variance for K2-114 and for the \cite{Soto:2018} RVs for K2-237 were bounded to $\pm300$ m/s, and for K2-98 the variance bounds were $\pm100$ m/s for the FIES RVs, and $\pm4$ m/s for HARPS and HARPS-N. For the HARPS RVs of K2-265, we removed three clear outliers that were included in the discovery paper based on visual inspection\footnote{All parameters were within uncertainties when compared to an earlier fit including the outliers.}.

\section{Global Fits} 
\label{sec:GlobalModel}

\begin{figure*}
    \centering
    \includegraphics[width=0.85\textwidth]{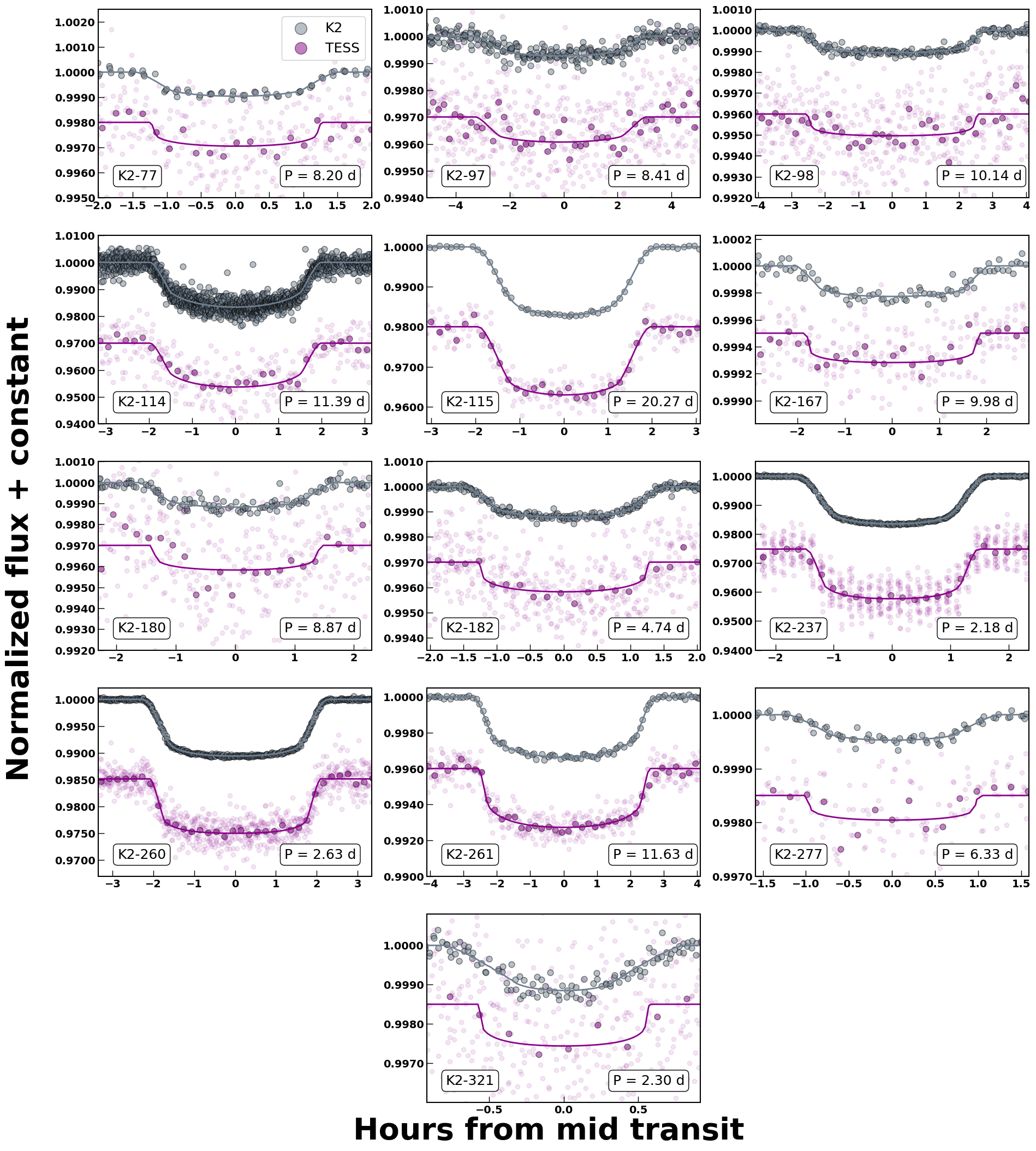}
    \caption{\ktwo (gray) and \tess (purple) transits for all systems where \tess added significant value to the ephemeris projection. The phase-folded lightcurves include all data available across the \ktwo campaigns and \tess sectors for each system, and have the best-fit model from {\tt EXOFASTv2} overlaid (see \citealt{Eastman:2013,Eastman:2019,Eastman:2017} for how this is calculated). The system \ktwo identifier and orbital period of the planet are displayed in each subplot. The \tess lightcurves are shown binned to 12 minutes, and the \ktwo lightcurves are unbinned. For K2-237, the discreteness of the points is likely due to the period being an integer multiple of the exposure time.}
    \label{fig:k2+tess_transits}
\end{figure*}

To analyze the wealth of data for these 26 known \ktwo exoplanet systems, we used {\tt EXOFASTv2} \citep{Eastman:2013,Eastman:2019,Eastman:2017} to perform global fits for our sample. {\tt EXOFASTv2} is an exoplanet fitting software package that uses MCMC sampling to simultaneously fit parameters for both the planets and host star. The \ktwo and \tess photometric observations (Figures \ref{fig:k2+tess_transits} and \ref{fig:k2_transits}), along with any archival RVs (Figure \ref{fig:RVs}), were jointly analyzed to obtain best-fit parameters for planets and host stars. 

\subsection{Stellar parameters}
\label{sec:star_params}
To characterize the host stars within each fit, we placed a uniform prior from 0 to an upper bound on  line-of-sight extinction ($A_v$) from \cite{Schlegel:1998} and \cite{Schlafly:2011}, and Gaussian priors on metallicity ([Fe/H]) and parallax (using Gaia EDR3 and accounting for the small systematic offset reported; \citealt{Gaia:2016,GaiaEDR3:2021,Lindegren:2021}). This also included the spectral energy distribution (SED) photometry as reported by Gaia DR2 \citep{GaiaDR2:2018}, WISE \citep{Cutri:2012} and 2MASS \citep{Cutri:2003}. These values are collated in Table \ref{tab:literature}, and all priors are listed in Tables \ref{tab:med1}-\ref{tab:med5}. We excluded the WISE4 SED values for three systems that had this photometric measurement (K2-115, K2-225 and K2-237) due to the large uncertainties, and as there was a $\gtrsim2\sigma$ discrepancy with the stellar model. Two other systems (K2-167 and K2-277) had WISE4 measurements that we used in the fits; these are consistent with the stellar models, but still have relatively large uncertainties. Within the {\tt EXOFASTv2} global fit, the MESA Isochrones and Stellar Tracks (MIST) stellar evolution models \citep{Paxton:2011, Paxton:2013, Paxton:2015, Choi:2016, Dotter:2016} are used as the base isochrone to better constrain the host star's parameters.

\startlongtable
\begin{deluxetable*}{lcccccccccc}
\centerwidetable
\tabletypesize{\scriptsize}
\tablecaption{Literature Values.}
\tablehead{\omit}
\startdata
\colhead{Param.} &  \colhead{Description}  & \colhead{K2-7}  & \colhead{K2-54} & \colhead{K2-57} & \colhead{K2-77} & \colhead{K2-97} & \colhead{K2-98} & \colhead{K2-114} & \colhead{K2-115} & \\
\hline
$\alpha_\mathrm{J2016}$ & Right ascension (R.A.) & 11:08:22.4996 &  22:32:12.9990 & 22:50:46.0386
 & 03:40:54.8458 & 08:31:03.0808 & 08:25:57.1702 & 08:31:31.8984  & 08:26:12.8406 & \\
$\delta_\mathrm{J2016}$ & Declination (Dec.) & -01:03:57.0898  & -17:32:38.6338 & -14:04:12.0152 & +12:34:20.7938 & +10:50:51.2025 & +11:30:39.9313 & +11:55:20.1168 & +12:16:54.6527 & \\
\\
$G$ & Gaia DR2 $G$ mag &	13.057 $\pm$ 0.020		&	---	&	14.104 $\pm$ 0.020	&	11.920 $\pm$ 0.020	&	12.306 $\pm$ 0.020	&	12.040 $\pm$ 0.020	&	14.275 $\pm$ 0.020	&	13.200 $\pm$ 0.020& 	\\
$G_\mathrm{Bp}$ & Gaia DR2 $B_P$ mag &	13.404 $\pm$ 0.020	&	---	&	14.781 $\pm$ 0.020	&	12.485 $\pm$ 0.020	&	12.895 $\pm$ 0.020	&	12.314 $\pm$ 0.020	&	14.806 $\pm$ 0.020	&	13.556 $\pm$ 0.020&  \\
$G_\mathrm{Rp}$ & Gaia DR2 $R_P$ mag & 	12.552 $\pm$ 0.020	&	---	&	13.326 $\pm$ 0.020	&	11.236 $\pm$ 0.020	&	11.601 $\pm$ 0.020	&	11.616 $\pm$ 0.020	&	13.615 $\pm$ 0.020	&	12.689 $\pm$ 0.020& \\ 
$T$ & \textit{TESS} mag &  	12.612 $\pm$ 0.008 &  --- & 13.383 $\pm$ 0.006 & 11.287 $\pm$ 0.006 & 11.652 $\pm$ 0.007 & 11.672 $\pm$ 0.008 & 13.667 $\pm$ 0.008 & 12.746 $\pm$ 0.006&  \\
$J$ & 2MASS $J$ mag &	11.952 $\pm$ 0.022	&	---	&	12.350 $\pm$ 0.024	&	10.384 $\pm$ 0.020	&	10.694 $\pm$ 0.023	&	11.124 $\pm$ 0.022	&	12.835 $\pm$ 0.020	&	12.108 $\pm$ 0.021& \\
$H$ & 2MASS $H$ mag &	11.628 $\pm$ 0.023	&	---	&	11.761 $\pm$ 0.022	&	9.910 $\pm$ 0.023	&	10.177 $\pm$ 0.023	&	10.905 $\pm$ 0.025	&	12.386 $\pm$ 0.030	&	11.760 $\pm$ 0.022 & \\
$K_S$ & 2MASS $K_S$ mag &	11.564 $\pm$ 0.021	&	---	&	11.645 $\pm$ 0.023	&	9.799 $\pm$ 0.020	&	10.035 $\pm$ 0.021	&	10.869 $\pm$ 0.028	&	12.304 $\pm$ 0.030	&	11.724 $\pm$ 0.020&  \\
WISE1 & WISE1 mag & 	11.527 $\pm$ 0.030	& --- &	11.586 $\pm$ 0.030	&	9.733 $\pm$ 0.030	&	9.990 $\pm$ 0.030	&	10.823 $\pm$ 0.030	&	12.230 $\pm$ 0.030	& 	11.658 $\pm$ 0.030&  \\
WISE2 & WISE2 mag & 	11.572 $\pm$ 0.030	&	 --- &	11.639 $\pm$ 0.030	&	9.790 $\pm$ 0.030	&	10.090 $\pm$ 0.030	&	10.856 $\pm$ 0.030	&	12.326 $\pm$ 0.030	&	11.700 $\pm$ 0.030 & \\
WISE3 & WISE3 mag &	11.554 $\pm$ 0.233	& ---	&	11.506 $\pm$ 0.217	&	9.773 $\pm$ 0.054	&	10.026 $\pm$ 0.088	&	10.678 $\pm$ 0.108	& --- &	11.723 $\pm$ 0.249 & \\
WISE4 & WISE4 mag &---&---&---&---&---&---&---& ---& \\
\\
$\mu_\alpha$ &Gaia p.m. in R.A. & -4.657 $\pm$ 0.016 & -5.018 $\pm$0.021 & 24.311 $\pm$ 0.022 & 22.425 $\pm$ 0.025 & -1.239 $\pm$ 0.018 & -16.165 $\pm$  0.014  & -13.062 $\pm$  0.022 & 15.557	$\pm$  0.018 & \\ 
$\mu_\delta$ & Gaia p.m. in Dec.  & -23.647$\pm$ 0.012 &  -9.804 $\pm$ 0.018	& -25.298 $\pm$ 0.019 & -37.908 $\pm$ 0.015 & -6.694 $\pm$0.013 & -9.401 $\pm$  0.010 & -2.472 $\pm$ 0.016 & -21.630$\pm$ 0.012 & \\
$\pi$ & Gaia parallax (mas) & 1.451 $\pm$ 0.028  & 5.782 $\pm$ 0.033  & 3.818 $\pm$ 0.029 & 7.111 $\pm$ 0.043 & 1.241 $\pm$ 0.063 & 1.950 $\pm$ 0.042 & 2.130 $\pm$ 0.036 & 2.497 $\pm$ 0.025 & \\
\hline
\colhead{Param.} & \colhead{K2-147}   & \colhead{K2-167} & \colhead{K2-180} & \colhead{K2-181} & \colhead{K2-182} & \colhead{K2-203} & \colhead{K2-204} & \colhead{K2-208} & \colhead{K2-211}&  \\
\hline
$\alpha_\mathrm{J2016}$ &  19:35:19.9267 & 22:26:18.2722 & 08:25:51.4492 & 08:30:12.9870 & 08:40:43.2088 & 00:51:05.6854 & 01:09:31.8015 & 01:23:06.9545 & 01:24:25.4797 & \\
$\delta_\mathrm{J2016}$  & -28:29:54.5839  & -18:00:42.0516 & +10:14:47.6330 & +10:54:36.5034 & +10:58:58.6242 & -01:11:45.1837 & -00:31:03.9292 & +00:53:20.4074 &  +01:42:17.6712 &  \\
\\
$G$ &	---	&	8.104 $\pm$ 0.020	&	12.404 $\pm$ 0.020	&	12.562 $\pm$ 0.020	&	11.720 $\pm$ 0.020	&	12.122 $\pm$ 0.020	&	12.889 $\pm$ 0.020	&	12.314 $\pm$ 0.020	&	12.933 $\pm$ 0.020	&  \\
$G_\mathrm{Bp}$ &	---	&	8.402 $\pm$ 0.020	&	12.817 $\pm$ 0.020	&	12.945 $\pm$ 0.020	&	12.190 $\pm$ 0.020	&	12.614 $\pm$ 0.020	&	13.223 $\pm$ 0.020	&	12.702 $\pm$ 0.020	&	13.399 $\pm$ 0.020& 	\\
$G_\mathrm{Rp}$ &---	&	7.689 $\pm$ 0.020	&	11.839 $\pm$ 0.020	&	12.036 $\pm$ 0.020	&	11.122 $\pm$ 0.020	&	11.493 $\pm$ 0.020	&	12.408 $\pm$ 0.020	&	11.781 $\pm$ 0.020	&	12.333 $\pm$ 0.020& 	\\
$T$ &  --- & 7.728 $\pm$ 0.006 & 11.896 $\pm$ 0.006 & 12.087 $\pm$ 0.006 & 11.170 $\pm$ 0.006 &  11.547 $\pm$ 0.006 & 12.461 $\pm$ 0.006 &  11.833 $\pm$ 0.006 & 12.383 $\pm$ 0.007 &  \\
$J$& 	---	&	7.202 $\pm$ 0.021	&	11.146 $\pm$ 0.023	&	11.438 $\pm$ 0.022	&	10.408 $\pm$ 0.021	&	10.773 $\pm$ 0.024	&	11.839 $\pm$ 0.021	&	11.164 $\pm$ 0.026	&	11.624 $\pm$ 0.024& \\
$H$ &	---	&	6.974 $\pm$ 0.038	&	10.747 $\pm$ 0.026	&	11.082 $\pm$ 0.021	&	9.994 $\pm$ 0.022	&	10.281 $\pm$ 0.026	&	11.569 $\pm$ 0.026	&	10.824 $\pm$ 0.022	&	11.205 $\pm$ 0.022& 	\\
$K_S$&	---	&	6.887 $\pm$ 0.034	&	10.677 $\pm$ 0.026	&	11.026 $\pm$ 0.021	&	9.913 $\pm$ 0.023	&	10.206 $\pm$ 0.023	&	11.478 $\pm$ 0.021	&	10.746 $\pm$ 0.020	&	11.104 $\pm$ 0.024	& \\
WISE1& 	---	&	6.810 $\pm$ 0.055	&	10.619 $\pm$ 0.030	&	10.999 $\pm$ 0.030	&	9.845 $\pm$ 0.030	&	10.145 $\pm$ 0.030	&	11.468 $\pm$ 0.030	&	10.697 $\pm$ 0.030	&	11.074 $\pm$ 0.030& \\
WISE2 &	---	&	6.866 $\pm$ 0.030	&	10.667 $\pm$ 0.030	&	11.062 $\pm$ 0.030	&	9.917 $\pm$ 0.030	&	10.217 $\pm$ 0.030	&	11.507 $\pm$ 0.030	&	10.739 $\pm$ 0.030	&	11.128 $\pm$ 0.030	& \\
WISE3 & ---	&	6.906 $\pm$ 0.030	&	10.599 $\pm$ 0.099	&	11.041 $\pm$ 0.205	&	9.896 $\pm$ 0.054	&	10.100 $\pm$ 0.083	&	11.279 $\pm$ 0.173	&	10.645 $\pm$ 0.089	&	10.933 $\pm$ 0.093& 	\\ 
WISE4		& --- & 6.917 $\pm$ 0.100	& ---	& --- & ---	& ---	& ---	& ---	& ---  & \\ 
\\
$\mu_\alpha$  & -31.399	$\pm$0.016 & 73.590$\pm$0.028 &  97.243$\pm$0.013 & 16.936$\pm$0.014  &  -65.130$\pm$0.029 &  -11.103$\pm$0.021 & -3.598$\pm$0.026 & -26.433$\pm$0.019  & 48.694$\pm$0.022&  \\
$\mu_\delta$  & -147.502$\pm$0.015 & -114.502$\pm$0.024  & -89.214$\pm$0.010  & -33.182$\pm$0.012 & 1.544$\pm$0.022 &  0.450$\pm$0.020 & -30.522$\pm$0.018 & -39.656$\pm$0.014 &  -10.262$\pm$0.017& \\
$\pi$  & 11.027 $\pm$ 0.033 & 12.457 $\pm$ 0.071 &  4.936 $\pm$ 0.041 & 2.805 $\pm$ 0.040 & 6.510 $\pm$ 0.052 & 5.937 $\pm$ 0.056  & 1.840 $\pm$ 0.052 & 3.859 $\pm$ 0.048 & 3.604 $\pm$ 0.055 & \\
\hline
\colhead{Param.} & \colhead{K2-225} & \colhead{K2-226} & \colhead{K2-237} &  \colhead{K2-250} & \colhead{K2-260} & \colhead{K2-261} & \colhead{K2-265} & \colhead{K2-277} & \colhead{K2-321} & \colhead{Ref.}\\
\hline
$\alpha_\mathrm{J2016}$ & 12:26:09.8617 & 12:14:34.9587 & 16:55:04.5232 &  12:20:07.5686 &  05:07:28.1596 & 10:52:07.7541 &  22:48:07.5960 & 13:28:03.8821 &10:25:37.3214  & 1 \\
$\delta_\mathrm{J2016}$  & -09:37:29.3675 & -09:33:45.4617 &-28:42:38.1039  &  -08:58:32.6688 & +16:52:03.6985 & +00:29:35.3793  & -14:29:41.2159 &  -15:56:16.7278 &  +02:30:49.9241 & 1 \\
\\
$G$ &	11.520 $\pm$ 0.020 & 12.092 $\pm$ 0.020	&	11.467 $\pm$ 0.020	&	13.973 $\pm$ 0.020	&	12.467 $\pm$ 0.020	&	10.459 $\pm$ 0.020	&	10.928 $\pm$ 0.020	&	10.121 $\pm$ 0.020	&	---	& 2 \\
$G_\mathrm{Bp}$ &	11.929 $\pm$ 0.020 &	12.545 $\pm$ 0.020	&	11.776 $\pm$ 0.020	&	14.484 $\pm$ 0.020	&	12.798 $\pm$ 0.020	&	10.872 $\pm$ 0.020	&	11.337 $\pm$ 0.020	&	10.495 $\pm$ 0.020	&---& 2 \\
$G_\mathrm{Rp}$ &	10.984 $\pm$ 0.020 &	11.492 $\pm$ 0.020	&	11.013 $\pm$ 0.020	&	13.324 $\pm$ 0.020	&	11.974 $\pm$ 0.020	&	9.917 $\pm$ 0.020	&	10.364 $\pm$ 0.020	&	9.624 $\pm$ 0.020	&	---& 2 \\
$T$ & 11.028 $\pm$ 0.007 &  11.547 $\pm$ 0.006 &  11.066 $\pm$ 0.006 & 13.379 $\pm$ 0.006 &  	12.036 $\pm$ 0.007 & 9.962 $\pm$ 0.007 & 10.422 $\pm$ 0.008 &  9.666 $\pm$ 0.006 & --- & 3\\
$J$ 	&	10.362 $\pm$ 0.023 & 	10.697 $\pm$ 0.023	&	10.508 $\pm$ 0.023	&	12.539 $\pm$ 0.026	&	11.400 $\pm$ 0.023	&	9.337 $\pm$ 0.030	&	9.726 $\pm$ 0.026	&	9.081 $\pm$ 0.034	&---	& 4 \\
$H$ &	10.046 $\pm$ 0.021 &	10.307 $\pm$ 0.023	&	10.268 $\pm$ 0.022	&	12.078 $\pm$ 0.022	&	11.189 $\pm$ 0.032	&	8.920 $\pm$ 0.042	&	9.312 $\pm$ 0.022	&	8.748 $\pm$ 0.071	&---	& 4  \\
$K_S$ &	9.954 $\pm$ 0.023 &	10.223 $\pm$ 0.023	&	10.217 $\pm$ 0.023	&	12.016 $\pm$ 0.024	&	11.093 $\pm$ 0.021	&	8.890 $\pm$ 0.022	&	9.259 $\pm$ 0.027	&	8.687 $\pm$ 0.024	&---	& 4 \\
WISE1  &	9.915 $\pm$ 0.030 &	10.166 $\pm$ 0.030	&	10.105 $\pm$ 0.030	&	11.878 $\pm$ 0.030	&	11.039 $\pm$ 0.030	&	8.828 $\pm$ 0.030	&	9.178 $\pm$ 0.030	&	8.630 $\pm$ 0.030	&---	& 5 \\
WISE2 &	9.978 $\pm$ 0.030 &	10.204 $\pm$ 0.030	&	10.129 $\pm$ 0.030	&	11.971 $\pm$ 0.030	&	11.036 $\pm$ 0.030	&	8.897 $\pm$ 0.030	&	9.213 $\pm$ 0.030	&	8.675 $\pm$ 0.030	&	---	& 5 \\
WISE3 &	9.932 $\pm$ 0.057 &	10.118 $\pm$ 0.083	&	9.972 $\pm$ 0.077	&	11.534 $\pm$ 0.259	&	10.895 $\pm$ 0.129	&	8.819 $\pm$ 0.031	&	9.162 $\pm$ 0.040	&	8.649 $\pm$ 0.030	&	---	& 5 \\
WISE4& ---	& ----	&	 --- &	---	& ---	& --- & ---  &  	8.418 $\pm$ 0.261	& --- &	5	\\
\\
$\mu_\alpha$ & -38.138$\pm$0.023 &  -22.324$\pm$0.023 & -8.568$\pm$0.035 & -50.359$\pm$0.030 & 0.646$\pm$0.018 &  -23.709$\pm$0.020 & 30.138$\pm$0.020 & -98.828$\pm$0.024  & 50.920$\pm$0.025 &  1 \\
$\mu_\delta$  & -8.190$\pm$0.017 & 2.109$\pm$0.021 & -5.625$\pm$0.022 &  9.874$\pm$0.018 &  -6.034$\pm$0.013 & -43.888$\pm$0.017  & -23.359$\pm$0.016 & -35.889$\pm$0.016 &  -109.167$\pm$0.022 &  1 \\
$\pi$ & 2.796 $\pm$ 0.045 & 4.807 $\pm$ 0.074 & 3.298 $\pm$ 0.071 & 2.476 $\pm$ 0.036 & 1.498 $\pm$ 0.043 &  4.685 $\pm$ 0.043 &  7.189 $\pm$ 0.051 &  8.842 $\pm$ 0.062 & 13.009 $\pm$ 0.053  &  1 \\
\enddata
 \begin{flushleft} 
  \footnotesize{
    \textbf{Notes.} The uncertainties of the photometry have a systematic error floor applied. The SEDs were not used for K2-54, K2-147 and K2-321 (see \S \ref{sec:GlobalModel}).\\
    Proper motions taken from the Gaia EDR3 archive and are in J2016. Parallaxes from Gaia EDR3 have a correction applied according to \cite{Lindegren:2021}. Parallax for K2-277 is from Gaia DR2 and has been corrected according to \cite{Lindegren:2018}. \\
    References: 1 - \cite{GaiaEDR3:2021}, 2 - \cite{GaiaDR2:2018}, 3 - \cite{Stassun:2018_TIC}, 4 - \cite{Cutri:2003}, 5 - \cite{Cutri:2012}
    }
\end{flushleft}
\label{tab:literature}
\end{deluxetable*}

\subsection{Low-mass stars}
\label{sec:lowm_stars}

Stellar evolutionary models struggle to constrain low-mass stars ($\lesssim$ 0.6 $M_\mathrm{\odot}$; \citealt{Mann:2015}) and are thus unreliable. For the three systems that fell into this category (K2-54, K2-147 and K2-321), we used the equations from \cite{Mann:2015,Mann:2019} that relate the apparent magnitude in the $K_S$ band ($M_{K_s}$) to $M_*$ and $R_*$ to set a starting point with wide $5\%$ Gaussian priors for these parameters. We excluded the SEDs from these fits and did not use the MIST models, fitting only the lightcurves (these systems did not have RV measurements). For this reason, we caution that the stellar parameters for these systems are unreliable. We also did not use the limb-darkening tables from \cite{Claret:2017} for the low-mass stars, as is the default in {\tt EXOFASTv2} for fitting the $u_1$ and $u_2$ coefficients, but rather placed starting points based on tables from \cite{Claret:2011} \citep{Eastman:2013} with a conservative Gaussian prior of 0.2 \citep{Patel:2022}.

\subsection{Contamination}
\label{sec:contam}

For systems with \tess contamination ratios specified in the \tess input catalog (TICv8, \citealp{Stassun:2018_TIC}) and a clear transit detected in both \ktwo and \tess, we fit for a dilution term\footnote{The starting point for dilution is calculated as D=C/(1+C).} on the \tess photometry with a 10\% Gaussian prior. This accounts for any nearby sources that may contribute flux to the target aperture that were unknown at the time the \tess Input Catalog was created. Although the \tess PDCSAP lightcurves are corrected for contamination, fitting the dilution allows an independent check on the contamination ratio correction performed by the SPOC pipeline. Fitting a dilution term for only the \tess photometry assumes the \ktwo aperture has been correctly decontaminated or is comparatively uncontaminated, which is based on \ktwo having a significantly smaller pixel scale than \tess (4" and 21" for \ktwo and \tess, respectively). However, it is possible that there is still a level of contamination within the \ktwo aperture that might be identified through high-resolution imaging. We checked the \ktwo aperture for all of our targets to identify any major sources of contamination from the \textit{Gaia} EDR3 catalog. We define contaminants as having flux ratios with the target star that are much larger than the uncertainties of the transit depth. To correct for the contaminating light, we followed the method from \cite{Rampalli:2019} to account for the fraction of the flux within the aperture that belonged to our targets ($F_\mathrm{star}$) as opposed to the contaminating stars based on the \textit{Gaia} G-band fluxes. We found significant contamination for K2-54 ($F_\mathrm{star}\approx0.56$) and K2-237 ($F_\mathrm{star}\approx0.98$; the latter was originally discussed in \citealt{Ikwut-Ukwa-SynergyI:2020}).  Several other systems had potential faint contaminants, however, the global fit for the system with the next highest level of contamination (K2-250; $F_\mathrm{star}\approx0.98$) did not change within uncertainties before and after flux correction, so we did not apply corrections to any systems other than K2-54 and K2-237.

\subsection{Global fits}
\label{sec:global_fits}

We ran a short preliminary fit for each system to identify any potential issues, e.g. particularly shallow transits, and then ran a final fit to convergence. For a fit to be accepted as converged, we adopted the default {\tt EXOFASTv2} criteria of $T_Z > 1000$, where $T_Z$ is the number of independent draws, and a slightly loose Gelman-Rubin value of $<1.02$ due to some transits being very shallow in \tess, resulting is long runtimes for the global fits. Within {\tt EXOFASTv2}, we opted to reject all flat and negative transit models, which ensured a more reliable recovery of marginal transits \citep{Eastman:2019}. We did not fit for transit timing variations, but plan to explore this in future papers.

\begin{figure*}
    \centering
    \includegraphics[width=0.85\textwidth]{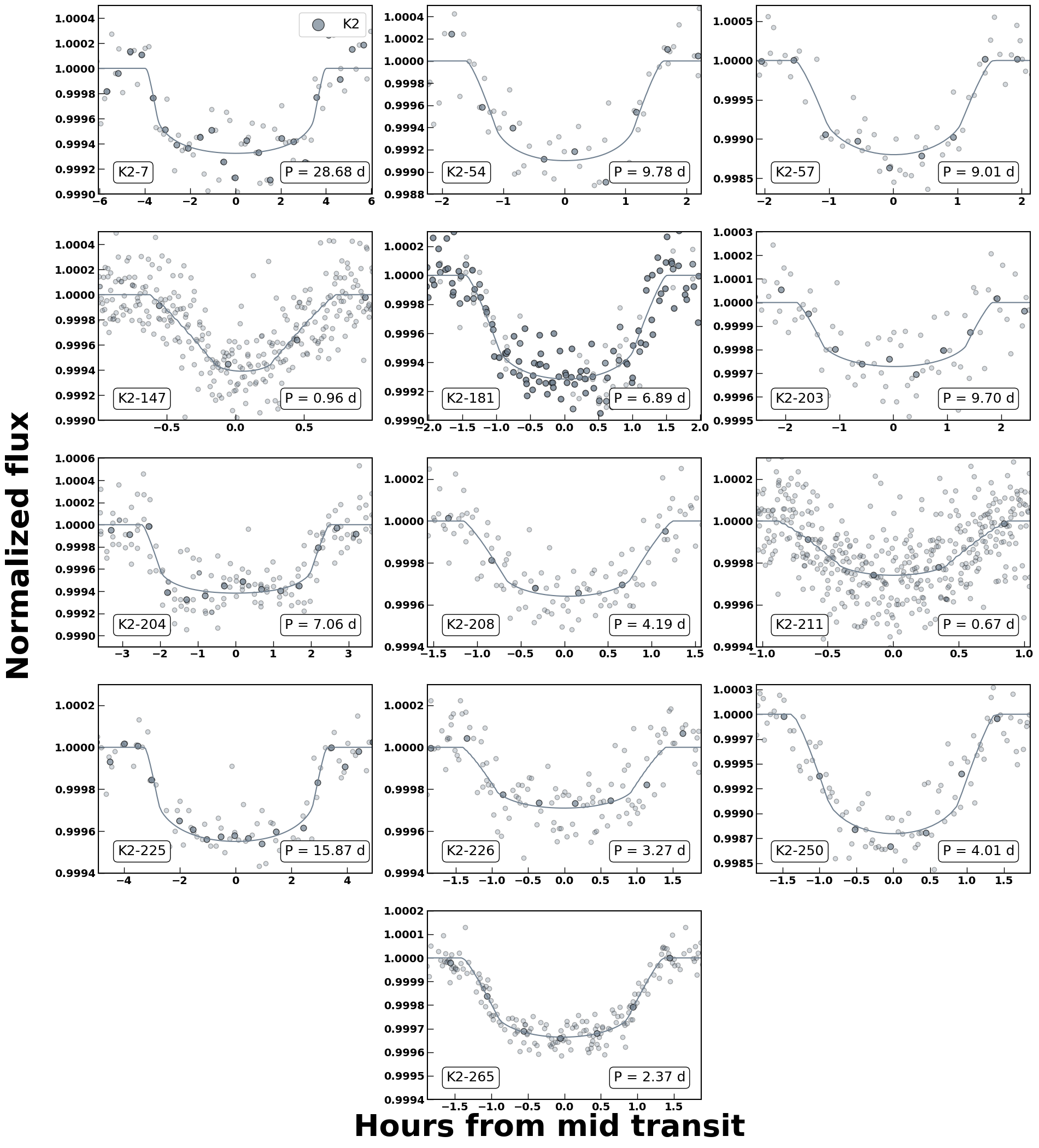}
    \caption{\ktwo transits for systems that were not recoverable in \tess lightcurves. The darker points are binned to 30 minutes, and the {\tt EXOFASTv2} best-fit model is shown.}.
    \label{fig:k2_transits}
\end{figure*}

Shallow transits clearly detected in \ktwo were not always evident in the \tess lightcurves as the latter are necessarily noisier due to the smaller collecting area of the telescope (see \S \ref{subsec:k2vtess} for discussion). For these systems we ran a \ktwo-only fit to convergence and a short preliminary fit (Gelman-Rubin of $\sim$ 1.1, $T_z \sim$ 100). To assess whether it was advantageous to include the \tess lightcurves, we required certain criteria be met before running the \ktwo and \tess fit to convergence. Firstly, we compared the improvement on uncertainties for parameters such as period and $T_C$, and projected these to the year 2030. If the uncertainties were notably smaller when including \tess data, we continued by visually inspecting the transits modelled by {\tt EXOFASTv2}. For extremely marginal transits, we further binned the phased lightcurves to determine whether the transit was indeed visible. If the transit in \tess was still not obvious, we inspected the probability distribution functions (PDFs) output by {\tt EXOFASTv2} for clearly non-Gaussian distributions for key parameters, particularly period. If the period was not well-constrained (e.g. multimodal) even with the increased baseline of \tess, we excluded the \tess lightcurve from the fit. A multimodal period indicates that the MCMC identified different transit solutions based on the \tess data, implying that the \tess transits are not securely enough detected to update the ephemeris.

We ultimately excluded any \tess lightcurves where the transit has SNR $\lesssim 7$. As these are all previously confirmed planets, we adopted a less conservative SNR for bona fide transits in \tess compared to what is required for initial planet verification. This SNR threshold was chosen because the first system below this cut (K2-265, SNR = 6.0) had a multimodal posterior for period, and all other systems with lower SNR exhibited similar issues. Conversely, the system just above this threshold (K2-277, SNR = 8.1) has a broad but Gaussian period posterior, with no other systems above this SNR having unreliable PDFs. 

Using this threshold, 13 of the 26 systems did not have recoverable \tess transits, so these were globally fit using only their \ktwo lightcurves (Figure \ref{fig:k2_transits}). While these systems will not have as significant improvement on their ephemerides, we still provide the updated parameters to include them in our final catalog of self-consistent parameters.

\section{Results and Discussion}
\label{sec:discussion}

We updated the system parameters for 26 single-planet systems discovered by \ktwo and reobserved by \tess, four of which were part of the pilot study for the \textit{K2} \& \textit{TESS} Synergy (K2-114, K2-167, K2-237 and K2-261; \citealt{Ikwut-Ukwa-SynergyI:2020}). Tables \ref{tab:med1}-\ref{tab:rv_params} contain parameters from the global fits. Here we address any points of interest for individual systems and for the sample as a whole.

\subsection{Ephemeris improvement}
\begin{figure*}
    \centering
    \includegraphics[width=0.95\textwidth]{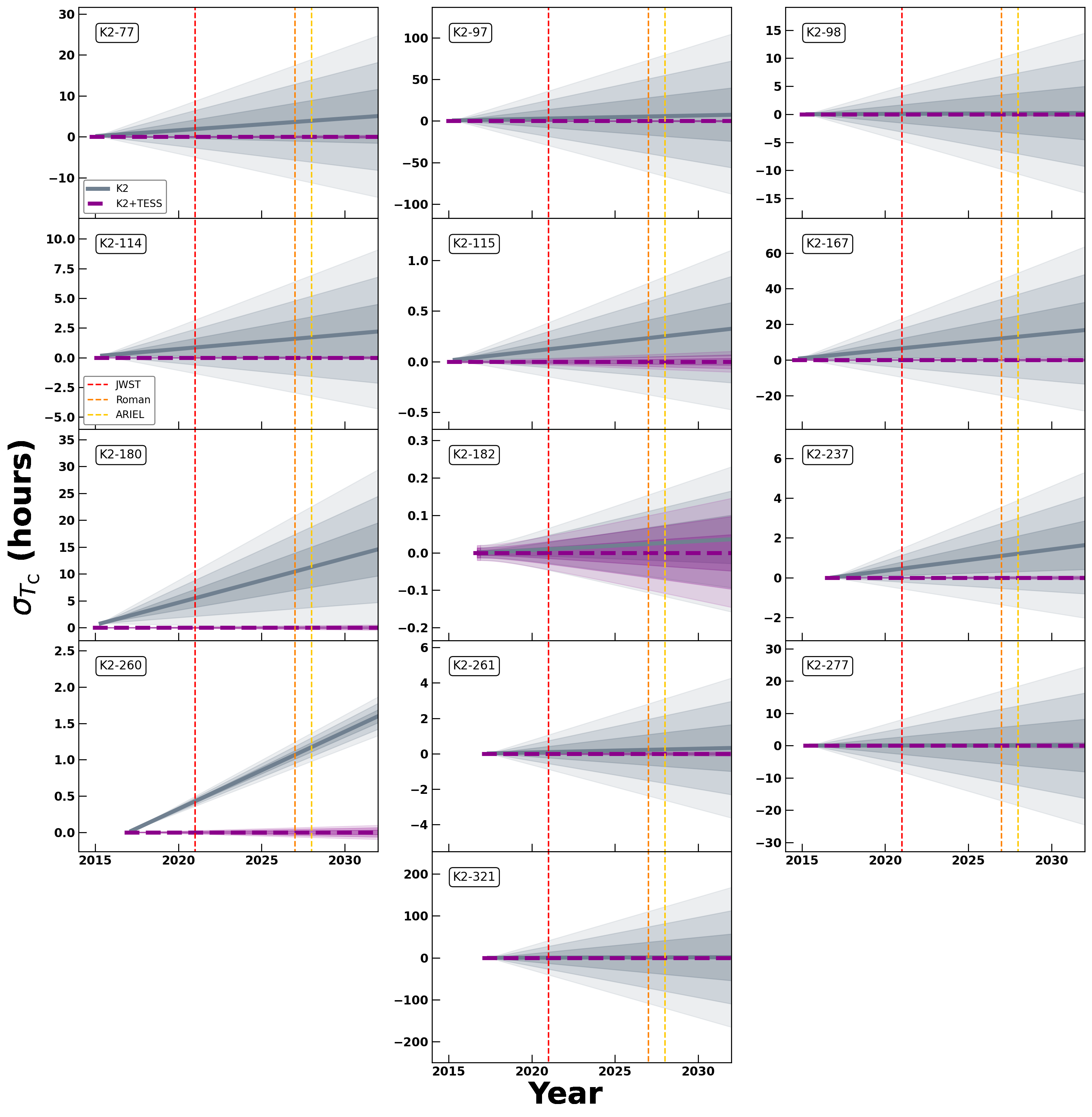}
    \caption{Projected uncertainties for transit times ($\sigma_{T_c}$) for systems with transits detected in both \ktwo and \tess. The shaded regions represent the 1, 2 and 3 $\sigma$ uncertainties, where gray is the uncertainty from the \ktwo ephemerides listed in Table \ref{tab:obs} and purple is our updated version using {\tt EXOFASTv2}. The vertical dashed lines show the expected or actual launch years for missions for which these systems would be prospective targets (\textit{JWST}: red, \textit{NGRST}: orange, \textit{ARIEL}: yellow). Note the y-axis scale is different in each subplot.}
    \label{fig:ephemerides-k2+tess}
\end{figure*}

\begin{figure*}
    \centering
    \includegraphics[width=0.9\textwidth]{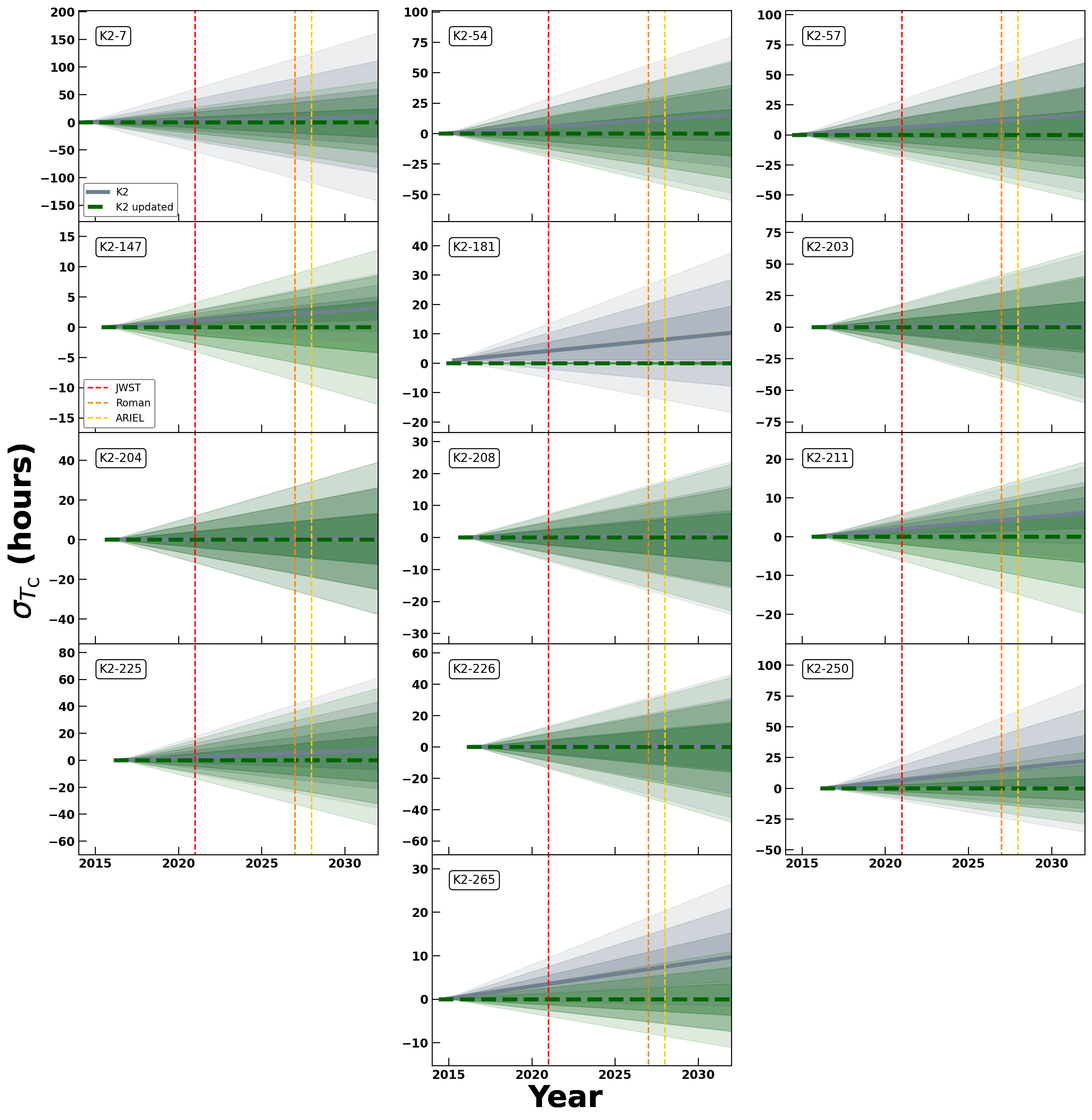}
    \caption{Same as Figure \ref{fig:ephemerides-k2+tess} but for systems with transits only detectable in the \ktwo lightcurves. The shaded regions represent the 1, 2 and 3 $\sigma$ uncertainties, where gray is the uncertainty from the \ktwo ephemerides listed in Table \ref{tab:obs} and green is our updated version using {\tt EXOFASTv2}. The vertical dashed lines show the expected or actual launch years for missions for which these systems would be prospective targets (\textit{JWST}: red, \textit{NGRST}: orange, \textit{ARIEL}: yellow). The ephemeris for K2-181 is significantly improved due to the inclusion of data from \ktwo Campaign 18.}
    \label{fig:ephemerides-k2}
\end{figure*}

As addressed in \S \ref{sec:intro}, a major incentive for refitting all \ktwo and \tess systems is to update their ephemerides to provide the community with accurate transit times for observing with existing and upcoming facilities. Figures \ref{fig:ephemerides-k2+tess} and \ref{fig:ephemerides-k2} show the projected transit timing uncertainties for our sample extrapolated to 2035, with markers indicating the expected launches for ongoing and future missions. The uncertainties on the transit times are calculated by standard error propagation,
\begin{equation}
    \sigma_{t_\mathrm{trans}} = \sqrt{\sigma_{T_0}^2 + (n_\mathrm{trans} \times \sigma_P)^2}
\end{equation}
where $\sigma_{t_\mathrm{trans}}$ is the uncertainty on future transit time, $\sigma_{T_0}$ is the uncertainty on the fitted optimal time of conjunction,  $n_\mathrm{trans}$ is the number of transits that occurred between timestamps and $\sigma_P$ is uncertainty on the period. For the future transit times using the results of the {\tt EXOFASTv2} global fits, we used the optimal time of conjunction in order to minimize the covariance between $T_C$ and $P$. However, $T_0$ is not generally available for the \ktwo discovery parameters, so for the projected uncertainties on transit times for the original \ktwo values we used $T_C$.

As expected, systems for which we excluded the \tess data due to shallow transits were not improved on the same scale as those with significant \tess transits. For the \ktwo and \tess systems, the updated global fits were able to reduce most uncertainties from hours to minutes within the scope of some of the major facilities in the near future (Figure \ref{fig:ephemerides-k2+tess}). For the 13 systems with detected \tess transits, the average 3$\sigma$ uncertainty on the future transit time by the year 2030 was reduced from 26.7 to 0.35 hours (Table \ref{tab:eph_comp_k2+tess}).

\begin{deluxetable}{lllll}
\caption{Ephemerides as of discovery compared to our updated values for systems with \ktwo and \tess transits, with the 3$\sigma$ uncertainty on future transit time by the year 2030.}

\tabletypesize{\scriptsize}
\tablecolumns{4}
\tablewidth{0pt}
\renewcommand{\arraystretch}{1.4}
\tablehead{\colhead{} & \colhead{$P$ (days)} & \colhead{$T_c$ (BJD)} & $3\sigma_{2030}$ & TSM}
\startdata
 \multicolumn{4}{c}{\textbf{K2-77}} \\ 
Discovery & $8.199814^{+0.000364}_{-0.000367}$ &  $2457070.806480^{+0.001511}_{-0.001449}$   & 17.4 hr \\
Updated &  $8.2000844^{+0.0000086}_{-0.0000073}$
 & $2457316.80766^{+0.00099}_{-0.00096}$  & 22 min & 27.3 \\ \hline
 \multicolumn{4}{c}{\textbf{K2-97}} \\ 
Discovery &  $8.406726^{+0.001863}_{-0.001827}$ & $2457142.04977^{+0.00888}_{-0.00854}$  & 84.8 hr \\
Updated & $8.407115\pm0.000023$  & $2457722.1447^{+0.0027}_{-0.0026}$ &  58 min & --- \\ \hline
\multicolumn{4}{c}{\textbf{K2-98}} \\ 
Discovery &   $10.13675\pm0.00033$ & $2457145.9807\pm0.0012$   & 12.6 hr \\
Updated & $10.1367349^{+0.0000094}_{-0.0000092}$ & $2457662.95321^{+0.00077}_{-0.00074}$ & 19 min & 13.5 \\ \hline
\multicolumn{4}{c}{\textbf{K2-114}} \\ 
Discovery & $11.39109^{+0.00018}_{-0.00017}$ & $2457174.49729\pm0.00033$ & 5.9 hr \\
Updated & $11.3909310^{0.0000031}_{-0.0000032}$ & $2457687.08869\pm0.00016$ & 6 min  & --- \\ \hline
\multicolumn{4}{c}{\textbf{K2-115}} \\ 
Discovery & $20.273034^{+0.000036}_{-0.000037}$ & $2457157.15701\pm0.00025$  & 42 min \\
Updated & $20.2729914\pm0.0000050$ & $2457522.07014\pm0.00017$ & 5 min & --- \\ \hline
\multicolumn{4}{c}{\textbf{K2-167}} \\ 
Discovery & $9.977481^{+0.001039}_{-0.001007}$ & $2456979.936780^{+0.002518}_{-0.002443}$  & 40.8 hr\\
Updated & $9.978541^{+0.000023}_{-0.000019}$ & $2457299.2465^{0.0022}_{0.0023}$ & 48 min & 46.1  \\ \hline
\multicolumn{4}{c}{\textbf{K2-180}} \\ 
Discovery &  $8.8665\pm0.0003$ & $2457143.390\pm0.002$  & 13.0 hr \\
Updated & $8.865663^{0.000011}_{0.000010}$ & $22457489.15656^{+0.00078}_{-0.00076}$ & 26 min & 15.1 \\ \hline
\multicolumn{4}{c}{\textbf{K2-182}} \\ 
Discovery &  $4.7369683\pm0.0000023$ & $2457719.11517\pm0.00028$  & 10 min  \\
Updated & $4.7369696\pm0.0000017$ & $2457652.79755^{+0.00027}_{-0.00028}$ & 8 min & 15.4
\\ \hline
\multicolumn{4}{c}{\textbf{K2-237}} \\ 
Discovery &   $2.18056\pm0.00002$ & $2457684.8101\pm0.0001$  & 3.2 hr \\
Updated & $2.18053332\pm0.00000054$ & $2457706.61618^{+0.00003}_{-0.00003}$ & 5 min & --- \\ \hline
\multicolumn{4}{c}{\textbf{K2-260}} \\ 
Discovery &  $2.6266657\pm0.0000018$ & $2457820.738135\pm0.00009$  & 14 min \\
Updated & $2.62669762\pm0.00000066$ & $2457894.284876^{+0.000060}_{-0.000059}$ & 5 min & --- \\ \hline
\multicolumn{4}{c}{\textbf{K2-261}} \\ 
Discovery &  $11.63344\pm0.00012$ & $2457906.84084^{+0.00054}_{-0.00067}$  & 3.4 hr \\
Updated & $11.6334681\pm0.0000044$ & $2458151.14394^{+0.00026}_{-0.00030}$  & 7 min & 85.5 \\ \hline
\multicolumn{4}{c}{\textbf{K2-277}} \\ 
Discovery & $6.326763^{+0.000355}_{-0.000361}$ & $2457221.22958^{+0.00221}_{-0.00217}$  & 21.5 hr \\
Updated & $6.326768^{+0.000015}_{-0.000012}$ & $2457303.4771\pm0.0010$ & 48 min & 35.6 \\ \hline
\multicolumn{4}{c}{\textbf{K2-321}} \\ 
Discovery &  $2.298\pm0.001$ & $2457909.17$  & 144.0 hr \\
Updated & $2.2979749^{+0.0000017}_{-0.0000019}$ & $2458141.26759^{+0.00064}_{-0.00068}$ & 15 min & --- \\ \hline
\enddata
\begin{flushleft}
 \footnotesize{ \textbf{\textsc{Notes:}}
The discovery values are taken from the \ktwo references listed in Table {\ref{tab:obs}}. The $T_C$ for the updated values is $T_0$ as determined by our global fits.
}
\end{flushleft}
\label{tab:eph_comp_k2+tess}
\end{deluxetable}

\begin{deluxetable}{lllll}
\caption{Ephemerides as of discovery compared to our updated values for systems with only \ktwo transits, with the 3$\sigma$ uncertainty on future transit time by the year 2030.}

\tabletypesize{\scriptsize}
\tablecolumns{4}
\tablewidth{0pt}
\renewcommand{\arraystretch}{1.4}
\tablehead{\colhead{} & \colhead{$P$ (days)} & \colhead{$T_c$ (BJD)} & $3\sigma_{2030}$ & TSM}
\startdata
 \multicolumn{4}{c}{\textbf{K2-7}} \\ 
Discovery & $28.67992\pm0.00947$ & $2456824.6155\pm0.0149$ & 135.0 hr \\
Updated & $28.6781^{+0.0046}_{-0.0051}$ & $2456853.2946^{+0.0046}_{-0.0042}$ & 68.8 hr & 5.9 \\ \hline
 \multicolumn{4}{c}{\textbf{K2-54}} \\ 
Discovery & $9.7843\pm0.0014$ & $2456982.9360\pm0.0053$ & 56.8 hr \\
Updated & $9.7833^{+0.0013}_{-0.0012}$ & $2457002.5042\pm0.0029$ & 50.6 hr & --- \\ \hline
\multicolumn{4}{c}{\textbf{K2-57}} \\ 
Discovery & $9.0063\pm0.0013$ & $2456984.3360\pm0.0048$ & 57.3 hr \\
Updated & $9.0073^{+0.0012}_{-0.0011}$ & $2457011.3568\pm0.0023$ & 50.4 hr & 10.8 \\ \hline
\multicolumn{4}{c}{\textbf{K2-147}} \\ 
Discovery & $0.961918\pm0.000013$ & $2457327.91683^{+0.00089}_{-0.00100}$ & 5.0 hr \\
Updated & $0.961939\pm0.000029$ & $2457343.30907^{+0.00100}_{-0.00099}$ & 11.2 hr & --- \\ \hline
\multicolumn{4}{c}{\textbf{K2-181}} \\ 
Discovery & $6.894252^{+0.000430}_{-0.000426}$ & $2457143.793550^{+0.002559}_{-0.002528}$ & 23.9 hr \\
Updated & $6.893813\pm0.000011$ & $2457778.0262\pm0.0012$ & 0.5 hr & 14.6 \\ \hline
\multicolumn{4}{c}{\textbf{K2-203}} \\ 
Discovery & $9.695101^{+0.001285}_{-0.001334}$ & $2457396.638780^{+0.005765}_{-0.005844}$ & 49.7 hr \\
Updated & $9.6952\pm0.0014$ & $2457435.4189^{+0.0037}_{-0.0036}$ & 52.7 hr  & 1.3 \\ \hline
\multicolumn{4}{c}{\textbf{K2-204}} \\ 
Discovery & $7.055784^{+0.000650}_{-0.000641}$ & $2457396.50862^{+0.00372}_{-0.00376}$ & 33.6 hr \\
Updated & $7.05576^{+0.00066}_{-0.00064}$ & $2457431.7872\pm0.0022$ & 33.6 hr & 11.1 \\ \hline
\multicolumn{4}{c}{\textbf{K2-208}} \\
Discovery   & $4.190948^{0.000230}_{-0.000248}$ & $2457396.51164^{+0.00248}_{-0.00235}$ & 21.0 hr \\
Updated & $4.19097\pm0.00023$ & $2457430.0390\pm0.0016$ & 20.0 hr & 12.9 \\    \hline
\multicolumn{4}{c}{\textbf{K2-211}} \\ 
Discovery    & $0.669532\pm0.000019$ & $2457395.82322\pm0.00160$ & 10.4 hr \\
Updated    & $0.669561^{+0.000031}_{-0.000032}$ & $2457432.6479\pm0.0013$ & 17.2 hr & 2.1 \\ \hline
\multicolumn{4}{c}{\textbf{K2-225}} \\
Discovery & $15.871455^{+0.002113}_{-0.001670}$ & $2457587.368230^{+0.004034}_{-0.004872}$ & 42.2 hr \\
Updated & $15.8723^{+0.0021}_{-0.0019}$ & $2457619.1111^{+0.0031}_{-0.0030}$ & 44.3 hr & 11.1 \\ \hline
\multicolumn{4}{c}{\textbf{K2-226}} \\ 
Discovery & $3.271106^{+0.000367}_{-0.000369}$ & $2457584.026130^{+0.004436}_{-0.004366}$ & 39.8 hr \\ 
Updated & $3.27109^{+0.00036}_{-0.00039}$ & $2457620.0082\pm0.0020$ & 40.3 hr & 14.6 \\  \hline
\multicolumn{4}{c}{\textbf{K2-250}} \\ 
Discovery & $4.01457^{+0.00062}_{-0.00057}$ & $2457584.1212^{+0.0061}_{-0.0066}$ & 52.5 hr \\
Updated & $4.01392\pm0.00029$ & $2457620.2535\pm0.0015$ & 25.4 hr  & 13.8 \\  \hline
\multicolumn{4}{c}{\textbf{K2-265}} \\ 
Discovery & $2.369172\pm0.000089$ & $2456981.6431\pm0.0016$ & 14.9 hr \\
Updated & $2.369020^{+0.000058}_{-0.000059}$ & $2457017.18078^{+0.00055}_{-0.00054}$ & 9.8 hr & 15.7 \\ \hline
\enddata
\begin{flushleft}
 \footnotesize{ \textbf{\textsc{Notes:}}
The discovery values are taken from the \ktwo references listed in Table {\ref{tab:obs}}. The $T_C$ for the updated values is $T_0$ as determined by our global fits.
}
\end{flushleft}
\label{tab:eph_comp_k2}
\end{deluxetable}

Systems for which we only included the \ktwo lightcurves had significantly less improvement on the precision of predicted transit times. However, the ephemeris for K2-181 was considerably refined due to the addition of data from \ktwo Campaign 18, which was not included in any previous analysis of this system. Excluding K2-181, there was a slight reduction of the average 3$\sigma$ uncertainty from 43.2 to 35.6 hours (Table \ref{tab:eph_comp_k2}). The small improvement for some systems is likely due to using optimized \ktwo lightcurves obtained from the pipeline described in \S \ref{sec:k2}, in conjunction with our fits including both the planet and the host star.

For systems with RV measurements, our ephemeris comparison uses uncertainties taken from previous analyses that included the RVs along with the \ktwo data. The uncertainties for systems without RVs are taken from the most recent study that included lightcurves from \ktwo. There are a handful of exceptions to this rule: for K2-77 we use the values from \cite{Mayo:2018} as \cite{Gaidos:2017} only has three RV measurements which is insufficient for our {\tt EXOFASTv2} fits; for K2-97, we use the values from \cite{Livingston:2018-K2C5-8} as no $T_c$ was presented in the analysis by \cite{Grunblatt:2018} that included RVs; for K2-237 we use the less precise values from \cite{Soto:2018} which are consistent with our results, rather than from \cite{Smith:2019} which have a $\sim 4\sigma$ discrepancy with our findings (this was also found in Paper I; \citealt{Ikwut-Ukwa-SynergyI:2020}).

As mentioned in \S \ref{sec:global_fits}, half of our sample did not have transits deep enough to be recovered by \tess. This presents a challenge for updating the transit times for these systems. If these systems are observed in future \tess sectors, it is possible that the SNR will increase sufficiently to include in a global fit. We will continue to monitor these and will include them in future releases, if this is the case.

\subsubsection{K2-167}

We note the use of an errant stellar metallicity prior used in the pilot study, where 0.45 instead of -0.45 (as reported by \citealt{Mayo:2018}) was used as the Gaussian center. While this may have affected the solutions of stellar and planetary parameters, it would not have significantly altered the ephemeris.

\subsubsection{K2-260}
\label{sec:k2-260}

There is a clear discrepancy between the previously published ephemeris and our updated version (see Figure \ref{fig:ephemerides-k2+tess}), well beyond a 3$\sigma$ level. To test whether this was an artifact of our global fit, we ran a fit using only the \ktwo lightcurves and compared the results to the original and \ktwo and \tess fits. Our \ktwo-only fit was consistent with our \ktwo and \tess ephemeris, and still in disagreement with the original results, suggesting that our updated fit provides the optimal ephemeris. It is possible that the original lightcurves introduced systematics in the discovery analysis, or the inclusion of additional follow-up data affected the ephemeris, but this is not clear. In any case, the consistency between our \ktwo-only and \ktwo and \tess ephemerides (and no other system showing similar issues) gives us confidence in our results.

\subsubsection{K2-261}
\label{sec:k2-261}

As discussed in the pilot study \citep{Ikwut-Ukwa-SynergyI:2020}, the PDFs for some stellar parameters (particularly age and mass) of K2-261 exhibit distinct bimodality that is likely due to the star being at a main sequence transition point (and not associated with the poor fits of shallow transits discussed in \S \ref{sec:global_fits}), causing difficulties with fitting the MIST isochrones to the data to constrain age. We followed the same procedure from \cite{Ikwut-Ukwa-SynergyI:2020}, splitting the posterior at the minimum probability for $M_*$ between the two Gaussian peaks (at $M_*$=1.19~$M_\odot$; see Figure 5 of \citealt{Ikwut-Ukwa-SynergyI:2020}) and extracting two separate solutions for each peak. We list both solutions in Table \ref{tab:med5}, however, we use the low-mass solution for all figures as this has the higher probability. The different stellar mass solutions do not affect the ephemeris projection for this planet.\\
~\\
\subsubsection{Comparison to pilot study}

The ephemerides were slightly improved for the four systems from the pilot study, the most significant being K2-167 (1.1 hours to 48 minutes) and K2-261 (30 minutes to 7 minutes). We did not expect to see major improvement because the baseline of new \tess sectors is relatively short compared to that of \ktwo and the \tess primary mission. 

\subsection{TSM}
\label{sec:TSM}

We calculated the transmission spectroscopy metric (TSM; \citealt{Kempton:2018}) for the planets in this sample to gauge the value of atmospheric follow-up (Tables \ref{tab:eph_comp_k2+tess} \& \ref{tab:eph_comp_k2}; Figure \ref{fig:configs}). As the TSM is dependent on stellar parameters, we excluded the three systems for which we did not fit the host star (K2-54, K2-147, K2-321; see \S\ref{sec:GlobalModel}). The TSM is only valid for planets with $R_p < 10~R_\oplus$, which removes a further five planets from this calculation (K2-97, K2-114, K2-115, K2-237 and K2-260). Only one system, K2-261, has a TSM above the threshold suggested by \cite{Kempton:2018}, and falls between the second and third quartile for the corresponding mass bin (see Table 1 of \citealt{Kempton:2018}.) Future work in this project to update ephemerides will prioritize planets with high TSMs relative to the entire \ktwo catalog.

\subsection{The sample}
\begin{figure}
    \centering
    \includegraphics[width=\linewidth]{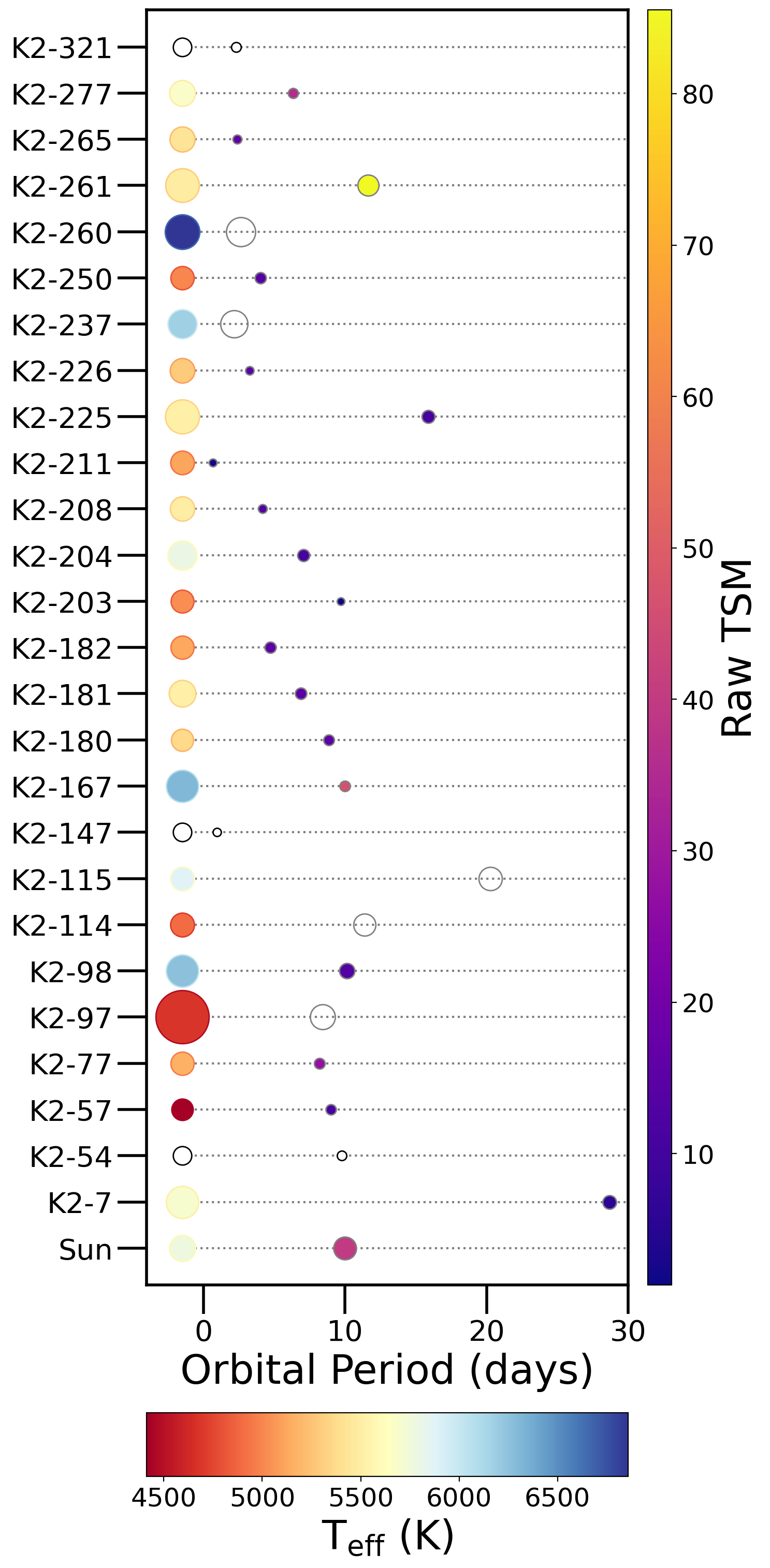}
    \caption{Architecture for each system showing the values from the global fits for the 26 systems in this analysis. The host stars are the left-most circles, with their temperatures indicated by color and relative radius shown by size. The right-most circles represent the planets, with size showing relative radius and color indicating their raw Transmission Spectroscopy Metric (TSM). The radius of the star and planet within each system is not scaled to each other. Systems for which we did not fit stellar parameters and planets that do not have a calculated TSM are represented by empty circles (see \S\ref{sec:TSM}). An example of the Sun hosting a Jupiter planet with a 10-day period and TSM of 40 is shown.}
    \label{fig:configs}
\end{figure}

\begin{figure*}
    \centering
    \includegraphics[width=\textwidth]{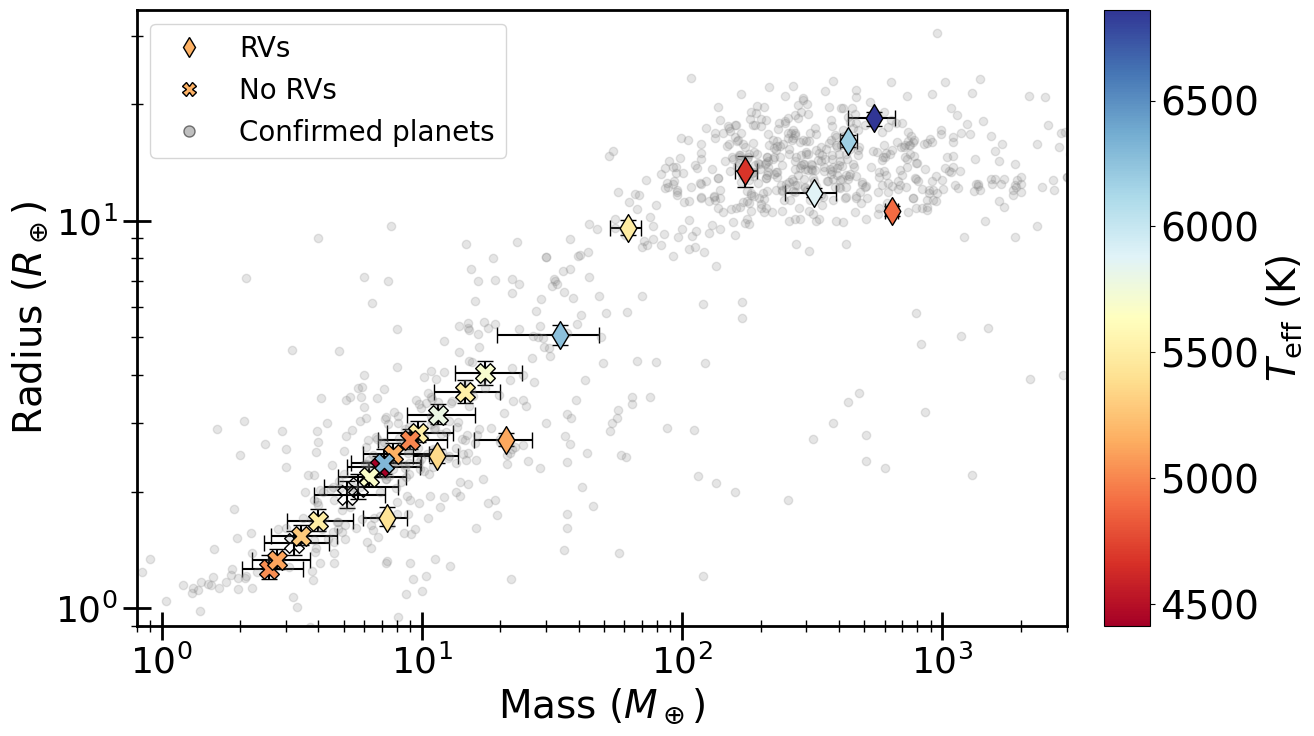}
    \caption{Radius versus mass for all confirmed exoplanets (gray; values taken from the NEA) and those in our work (using the median values from the {\tt EXOFASTv2} output). The 10 systems with planetary masses measured through RVs are indicated by diamonds, while the planets without RVs that have masses obtained from the \cite{Chen:2017} mass-radius relations are shown as crosses. The points are colored by the effective temperature of the host star, and are empty for the three systems without fitted stellar parameters.}
    \label{fig:massradius}
\end{figure*}

While the systems in this analysis span a broad range of stellar temperatures and planet masses, most planets have orbital periods $\lesssim$10 days and radii $\lesssim$5 $R_\oplus$ (Figures \ref{fig:configs} and \ref{fig:massradius}). Planet masses range from $2.6\sim639~M_\oplus$ and host stars include M dwarfs to F-type spectral classifications. This demonstrates the diversity of the original \ktwo sample as largely community-selected targets. Figure \ref{fig:massradius} shows how this sample compares to other known exoplanets.

\subsection{TTVs}
We did not fit for transit timing variations (TTVs) in this study. We would expect these to manifest as a significant change in ephemeris over time, whereas all of the systems studied here have updated ephemerides consistent to within $3\sigma$ of the original \ktwo ephemeris (except K2-260; see \S \ref{sec:k2-260}). Therefore, any TTVs that may be present are currently too small to detect for these systems. Differences in the ephemerides on the $1\sim3\sigma$ level are likely due to the addition of the \tess lightcurves.

\subsection{Candidate planets}

We note that a couple of the systems in our analysis have additional candidate planets (K2-203 and K2-211). However, we ignore these for the purpose of updating ephemerides of known exoplanets that are more likely future targets for missions such as JWST, but plan to revisit these in a future paper addressing multi-planet systems.

\subsection{\ktwo vs. \tess}
\label{subsec:k2vtess}

It is not surprising that relatively shallow \ktwo transits were not detected by \tess. \Kepler and \tess were designed to observe different stellar demographics, resulting in different photometric capabilities. \Kepler was built with the intent to explore the number of near-Earth-sized planets close to their respective habitable zones around distant stars with apparent magnitudes $\lesssim$ 16. The original \Kepler mission could reach a precision of $\sim$ 20 parts per million (ppm), which was generally the same for the \ktwo mission \citep{Vanderburg:2014,Vanderburg:2016b}. 

On the other hand, \tess is focused on nearby, brighter stars with magnitude $\lesssim$ 12. The precision of \tess has a floor at $\sim$20 ppm at 1 hour for the brightest stars with $T_\mathrm{mag}<4$, but is more realistically $\gtrsim$100 ppm for the majority of stars. Due to the all-sky nature of the \tess missions, observing sectors last on average 27 days for efficient sky coverage. \ktwo campaigns were around 80 days in duration, meaning the same targets may have $\sim$3 times as many transits observed by \ktwo.

While \tess may not be able to recover all \ktwo systems, the ones it can detect will have vastly improved ephemerides as demonstrated in Figure \ref{fig:ephemerides-k2+tess}. Our analysis indicates \tess transits with SNR $\gtrsim$ 7 are recoverable, and while this places a limit on the scope of this reanalysis, we can potentially gain access for reobservation of at least half of known \ktwo planets. It is possible that future \tess missions that reobserve the planets with currently marginal transits (SNR $\sim5-6$) will increase the SNR enough for a significant detection. However, for transits with SNR$\lesssim5$, it is unlikely that more \tess observations will result in recoverable transits.

\subsection{Future work}

With several major facilities able to characterize exoplanets in extensive detail planned to come online within the next decade, not having accurate and precise transit times is a relevant issue. The \ktwo \& \tess Synergy aims to solve the problem of degrading ephemerides for all \ktwo systems reobserved by \tess (with clearly detectable transits as shown by this effort). Assuming \tess will reobserve all \ktwo systems throughout its extended missions, we expect to be able to update the ephemerides for around half of \ktwo planets ($\sim$250 planets) with transits deep enough to be detected by \tess, based on this study. Over the next couple of years, we plan to reanalyze the remaining \ktwo systems with current \tess overlap, providing the updated parameters to the community. In future batches, we will place a focus on systems that are potentially well suited as JWST targets for atmospheric studies based on their TSMs. While we do not see strong evidence for TTVs in the current work, we will make note of this in future for any systems with significant change in ephemeris, particularly for known multi-planet systems where this would be more readily detectable.

\section{Conclusion}
\label{sec:conclusion}
Past efforts to create and analyze homogeneous populations of exoplanet parameters have led to great insight into major questions in planetary formation and evolution \citep{Wang:2014, Fulton:2017, Fulton:2018}. The \ktwo \& \tess Synergy is uniting NASA's planet hunting missions, and focuses on extending the scientific output of both telescopes by creating a self-consistent catalog for the \ktwo\ and \tess\ sample while providing the community with updated ephemerides to efficiently schedule future characterization observations with facilities like JWST \citep{Gardner:2006}. As well as refreshing stale ephemerides, this provides a uniform way of addressing any inconsistencies between the original \ktwo ephemeris and the updated value from \tess. In this paper, we have presented updated parameters for 26 single-planet systems originally discovered by \ktwo and more recently reobserved by \tess\ during its primary and extended missions. Following from the success of the pilot study \citep{Ikwut-Ukwa-SynergyI:2020}, we have significantly reduced the uncertainties on transit times for the 13 systems with transits detectable in \tess\ from hours down to minutes through the JWST operations window ($\sim$2030). Assuming the current sample is representative of the entire \ktwo catalog, we expect significant improvement on ephemerides for about half of the systems revisited by \tess, with the goal of a $\sim$250-system catalog of parameters that will be publicly available. As \tess\ continues to reobserve large portions of the entire sky during its current and possible future extended missions, there will be a well-suited opportunity to conduct this analysis on all known exoplanets, possibly leading to key insights into the evolutionary processes of exoplanets.

\software{{Lightkurve \citep{Lightkurve:2018}}, EXOFASTv2 \citep{Eastman:2013, Eastman:2019}}
\facilities{\TESS, \ktwo, Keck (HIRES), Lick Observatory 2.4 m (Levy), Nordic Optical 2.56 m (FIES), La Silla 3.6 m (HARPS), Telescopio Nazionale Gailieo 3.58 m (HARPS-N), La Silla 1.2 m (CORALIE), FLWO 1.5m (Tillinghast Reflector Echelle Spectrograph).}

\acknowledgements
We thank the anonymous referee for the constructive feedback that helped improve this manuscript. ET and JER acknowledge support for this project from NASA'S TESS Guest Investigator program (G04205, P.I. Rodriguez). This research has made use of SAO/NASA's Astrophysics Data System Bibliographic Services. This research has made use of the SIMBAD database, operated at CDS, Strasbourg, France. This work has made use of data from the European Space Agency (ESA) mission {\it Gaia} (\url{https://www.cosmos.esa.int/gaia}), processed by the {\it Gaia} Data Processing and Analysis Consortium (DPAC, \url{https://www.cosmos.esa.int/web/gaia/dpac/consortium}). Funding for the DPAC has been provided by national institutions, in particular the institutions participating in the {\it Gaia} Multilateral Agreement. This work makes use of observations from the LCO network. This research made use of Lightkurve, a Python package for \textit{Kepler} and \TESS\ data analysis.

Funding for the \TESS\ mission is provided by NASA's Science Mission directorate. We acknowledge the use of public \TESS\ Alert data from pipelines at the \TESS\ Science Office and at the \TESS\ Science Processing Operations Center. This research has made use of the Exoplanet Follow-up Observation Program website, which is operated by the California Institute of Technology, under contract with the National Aeronautics and Space Administration under the Exoplanet Exploration Program. Resources supporting this work were provided by the NASA High-End Computing (HEC) Program through the NASA Advanced Supercomputing (NAS) Division at Ames Research Center for the production of the SPOC data products. The data presented in this paper were obtained from the Mikulski Archive for Space Telescopes (MAST) at the Space Telescope Science Institute. The datasets and corresponding DOIs are \ktwo:\dataset[DOI:10.17909/T9WS3R]{https://doi.org/10.17909/T9WS3R}, \tess 20-second cadence: \dataset[DOI:10.17909/t9-st5g-3177]{https://doi.org/10.17909/t9-st5g-3177}, \tess 2-minute cadence: \dataset[DOI:10.17909/t9-nmc8-f686]{https://doi.org/10.17909/t9-nmc8-f686}, \tess-SPOC FFIs: \dataset[DOI:10.17909/t9-wpz1-8s54]{https://doi.org/10.17909/t9-wpz1-8s54}. \tess data from Sectors 2-46 were used in this analysis.

\bibliographystyle{apj}
\bibliography{refs}

\appendix

\section{Additional Tables}

\providecommand{\bjdtdb}{\ensuremath{\rm {BJD_{TDB}}}}
\providecommand{\feh}{\ensuremath{\left[{\rm Fe}/{\rm H}\right]}}
\providecommand{\teff}{\ensuremath{T_{\rm eff}}}
\providecommand{\teq}{\ensuremath{T_{\rm eq}}}
\providecommand{\ecosw}{\ensuremath{e\cos{\omega_*}}}
\providecommand{\esinw}{\ensuremath{e\sin{\omega_*}}}
\providecommand{\msun}{\ensuremath{\,M_\Sun}}
\providecommand{\rsun}{\ensuremath{\,R_\Sun}}
\providecommand{\lsun}{\ensuremath{\,L_\Sun}}
\providecommand{\mj}{\ensuremath{\,M_{\rm J}}}
\providecommand{\rj}{\ensuremath{\,R_{\rm J}}}
\providecommand{\me}{\ensuremath{\,M_{\rm E}}}
\providecommand{\re}{\ensuremath{\,R_{\rm E}}}
\providecommand{\fave}{\langle F \rangle}
\providecommand{\fluxcgs}{10$^9$ erg s$^{-1}$ cm$^{-2}$}
\startlongtable
\begin{deluxetable*}{lccccccc}
\centerwidetable
\tablecaption{Median values and 68\% confidence intervals.}
\tablehead{\omit}
\tabletypesize{\tiny}
\startdata
\multicolumn{2}{l}{Priors:}  & K2-7 & K2-54$^{\prime}$ & K2-57 & K2-77 & K2-97 & K2-98\\
\hline \\
$\pi$\dotfill &  Gaia Parallax (mas) \dotfill & $\mathcal{G}[1.45113, 0.02810]$ &  --- & $\mathcal{G}[3.81841,0.02910]$ & $\mathcal{G}[7.11086, 0.04290]$  & $\mathcal{G}[1.24110, 0.06320]$  & $\mathcal{G}[1.94998,0.04160]$ \\
$[{\rm Fe/H}]$\dotfill &  Metallicity (dex) \dotfill & $\mathcal{G}[-0.153, 0.24]^*$ & --- & $\mathcal{G}[-0.01, 0.20]^*$ & $\mathcal{G}[0.118, 0.080]$ & $\mathcal{G}[0.267, 0.080]$ & $\mathcal{G}[-0.104, 0.080]$  \\
 $A_V$ & V-band extinction (mag) \dotfill & $\mathcal{U}[0, 0.12741]$ & --- & $\mathcal{U}[0, 0.1209]$  &  $\mathcal{U}[0, 1.11693]$ & $\mathcal{U}[0,0.13733]$  &  $\mathcal{U}[0, 0.14477]$ \\
$D_T^{\star}$&  Dilution in \tess\  \dotfill & $\mathcal{G}[0, 0.00030159]$ & --- & --- & $\mathcal{G}[0, 0.00023943]$ & --- & $\mathcal{G}[0, 0.0018889]$  \\
\hline\hline
Parameter & Units & Values \\
Stellar Parameters: \\
~~~~$M_*$\dotfill &Mass (\msun)\dotfill & $1.056^{+0.098}_{-0.068}$ & \textcolor{red}{$0.615\pm 0.031$} & $0.699^{+0.034}_{-0.031}$ & $0.847^{+0.039}_{-0.038}$ & $1.17^{+0.19}_{-0.13}$ & $1.246^{+0.078}_{-0.11}$ \\
~~~~$R_*$\dotfill &Radius (\rsun)\dotfill & $1.533^{+0.082}_{-0.074}$ & \textcolor{red}{$0.643\pm 0.031$} & $0.671^{+0.023}_{-0.022}$ & $0.792^{+0.027}_{-0.025}$ & $4.14^{+0.28}_{-0.26}$ & $1.518^{+0.082}_{-0.076}$ \\
~~~~$L_*$\dotfill &Luminosity (\lsun)\dotfill & $2.24^{+0.12}_{-0.11}$ & \textcolor{red}{$0.095^{+0.034}_{-0.026}$} & $0.1541^{+0.0053}_{-0.0051}$ & $0.401^{+0.04}_{-0.037}$ & $7.36^{+0.83}_{-0.72}$ & $3.19^{+0.23}_{-0.21}$ \\
~~~~$F_{Bol}$\dotfill &Bolometric Flux (cgs)\dotfill & $1.504e-10^{+5.1e-12}_{-5.5e-12}$ & ---& $7.18e-11^{+2.2e-12}_{-2.1e-12}$ & $6.47e-10^{+6.3e-11}_{-5.9e-11}$ & $3.521e-10^{+9.9e-12}_{-1e-11}$ & $3.88e-10^{+2.2e-11}_{-2e-11}$ \\
~~~~$\rho_*$\dotfill &Density (cgs)\dotfill & $0.416^{+0.086}_{-0.074}$ & \textcolor{red}{$3.25^{+0.54}_{-0.45}$} & $3.26^{+0.32}_{-0.29}$ & $2.41^{+0.25}_{-0.24}$ & $0.0235^{+0.0065}_{-0.0048}$ & $0.499^{+0.1}_{-0.087}$ \\
~~~~$\log{g}$\dotfill &Surface gravity (cgs)\dotfill & $4.092^{+0.065}_{-0.062}$ & \textcolor{red}{$4.61\pm0.047$} & $4.629\pm 0.028$ & $4.569^{+0.03}_{-0.033}$ & $3.276^{+0.085}_{-0.077}$ & $4.169^{+0.06}_{-0.063}$ \\
~~~~$T_{\rm eff}$\dotfill &Effective Temperature (K)\dotfill & $5700.0\pm 140.0$ & \textcolor{red}{$3990.0\pm 300.0$} & $4413.0\pm 76.0$ & $5160.0\pm 130.0$ & $4673.0^{+95.0}_{-93.0}$ & $6260.0^{+170.0}_{-160.0}$ \\
~~~~$[{\rm Fe/H}]$\dotfill &Metallicity (dex)\dotfill & $0.049^{+0.120}_{-0.081}$ & \textcolor{red}{$-0.02^{+0.95}_{-0.99}$} & $0.08^{+0.2}_{-0.14}$ & $0.13^{+0.078}_{-0.076}$ & $0.328^{+0.079}_{-0.081}$ & $-0.044^{+0.064}_{-0.07}$ \\
~~~~$[{\rm Fe/H}]_{0}$\dotfill &Initial Metallicity$^{1}$ \dotfill & $0.103^{+0.100}_{-0.082}$ & ---& $0.07^{+0.18}_{-0.13}$ & $0.115^{+0.08}_{-0.078}$ & $0.261^{+0.074}_{-0.078}$ & $0.069^{+0.062}_{-0.065}$ \\
~~~~$Age$\dotfill &Age (Gyr)\dotfill & $8.6^{+2.5}_{-3.1}$ & ---& $6.8^{+4.7}_{-4.5}$ & $4.7^{+5.2}_{-3.4}$ & $7.5^{+4.0}_{-3.2}$ & $3.4^{+2.2}_{-1.4}$ \\
~~~~$EEP$\dotfill &Equal Evolutionary Phase$^{2}$ \dotfill & $447.5^{+8.7}_{-24.0}$ & ---& $327.0^{+14.0}_{-30.0}$ & $334.0^{+19.0}_{-37.0}$ & $499.6^{+8.7}_{-9.6}$ & $389.0^{+41.0}_{-42.0}$ \\
~~~~$A_V$\dotfill &V-band extinction (mag)\dotfill & $0.087^{+0.029}_{-0.047}$ & ---& $0.063^{+0.04}_{-0.042}$ & $0.59^{+0.14}_{-0.15}$ & $0.091^{+0.034}_{-0.051}$ & $0.088^{+0.041}_{-0.055}$ \\
~~~~$\sigma_{SED}$\dotfill & SED photometry error scaling\dotfill & $0.6^{+0.26}_{-0.15}$ & ---& $1.45^{+0.73}_{-0.39}$ & $0.9^{+0.4}_{-0.24}$ & $0.9^{+0.38}_{-0.24}$ & $1.13^{+0.54}_{-0.3}$ \\
~~~~$\varpi$\dotfill &Parallax (mas)\dotfill & $1.449\pm 0.028$ & ---& $3.818\pm 0.03$ & $7.107\pm 0.043$ & $1.222\pm 0.062$ & $1.951\pm 0.042$ \\
~~~~$d$\dotfill &Distance (pc)\dotfill & $689.0^{+14.0}_{-13.0}$ & ---& $261.9^{+2.1}_{-2.0}$ & $140.71^{+0.85}_{-0.84}$ & $818.0^{+44.0}_{-40.0}$ & $512.0\pm 11.0$ \\
Planetary Parameters:\\
~~~~$P$\dotfill &Period (days)\dotfill & $28.6781^{+0.0046}_{-0.0051}$ & $9.7833^{+0.0013}_{-0.0012}$ & $9.0073^{+0.0012}_{-0.0011}$ & $8.2000844^{+8.6e-06}_{-7.3e-06}$ & $8.407115\pm 2.3e-05$ & $10.1367349^{+9.4e-06}_{-9.2e-06}$ \\
~~~~$R_P$\dotfill &Radius (\rj)\dotfill & $0.360^{+0.027}_{-0.023}$ & $0.233^{+0.019}_{-0.017}$ & $0.206^{+0.015}_{-0.012}$ & $0.223^{+0.015}_{-0.011}$ & $1.2\pm 0.11$ & $0.452^{+0.028}_{-0.025}$ \\
~~~~$M_P$\dotfill &Mass (\mj)\dotfill & $0.055^{+0.021}_{-0.013}$ & $0.0264^{+0.0100}_{-0.0066}$ & $0.0217^{+0.0091}_{-0.0055}$ & $0.0244^{+0.0091}_{-0.0058}$ & $0.549^{+0.059}_{-0.046}$ & $0.107^{+0.044}_{-0.046}$ \\
~~~~$T_C$\dotfill &Time of conjunction$^{3}$ (\bjdtdb)\dotfill & $2456824.6164^{+0.0079}_{-0.0067}$ & $2456982.9376^{+0.0044}_{-0.0045}$ & $2456984.335^{+0.0037}_{-0.0039}$ & $2457070.8051\pm 0.001$ & $2457142.0537\pm 0.0031$ & $2457145.97972^{+0.00092}_{-0.00091}$ \\
~~~~$T_T$\dotfill &Time of minimum projected separation$^{4}$ (\bjdtdb)\dotfill & $2456824.6164^{+0.0076}_{-0.0064}$ & $2456982.9376^{+0.0043}_{-0.0044}$ & $2456984.335^{+0.0036}_{-0.0038}$ & $2457070.80511\pm 0.0008$ & $2457142.05\pm 0.0027$ & $2457145.97967\pm 0.00084$ \\
~~~~$T_0$\dotfill &Optimal conjunction Time$^{5}$ (\bjdtdb)\dotfill & $2456853.2946^{+0.0046}_{-0.0042}$ & $2457012.2875^{+0.0028}_{-0.0027}$ & $2457011.3568\pm 0.0023$ & $2457316.80766^{+0.00099}_{-0.00096}$ & $2457722.1447^{+0.0027}_{-0.0026}$ & $2457662.95321^{+0.00077}_{-0.00074}$ \\
~~~~$a$\dotfill &Semi-major axis (AU)\dotfill & $0.1867^{+0.0056}_{-0.0041}$ & $0.0761\pm0.0013$ & $0.0752^{+0.0012}_{-0.0011}$ & $0.0753\pm 0.0011$ & $0.0854^{+0.0044}_{-0.0033}$ & $0.0986^{+0.002}_{-0.0029}$ \\
~~~~$i$\dotfill &Inclination (Degrees)\dotfill & $89.08^{+0.58}_{-0.47}$ & $89.08^{+0.59}_{-0.5}$ & $88.95^{+0.67}_{-0.46}$ & $88.33^{+0.83}_{-0.37}$ & $75.5^{+1.6}_{-1.8}$ & $88.39^{+0.98}_{-0.74}$ \\
~~~~$e$\dotfill &Eccentricity \dotfill & $0.24^{+0.40}_{-0.17}$ & $0.24^{+0.40}_{-0.18}$ & $0.25^{+0.38}_{-0.17}$ & $0.29^{+0.37}_{-0.19}$ & $0.207^{+0.039}_{-0.038}$ & $0.119^{+0.15}_{-0.083}$ \\
~~~~$\omega_*$\dotfill &Argument of Periastron (Degrees)\dotfill & $-30.0^{+110.0}_{-130.0}$ & $-43.0^{+93.0}_{-120.0}$ & $-20.0^{+110.0}_{-140.0}$ & $-40.0^{+100.0}_{-130.0}$ & $68.4^{+9.0}_{-11.0}$ & $14.0^{+110.0}_{-98.0}$ \\
~~~~$T_{eq}$\dotfill &Equilibrium temperature$^{6}$ (K)\dotfill & $786.0^{+14.0}_{-14.0}$ & $560.0^{+45.0}_{-44.0}$ & $635.8^{+6.8}_{-6.7}$ & $806.0\pm 18.0$ & $1566.0^{+49.0}_{-50.0}$ & $1185.0^{+23.0}_{-21.0}$ \\
~~~~$\tau_{\rm circ}$\dotfill &Tidal circularization timescale (Gyr)\dotfill & $28000.0^{+35000.0}_{-27000.0}$ & $750.0^{+1000.0}_{-740.0}$ & $880.0^{+1000.0}_{-860.0}$ & $380.0^{+580.0}_{-380.0}$ & $4.5^{+3.2}_{-1.8}$ & $330.0^{+300.0}_{-220.0}$ \\
~~~~$K$\dotfill &RV semi-amplitude (m/s)\dotfill & $3.9^{+2.0}_{-1.1}$ & $3.8^{+2.1}_{-1.1}$ & $2.98^{+1.7}_{-0.83}$ & $3.09^{+1.5}_{-0.83}$ & $48.7\pm 2.3$ & $8.8^{+3.5}_{-3.8}$ \\
~~~~$R_P/R_*$\dotfill &Radius of planet in stellar radii \dotfill & $0.02408^{+0.0011}_{-0.0009}$ & $0.0373^{+0.0023}_{-0.0020}$ & $0.0315^{+0.0019}_{-0.0014}$ & $0.0288^{+0.0016}_{-0.0011}$ & $0.0298\pm 0.0011$ & $0.03049^{+0.0007}_{-0.00044}$ \\
~~~~$a/R_*$\dotfill &Semi-major axis in stellar radii \dotfill & $26.3^{+1.7}_{-1.6}$ & $25.4^{+1.3}_{-1.2}$ & $24.09^{+0.76}_{-0.73}$ & $20.46^{+0.68}_{-0.69}$ & $4.44^{+0.38}_{-0.33}$ & $13.94^{+0.9}_{-0.87}$ \\
~~~~$\delta$\dotfill &$\left(R_P/R_*\right)^2$ \dotfill & $0.00058^{+5.4e-05}_{-4.3e-05}$ & $0.00139^{+0.00018}_{-0.00014}$ & $0.00099^{+0.00012}_{-8.3e-05}$ & $0.000831^{+9.7e-05}_{-6e-05}$ & $0.000891^{+6.7e-05}_{-6.2e-05}$ & $0.00093^{+4.3e-05}_{-2.7e-05}$ \\
~~~~$Depth_{\rm K2}$\dotfill &Flux decrement at mid transit for \ktwo \dotfill & $0.000693\pm 4.5e-05$ & $0.00169^{+0.00019}_{-0.00017}$ & $0.001377^{+9.7e-05}_{-0.0001}$ & $0.000995^{+4e-05}_{-4.4e-05}$ & $0.000576^{+5.6e-05}_{-6.4e-05}$ & $0.00108^{+2.1e-05}_{-2e-05}$ \\
~~~~$Depth_{\rm TESS}$\dotfill &Flux decrement at mid transit for \tess \dotfill & ---& ---& ---& $0.00096\pm 3e-05$ & $0.000681^{+3.5e-05}_{-3.9e-05}$ & $0.001035\pm 2.4e-05$ \\
~~~~$\tau$\dotfill &Ingress/egress transit duration (days)\dotfill & $0.0085^{+0.0036}_{-0.0012}$ & $0.00476^{+0.0023}_{-0.00066}$ & $0.00382^{+0.0017}_{-0.00059}$ & $0.0042^{+0.0026}_{-0.0012}$ & $0.0348^{+0.0087}_{-0.0073}$ & $0.00729^{+0.0021}_{-0.00096}$ \\
~~~~$T_{14}$\dotfill &Total transit duration (days)\dotfill & $0.3159^{+0.0096}_{-0.0093}$ & $0.1162^{+0.0074}_{-0.0072}$ & $0.1075^{+0.0056}_{-0.006}$ & $0.106^{+0.0028}_{-0.0023}$ & $0.2613^{+0.0077}_{-0.0072}$ & $0.2134^{+0.0024}_{-0.0021}$ \\
~~~~$b$\dotfill &Transit Impact parameter \dotfill & $0.37^{+0.24}_{-0.25}$ & $0.37^{+0.26}_{-0.25}$ & $0.39^{+0.24}_{-0.26}$ & $0.55^{+0.19}_{-0.33}$ & $0.895^{+0.02}_{-0.026}$ & $0.37^{+0.19}_{-0.24}$ \\
~~~~$\rho_P$\dotfill &Density (cgs)\dotfill & $1.45^{+0.53}_{-0.35}$ & $2.55^{+0.95}_{-0.62}$ & $3.04^{+1.2}_{-0.74}$ & $2.69^{+0.97}_{-0.63}$ & $0.392^{+0.14}_{-0.1}$ & $1.4^{+0.71}_{-0.64}$ \\
~~~~$logg_P$\dotfill &Surface gravity \dotfill & $3.02^{+0.13}_{-0.11}$ & $3.08^{+0.13}_{-0.11}$ & $3.1^{+0.14}_{-0.12}$ & $3.08^{+0.13}_{-0.11}$ & $2.974^{+0.095}_{-0.091}$ & $3.12^{+0.16}_{-0.22}$ \\
~~~~$T_S$\dotfill &Time of eclipse (\bjdtdb)\dotfill & $2456810.3^{+6.7}_{-6.4}$ & $2456987.8\pm 2.2$ & $2456979.8\pm 2.1$ & $2457066.7^{+2.1}_{-2.0}$ & $2457138.26^{+0.14}_{-0.15}$ & $2457141.05^{+0.93}_{-0.61}$ \\
~~~~$T_{S,14}$\dotfill &Total eclipse duration (days)\dotfill & $0.305^{+0.060}_{-0.096}$ & $0.108^{+0.02}_{-0.037}$ & $0.105^{+0.022}_{-0.032}$ & $0.103^{+0.028}_{-0.03}$ & $0.0\pm 0.0$ & $0.217^{+0.036}_{-0.027}$ \\
~~~~$e\cos{\omega_*}$\dotfill & \dotfill& $0.0^{+0.37}_{-0.35}$ & $0.0^{+0.34}_{-0.35}$ & $0.0\pm 0.37$ & $0.0^{+0.4}_{-0.39}$ & $0.075^{+0.025}_{-0.027}$ & $0.021^{+0.14}_{-0.094}$ \\
~~~~$e\sin{\omega_*}$\dotfill & \dotfill& $-0.02^{+0.10}_{-0.27}$ & $-0.04^{+0.10}_{-0.29}$ & $-0.01^{+0.11}_{-0.26}$ & $-0.02^{+0.16}_{-0.3}$ & $0.191^{+0.045}_{-0.046}$ & $0.008^{+0.086}_{-0.1}$ \\
~~~~$M_P/M_*$\dotfill &Mass ratio \dotfill & $4.9e-05^{+2.0e-05}_{-1.3e-05}$ & $4.1E-05^{+1.7E-05}_{-1E-05}$ & $2.96e-05^{+1.2e-05}_{-7.5e-06}$ & $2.75e-05^{+1e-05}_{-6.6e-06}$ & $0.000444^{+2.9e-05}_{-3e-05}$ & $8.2e-05^{+3.3e-05}_{-3.5e-05}$ \\
~~~~$d/R_*$\dotfill &Separation at mid transit \dotfill & $24.2^{+4.3}_{-5.7}$ & $24.2^{+3.9}_{-5.3}$ & $21.9^{+3.3}_{-4.7}$ & $18.1^{+3.9}_{-4.1}$ & $3.57^{+0.38}_{-0.34}$ & $13.5^{+1.7}_{-2.0}$ \\
\\
\enddata
\label{tab:med1}
 \begin{flushleft} 
  \footnotesize{
    \textbf{Notes.} See Table 3 in \citet{Eastman:2019} for a detailed description of all parameters. Gaussian and uniform priors are indicated as $\mathcal{G\mathrm{[median, width]}}$ and $\mathcal{U\mathrm{[lower~bound, upper~bound]}}$, respectively. Metallicity priors are adopted from TRES spectra unless otherwise indicated. $^\star$ Gaussian priors were placed on dilution in \tess only for systems with a contamination listed on EXOFOP. ~$^\prime$ The stellar parameters from the global fit are not considered reliable as the SED was not included within this fit. $^*$ From \cite{Huber:2016}. \\
    $^1$The metallicity of the star at birth. $^2$Corresponds to static points in a star's evolutionary history. See \S2 in \citet{Dotter:2016}. $^3$Time of conjunction is commonly reported as the "transit time". $^4$Time of minimum projected separation is a more correct "transit time". $^5$Optimal time of conjunction minimizes the covariance between $T_C$ and Period. $^6$Assumes no albedo and perfect redistribution.
    }
\end{flushleft}
\end{deluxetable*}
\providecommand{\bjdtdb}{\ensuremath{\rm {BJD_{TDB}}}}
\providecommand{\feh}{\ensuremath{\left[{\rm Fe}/{\rm H}\right]}}
\providecommand{\teff}{\ensuremath{T_{\rm eff}}}
\providecommand{\teq}{\ensuremath{T_{\rm eq}}}
\providecommand{\ecosw}{\ensuremath{e\cos{\omega_*}}}
\providecommand{\esinw}{\ensuremath{e\sin{\omega_*}}}
\providecommand{\msun}{\ensuremath{\,M_\Sun}}
\providecommand{\rsun}{\ensuremath{\,R_\Sun}}
\providecommand{\lsun}{\ensuremath{\,L_\Sun}}
\providecommand{\mj}{\ensuremath{\,M_{\rm J}}}
\providecommand{\rj}{\ensuremath{\,R_{\rm J}}}
\providecommand{\me}{\ensuremath{\,M_{\rm E}}}
\providecommand{\re}{\ensuremath{\,R_{\rm E}}}
\providecommand{\fave}{\langle F \rangle}
\providecommand{\fluxcgs}{10$^9$ erg s$^{-1}$ cm$^{-2}$}
\startlongtable
\begin{deluxetable*}{lcccccc}
\clearpage
\newpage
\centering
\tablecaption{Median values and 68\% confidence intervals.}
\tablehead{\omit}
\tabletypesize{\tiny}
\startdata
\multicolumn{2}{l}{Priors:}  & K2-114 & K2-115 & K2-147$^\prime$ & K2-167 & K2-180\\
\hline \\
$\pi$ &  Gaia Parallax (mas) \dotfill & $\mathcal{G}[2.12963, 0.03560]$ & $\mathcal{G}[2.49675, 0.02480]$ & --- & $\mathcal{G}[12.45657, 0.07130]$ & $\mathcal{G}[4.93626, 0.04120]$  \\
$[{\rm Fe/H}]$  & Metallicity (dex) \dotfill  & $\mathcal{G}[0.401, 0.037]^\dagger$ & $\mathcal{G}[-0.23, 0.04]^*$  & ---  & $\mathcal{G}[-0.459, 0.080]$  & $\mathcal{G}[-0.588, 0.080]$ \\
 $A_V$ & V-band extinction (mag) \dotfill & $\mathcal{U}[0, 0.08928]$ & $\mathcal{U}[0, 1.302]$ & --- & $\mathcal{U}[0, 0.12431]$ & $\mathcal{U}[0, 0.08866]$ \\
$D_T^*$&  Dilution in \tess\  \dotfill & ---  & $\mathcal{G}[0,1.40623e-05]$  & ---  & $\mathcal{G}[0, 2.69843e-05]$  & $\mathcal{G}[0, 0.040635]$ \\
\hline\hline
Parameter & Units & Values \\
Stellar Parameters: \\
~~~~$M_*$\dotfill &Mass (\msun)\dotfill & $0.863^{+0.037}_{-0.031}$ & $0.918^{+0.039}_{-0.046}$ & \textcolor{red}{$0.563\pm 0.028$} & $1.084^{+0.1}_{-0.097}$ & $0.735^{+0.035}_{-0.029}$ \\
~~~~$R_*$\dotfill &Radius (\rsun)\dotfill & $0.832^{+0.022}_{-0.021}$ & $0.855^{+0.024}_{-0.021}$ & \textcolor{red}{$0.578\pm 0.028$} & $1.499^{+0.077}_{-0.071}$ & $0.719^{+0.025}_{-0.023}$ \\
~~~~$L_*$\dotfill &Luminosity (\lsun)\dotfill & $0.36\pm 0.015$ & $0.78^{+0.1}_{-0.071}$ & \textcolor{red}{$0.056^{+0.022}_{-0.017}$} & $3.21\pm 0.14$ & $0.386\pm 0.014$ \\
~~~~$F_{Bol}$\dotfill &Bolometric Flux (cgs)\dotfill & $5.18e-11\pm 1.4e-12$ & $1.55e-10^{+2e-11}_{-1.4e-11}$ & ---& $1.594e-08^{+6.6e-10}_{-6.7e-10}$ & $3.011e-10^{+9.6e-12}_{-9.7e-12}$ \\
~~~~$\rho_*$\dotfill &Density (cgs)\dotfill & $2.11^{+0.18}_{-0.17}$ & $2.08^{+0.16}_{-0.19}$ & \textcolor{red}{$4.11^{+0.71}_{-0.58}$} & $0.454^{+0.098}_{-0.086}$ & $2.79^{+0.31}_{-0.28}$ \\
~~~~$\log{g}$\dotfill &Surface gravity (cgs)\dotfill & $4.534^{+0.027}_{-0.026}$ & $4.538^{+0.024}_{-0.033}$ & \textcolor{red}{$4.665\pm 0.048$} & $4.122^{+0.066}_{-0.072}$ & $4.591^{+0.034}_{-0.033}$ \\
~~~~$T_{\rm eff}$\dotfill &Effective Temperature (K)\dotfill & $4899.0^{+59.0}_{-58.0}$ & $5870.0^{+170.0}_{-150.0}$ & \textcolor{red}{$3690.0\pm 300.0$} & $6310.0\pm 170.0$ & $5365.0\pm+92.0$ \\
~~~~$[{\rm Fe/H}]$\dotfill &Metallicity (dex)\dotfill & $0.42^{+0.036}_{-0.037}$ & $-0.198^{+0.043}_{-0.049}$ & \textcolor{red}{$-0.02^{+0.96}_{-1.0}$} & $-0.456\pm 0.081$ & $-0.578^{+0.079}_{-0.080}$ \\
~~~~$[{\rm Fe/H}]_{0}$\dotfill &Initial Metallicity$^{1}$ \dotfill & $0.39\pm 0.046$ & $-0.202^{+0.056}_{-0.058}$ & ---& $-0.253^{+0.075}_{-0.088}$ & $-0.531^{+0.085}_{-0.086}$ \\
~~~~$Age$\dotfill &Age (Gyr)\dotfill & $7.6^{+4.2}_{-4.6}$ & $2.2^{+3.5}_{-1.6}$ & ---& $5.3^{+2.5}_{-2.0}$ & $7.5^{+4.4}_{-4.7}$ \\
~~~~$EEP$\dotfill &Equal Evolutionary Phase$^{2}$ \dotfill & $348.0^{+23.0}_{-26.0}$ & $325.0^{+24.0}_{-39.0}$ & ---& $431.0^{+16.0}_{-39.0}$ & $345.0^{+16.0}_{-25.0}$ \\
~~~~$A_V$\dotfill &V-band extinction (mag)\dotfill & $0.05^{+0.027}_{-0.032}$ & $0.17^{+0.13}_{-0.11}$ & ---& $0.07^{+0.038}_{-0.045}$ & $0.051^{+0.027}_{-0.032}$ \\
~~~~$\sigma_{SED}$\dotfill & SED photometry error scaling\dotfill & $1.09^{+0.48}_{-0.29}$ & $1.13^{+0.5}_{-0.32}$ & ---& $1.65^{+0.59}_{-0.37}$ & $1.48^{+0.58}_{-0.35}$ \\
~~~~$\varpi$\dotfill &Parallax (mas)\dotfill & $2.122^{+0.035}_{-0.034}$ & $2.496\pm 0.025$ & ---& $12.456\pm 0.071$ & $4.937^{+0.042}_{-0.041}$ \\
~~~~$d$\dotfill &Distance (pc)\dotfill & $471.2^{+7.8}_{-7.6}$ & $400.7^{+4.0}_{-3.9}$ & ---& $80.29\pm 0.46$ & $202.6\pm 1.7$ \\
Planetary Parameters:\\
~~~~$P$\dotfill &Period (days)\dotfill & $11.390931^{+3.1e-06}_{-3.2e-06}$ & $20.2729914\pm 5e-06$ & $0.961939\pm 2.9e-05$ & $9.978541^{+2.3e-05}_{-1.9e-05}$ & $8.865663^{+1.1e-05}_{-1e-05}$ \\
~~~~$R_P$\dotfill &Radius (\rj)\dotfill & $0.945^{+0.029}_{-0.028}$ & $1.053^{+0.03}_{-0.028}$ & $0.1314^{+0.0099}_{-0.0088}$ & $0.211^{+0.019}_{-0.013}$ & $0.2200^{+0.0098}_{-0.0086}$ \\
~~~~$M_P$\dotfill &Mass (\mj)\dotfill & $2.01\pm 0.12$ & $1.01^{+0.22}_{-0.23}$ & $0.0101^{+0.0037}_{-0.0023}$ & $0.0224^{+0.0088}_{-0.0056}$ & $0.0359^{+0.0075}_{-0.0068}$ \\
~~~~$T_C$\dotfill &Time of conjunction$^{3}$ (\bjdtdb)\dotfill & $2457140.324^{+0.00023}_{-0.00022}$ & $2458495.17373\pm 0.0003$ & $2457301.9457^{+0.0016}_{-0.0017}$ & $2456979.9331\pm 0.0024$ & $2457143.3957^{+0.00089}_{-0.00088}$ \\
~~~~$T_T$\dotfill &Time of minimum projected separation$^{4}$ (\bjdtdb)\dotfill & $2457140.32397^{+0.00023}_{-0.00022}$ & $2458495.17376\pm 0.00028$ & $2457301.9457^{+0.0014}_{-0.0015}$ & $2456979.9331^{+0.0016}_{-0.0015}$ & $2457143.3957^{+0.00089}_{-0.00088}$ \\
~~~~$T_0$\dotfill &Optimal conjunction Time$^{5}$ (\bjdtdb)\dotfill & $2457687.08869\pm 0.00016$ & $2457522.07014\pm 0.00017$ & $2457343.30907^{+0.001}_{-0.00099}$ & $2457299.2465^{+0.0022}_{-0.0023}$ & $2457489.1566^{+0.00078}_{-0.00076}$ \\
~~~~$a$\dotfill &Semi-major axis (AU)\dotfill & $0.0944^{+0.0013}_{-0.0011}$ & $0.1415^{+0.002}_{-0.0024}$ & $0.01575^{+0.00026}_{-0.00027}$ & $0.0932^{+0.0028}_{-0.0029}$ & $0.0756^{+0.0012}_{-0.001}$ \\
~~~~$i$\dotfill &Inclination (Degrees)\dotfill & $89.16^{+0.18}_{-0.13}$ & $88.912^{+0.057}_{-0.075}$ & $83.6^{+3.6}_{-1.8}$ & $86.8^{+1.6}_{-1.1}$ & $89.15^{+0.51}_{-0.38}$ \\
~~~~$e$\dotfill &Eccentricity \dotfill & $0.079\pm 0.03$ & $0.063^{+0.061}_{-0.042}$ & $0.38^{+0.26}_{-0.22}$ & $0.48\pm 0.26$ & $0.075^{+0.080}_{-0.051}$ \\
~~~~$\omega_*$\dotfill &Argument of Periastron (Degrees)\dotfill & $-50.0^{+22.0}_{-13.0}$ & $137.0^{+46.0}_{-87.0}$ & $40.0^{+100.0}_{-150.0}$ & $140.0^{+120.0}_{-100.0}$ & $130.0^{+120.0}_{-110.0}$ \\
~~~~$T_{eq}$\dotfill &Equilibrium temperature$^{6}$ (K)\dotfill & $701.4^{+7.6}_{-7.7}$ & $696.0^{+19.0}_{-15.0}$ & $1077.0^{+93.0}_{-92.0}$ & $1220.0^{+20.0}_{-19.0}$ & $797.5^{+8.4}_{-8.5}$ \\
~~~~$\tau_{\rm circ}$\dotfill &Tidal circularization timescale (Gyr)\dotfill & $256.0^{+45.0}_{-37.0}$ & $960.0^{+310.0}_{-330.0}$ & $0.089^{+0.24}_{-0.084}$ & $270.0^{+1600.0}_{-270.0}$ & $1950.0^{+750.0}_{-690.0}$ \\
~~~~$K$\dotfill &RV semi-amplitude (m/s)\dotfill & $200.0\pm 10.0$ & $80.0^{+17.0}_{-18.0}$ & $3.44^{+1.4}_{-0.86}$ & $2.44^{+1.3}_{-0.72}$ & $4.36^{+0.90}_{-0.82}$ \\
~~~~$R_P/R_*$\dotfill &Radius of planet in stellar radii \dotfill & $0.1167^{+0.0013}_{-0.0014}$ & $0.12659^{+0.00075}_{-0.00084}$ & $0.0234^{+0.0012}_{-0.0011}$ & $0.01436^{+0.0011}_{-0.0005}$ & $0.03141^{+0.00069}_{-0.00055}$ \\
~~~~$a/R_*$\dotfill &Semi-major axis in stellar radii \dotfill & $24.4^{+0.67}_{-0.66}$ & $35.62^{+0.91}_{-1.1}$ & $5.86^{+0.32}_{-0.29}$ & $13.37\pm 0.9$ & $22.63^{+0.82}_{-0.79}$ \\
~~~~$\delta$\dotfill &$\left(R_P/R_*\right)^2$ \dotfill & $0.01361^{+0.00031}_{-0.00033}$ & $0.01603^{+0.00019}_{-0.00021}$ & $0.000545^{+5.9e-05}_{-4.8e-05}$ & $0.000206^{+3.1e-05}_{-1.4e-05}$ & $0.000987^{+4.4e-05}_{-3.4e-05}$ \\
~~~~$Depth_{\rm K2}$\dotfill &Flux decrement at mid transit for \ktwo \dotfill & $0.01856^{+0.00028}_{-0.00027}$ & $0.01732^{+0.00018}_{-0.00016}$ & $0.0006^{+4.7e-05}_{-4.3e-05}$ & $0.0002287^{+8.7e-06}_{-8.6e-06}$ & $0.00121^{+3.5e-05}_{-3.4e-05}$ \\
~~~~$Depth_{\rm TESS}$\dotfill &Flux decrement at mid transit for \tess \dotfill & $0.0171^{+0.00029}_{-0.0003}$ & $0.01695\pm 0.00015$ & ---& $0.0002225^{+1.2e-05}_{-1e-05}$ & $0.001152\pm3.4e-05$ \\
~~~~$\tau$\dotfill &Ingress/egress transit duration (days)\dotfill & $0.0199\pm 0.0014$ & $0.0299^{+0.0014}_{-0.0015}$ & $0.00114^{+0.00083}_{-0.00028}$ & $0.00301^{+0.0035}_{-0.00089}$ & $0.00403^{+0.00081}_{-0.00041}$ \\
~~~~$T_{14}$\dotfill &Total transit duration (days)\dotfill & $0.1654\pm 0.0012$ & $0.1614\pm 0.0011$ & $0.0376\pm 0.0025$ & $0.15^{+0.0045}_{-0.0042}$ & $0.1187^{+0.0020}_{-0.0019}$ \\
~~~~$b$\dotfill &Transit Impact parameter \dotfill & $0.378^{+0.061}_{-0.085}$ & $0.655^{+0.018}_{-0.021}$ & $0.5^{+0.24}_{-0.34}$ & $0.55^{+0.26}_{-0.36}$ & $0.33^{+0.17}_{-0.20}$ \\
~~~~$\rho_P$\dotfill &Density (cgs)\dotfill & $2.96^{+0.34}_{-0.31}$ & $1.07^{+0.26}_{-0.27}$ & $5.5^{+1.9}_{-1.3}$ & $2.89^{+1.1}_{-0.69}$ & $4.16^{+1.10}_{-0.92}$ \\
~~~~$logg_P$\dotfill &Surface gravity \dotfill & $3.748^{+0.037}_{-0.038}$ & $3.356^{+0.088}_{-0.12}$ & $3.16^{+0.13}_{-0.1}$ & $3.09^{+0.13}_{-0.11}$ & $3.263^{+0.089}_{-0.099}$ \\
~~~~$T_S$\dotfill &Time of eclipse (\bjdtdb)\dotfill & $2457134.99^{+0.11}_{-0.12}$ & $2458505.03^{+0.62}_{-0.89}$ & $2457301.47\pm 0.26$ & $2456974.9^{+3.2}_{-3.3}$ & $2457147.81^{+0.43}_{-0.36}$ \\
~~~~$T_{S,14}$\dotfill &Total eclipse duration (days)\dotfill & $0.1493^{+0.009}_{-0.0079}$ & $0.165^{+0.0043}_{-0.0038}$ & $0.0399^{+0.019}_{-0.0073}$ & $0.151^{+0.094}_{-0.054}$ & $0.121^{+0.014}_{-0.011}$ \\
~~~~$e\cos{\omega_*}$\dotfill & \dotfill& $0.049^{+0.015}_{-0.017}$ & $-0.022^{+0.048}_{-0.069}$ & $0.0\pm 0.43$ & $-0.0^{+0.5}_{-0.52}$ & $-0.003^{+0.076}_{-0.063}$ \\
~~~~$e\sin{\omega_*}$\dotfill & \dotfill& $-0.058^{+0.036}_{-0.034}$ & $0.024^{+0.04}_{-0.028}$ & $0.09^{+0.19}_{-0.21}$ & $0.1^{+0.25}_{-0.34}$ & $0.011^{+0.059}_{-0.062}$ \\
~~~~$M_P/M_*$\dotfill &Mass ratio \dotfill & $0.00222\pm 0.00011$ & $0.00106^{+0.00022}_{-0.00024}$ & $1.72e-05^{+6.4e-06}_{-4e-06}$ & $1.98e-05^{+8.4e-06}_{-5.3e-06}$ & $4.66e-05^{+9.6e-06}_{-8.9e-06}$ \\
~~~~$d/R_*$\dotfill &Separation at mid transit \dotfill & $25.7\pm 1.4$ & $34.5^{+1.9}_{-2.5}$ & $4.3^{+1.5}_{-1.1}$ & $8.7^{+4.7}_{-2.8}$ & $22.3^{+1.8}_{-1.9}$ \\
\enddata
\label{tab:med2}
 \begin{flushleft} 
  \footnotesize{
    \textbf{Notes.} See Table 3 in \citet{Eastman:2019} for a detailed description of all parameters. Gaussian and uniform priors are indicated as $\mathcal{G\mathrm{[median, width]}}$ and $\mathcal{U\mathrm{[lower~bound, upper~bound]}}$, respectively. Metallicity priors are adopted from TRES spectra unless otherwise indicated. $^\star$ Gaussian priors were placed on dilution in \tess only for systems with a contamination listed on EXOFOP. ~$^\prime$ The stellar parameters from the global fit are not considered reliable as the SED was not included within this fit. $^\dagger$ From \cite{Ikwut-Ukwa-SynergyI:2020}. ~$^*$ From \cite{Shporer:2017}. \\
    $^1$The metallicity of the star at birth. $^2$Corresponds to static points in a star's evolutionary history. See \S2 in \citet{Dotter:2016}. $^3$Time of conjunction is commonly reported as the "transit time". $^4$Time of minimum projected separation is a more correct "transit time". $^5$Optimal time of conjunction minimizes the covariance between $T_C$ and Period. $^6$Assumes no albedo and perfect redistribution.
    }
\end{flushleft}
\end{deluxetable*}
\providecommand{\bjdtdb}{\ensuremath{\rm {BJD_{TDB}}}}
\providecommand{\feh}{\ensuremath{\left[{\rm Fe}/{\rm H}\right]}}
\providecommand{\teff}{\ensuremath{T_{\rm eff}}}
\providecommand{\teq}{\ensuremath{T_{\rm eq}}}
\providecommand{\ecosw}{\ensuremath{e\cos{\omega_*}}}
\providecommand{\esinw}{\ensuremath{e\sin{\omega_*}}}
\providecommand{\msun}{\ensuremath{\,M_\Sun}}
\providecommand{\rsun}{\ensuremath{\,R_\Sun}}
\providecommand{\lsun}{\ensuremath{\,L_\Sun}}
\providecommand{\mj}{\ensuremath{\,M_{\rm J}}}
\providecommand{\rj}{\ensuremath{\,R_{\rm J}}}
\providecommand{\me}{\ensuremath{\,M_{\rm E}}}
\providecommand{\re}{\ensuremath{\,R_{\rm E}}}
\providecommand{\fave}{\langle F \rangle}
\providecommand{\fluxcgs}{10$^9$ erg s$^{-1}$ cm$^{-2}$}
\startlongtable
\begin{deluxetable*}{lcccccc}
\clearpage
\newpage
\centering
\tablecaption{Median values and 68\% confidence intervals.}
\tablehead{\omit}
\tabletypesize{\tiny}
\startdata
\multicolumn{2}{l}{Priors:}  & K2-181 & K2-182 & K2-203 & K2-204 & K2-208\\
\hline \\
$\pi$ &  Gaia Parallax (mas) \dotfill & $\mathcal{G}[2.80448, 0.04020]$ & $\mathcal{G}[6.50953, 0.05160]$ & $\mathcal{G}[5.93701, 0.05600]$ & $\mathcal{G}[1.83974, 0.05160]$ & $\mathcal{G}[3.85946, 0.04790]$  \\
$[{\rm Fe/H}]$  & Metallicity (dex) \dotfill  & $\mathcal{G}[0.416, 0.080]$ & $\mathcal{G}[-0.006, 0.080]$  & $\mathcal{G}[-0.073, 0.080]$  & $\mathcal{G}[0.064, 0.080]$  & $\mathcal{G}[-0.116, 0.080]$ \\
 $A_V$ & V-band extinction (mag) \dotfill & $\mathcal{U}[0, 0.10664]$ & $\mathcal{U}[0, 0.086490]$ & $\mathcal{U}[0, 0.13733]$ & $\mathcal{U}[0, 0.10416]$ & $\mathcal{U}[0, 0.08556]$ \\
$D_T^*$&  Dilution in \tess\  \dotfill & ---  & $\mathcal{G}[0, 0.00028648]$  & ---  & ---  & --- \\
\hline\hline
Parameter & Units & Values \\
Stellar Parameters: \\
~~~~$M_*$\dotfill &Mass (\msun)\dotfill & $1.022^{+0.060}_{-0.058}$ & $0.823^{+0.036}_{-0.033}$ & $0.793^{+0.033}_{-0.029}$ & $1.076^{+0.092}_{-0.086}$ & $0.881^{+0.045}_{-0.042}$ \\
~~~~$R_*$\dotfill &Radius (\rsun)\dotfill & $1.04^{+0.043}_{-0.039}$ & $0.789^{+0.025}_{-0.023}$ & $0.765^{+0.024}_{-0.023}$ & $1.253^{+0.055}_{-0.053}$ & $0.872^{+0.033}_{-0.03}$ \\
~~~~$L_*$\dotfill &Luminosity (\lsun)\dotfill & $0.903^{+0.038}_{-0.036}$ & $0.388^{+0.013}_{-0.012}$ & $0.337\pm 0.013$ & $1.58^{+0.11}_{-0.1}$ & $0.629^{+0.027}_{-0.026}$ \\
~~~~$F_{Bol}$\dotfill &Bolometric Flux (cgs)\dotfill & $2.267e-10^{+6.9e-12}_{-6.4e-12}$ & $5.26e-10\pm 1.5e-11$ & $3.79e-10\pm 1.3e-11$ & $1.692e-10^{+6.2e-12}_{-5.7e-12}$ & $2.992e-10^{+1.1e-11}_{-9.9e-12}$ \\
~~~~$\rho_*$\dotfill &Density (cgs)\dotfill & $1.28^{+0.19}_{-0.18}$ & $2.37^{+0.23}_{-0.22}$ & $2.5^{+0.26}_{-0.23}$ & $0.77^{+0.14}_{-0.11}$ & $1.88^{+0.24}_{-0.23}$ \\
~~~~$\log{g}$\dotfill &Surface gravity (cgs)\dotfill & $4.413^{+0.047}_{-0.050}$ & $4.56^{+0.029}_{-0.032}$ & $4.57\pm 0.031$ & $4.273^{+0.056}_{-0.054}$ & $4.503^{+0.04}_{-0.042}$ \\
~~~~$T_{\rm eff}$\dotfill &Effective Temperature (K)\dotfill & $5520.0^{+100.0}_{-110.0}$ & $5128.0^{+79.0}_{-80.0}$ & $5026.0^{+82.0}_{-81.0}$ & $5783.0^{+100.0}_{-98.0}$ & $5500.0\pm 100.0$ \\
~~~~$[{\rm Fe/H}]$\dotfill &Metallicity (dex)\dotfill & $0.385\pm 0.057$ & $0.022^{+0.077}_{-0.066}$ & $-0.021^{+0.064}_{-0.041}$ & $0.21^{+0.18}_{-0.19}$ & $-0.064\pm 0.054$ \\
~~~~$[{\rm Fe/H}]_{0}$\dotfill &Initial Metallicity$^{1}$ \dotfill & $0.366^{+0.058}_{-0.060}$ & $0.025^{+0.077}_{-0.071}$ & $-0.01^{+0.063}_{-0.057}$ & $0.23^{+0.15}_{-0.16}$ & $-0.048\pm 0.063$ \\
~~~~$Age$\dotfill &Age (Gyr)\dotfill & $5.9^{+4.5}_{-3.8}$ & $6.2^{+4.9}_{-4.2}$ & $6.9^{+4.6}_{-4.5}$ & $6.5^{+3.7}_{-3.1}$ & $6.1^{+4.8}_{-4.1}$ \\
~~~~$EEP$\dotfill &Equal Evolutionary Phase$^{2}$ \dotfill & $374.0^{+36.0}_{-43.0}$ & $341.0^{+19.0}_{-31.0}$ & $341.0^{+15.0}_{-29.0}$ & $413.0^{+20.0}_{-44.0}$ & $349.0^{+35.0}_{-30.0}$ \\
~~~~$A_V$\dotfill &V-band extinction (mag)\dotfill & $0.046^{+0.036}_{-0.030}$ & $0.05^{+0.026}_{-0.032}$ & $0.073^{+0.044}_{-0.047}$ & $0.063^{+0.029}_{-0.038}$ & $0.049^{+0.025}_{-0.032}$ \\
~~~~$\sigma_{SED}$\dotfill & SED photometry error scaling\dotfill & $0.341^{+0.140}_{-0.085}$ & $1.1^{+0.45}_{-0.27}$ & $0.8^{+0.33}_{-0.2}$ & $1.01^{+0.46}_{-0.27}$ & $0.97^{+0.41}_{-0.25}$ \\
~~~~$\varpi$\dotfill &Parallax (mas)\dotfill & $2.802\pm 0.04$ & $6.508\pm 0.052$ & $5.935^{+0.055}_{-0.056}$ & $1.83\pm 0.052$ & $3.857\pm 0.048$ \\
~~~~$d$\dotfill &Distance (pc)\dotfill & $356.9^{+5.2}_{-5.1}$ & $153.7\pm 1.2$ & $168.5\pm 1.6$ & $546.0^{+16.0}_{-15.0}$ & $259.3\pm 3.2$ \\
Planetary Parameters:\\
~~~~$P$\dotfill &Period (days)\dotfill & $6.893813\pm 1.1e-05$ & $4.7369696\pm 1.7e-06$ & $9.6952\pm 0.0014$ & $7.055908^{+5.8e-05}_{-7.4e-05}$ & $4.19097\pm 0.00023$ \\
~~~~$R_P$\dotfill &Radius (\rj)\dotfill & $0.253^{+0.019}_{-0.014}$ & $0.242^{+0.0099}_{-0.0081}$ & $0.1129^{+0.0091}_{-0.0071}$ & $0.283^{+0.016}_{-0.014}$ & $0.1494^{+0.011}_{-0.0082}$ \\
~~~~$M_P$\dotfill &Mass (\mj)\dotfill & $0.0304^{+0.0110}_{-0.0073}$ & $0.066^{+0.017}_{-0.016}$ & $0.0081^{+0.0029}_{-0.0017}$ & $0.0365^{+0.014}_{-0.0089}$ & $0.0125^{+0.0046}_{-0.003}$ \\
~~~~$T_C$\dotfill &Time of conjunction$^{3}$ (\bjdtdb)\dotfill & $2457143.7954^{0.0017}_{0.0016}$ & $2457145.9418\pm 0.00033$ & $2457396.6382^{+0.0067}_{-0.0065}$ & $2457396.5078^{+0.0022}_{-0.0023}$ & $2457396.5113\pm 0.0026$ \\
~~~~$T_T$\dotfill &Time of minimum projected separation$^{4}$ (\bjdtdb)\dotfill & $2457143.7954\pm 0.0014$ & $2457145.94181\pm 0.00032$ & $2457396.6382^{+0.0066}_{-0.0064}$ & $2457396.5078\pm 0.002$ & $2457396.5113^{+0.0024}_{-0.0023}$ \\
~~~~$T_0$\dotfill &Optimal conjunction Time$^{5}$ (\bjdtdb)\dotfill & $2457778.0262\pm 0.0012$ & $2457652.79755^{+0.00027}_{-0.00028}$ & $2457435.4189^{+0.0037}_{-0.0036}$ & $2457452.955\pm 0.0022$ & $2457430.039\pm 0.0016$ \\
~~~~$a$\dotfill &Semi-major axis (AU)\dotfill & $0.0714\pm 0.0014$ & $0.05174^{+0.00075}_{-0.0007}$ & $0.0824^{+0.0011}_{-0.001}$ & $0.0738\pm 0.002$ & $0.04878^{+0.00082}_{-0.00079}$ \\
~~~~$i$\dotfill &Inclination (Degrees)\dotfill & $87.4^{+1.5}_{-0.71}$ & $88.91^{+0.71}_{-0.69}$ & $89.12^{+0.59}_{-0.57}$ & $88.76^{+0.86}_{-1.0}$ & $86.94^{+1.8}_{-0.76}$ \\
~~~~$e$\dotfill &Eccentricity \dotfill & $0.40^{+0.30}_{-0.22}$ & $0.071^{+0.11}_{-0.051}$ & $0.23^{+0.42}_{-0.18}$ & $0.28^{+0.37}_{-0.19}$ & $0.37^{+0.31}_{-0.21}$ \\
~~~~$\omega_*$\dotfill &Argument of Periastron (Degrees)\dotfill & $20.0^{+110.0}_{-180.0}$ & $-160.0^{+110.0}_{-130.0}$ & $-124.0^{+100.0}_{-74.0}$ & $-88.0^{+61.0}_{-63.0}$ & $-210.0^{+160.0}_{-110.0}$ \\
~~~~$T_{eq}$\dotfill &Equilibrium temperature$^{6}$ (K)\dotfill & $1015.0\pm 13.0$ & $965.8^{+10.0}_{-9.8}$ & $738.4^{+8.0}_{-7.9}$ & $1148.0^{+25.0}_{-24.0}$ & $1122.0^{+15.0}_{-14.0}$ \\
~~~~$\tau_{\rm circ}$\dotfill &Tidal circularization timescale (Gyr)\dotfill & $61.0^{+170.0}_{-60.0}$ & $153.0^{+64.0}_{-62.0}$ & $11000.0^{+12000.0}_{-11000.0}$ & $120.0^{+160.0}_{-120.0}$ & $49.0^{+110.0}_{-48.0}$ \\
~~~~$K$\dotfill &RV semi-amplitude (m/s)\dotfill & $3.7^{+1.8}_{-1.0}$ & $9.1\pm 2.3$ & $0.99^{+0.52}_{-0.25}$ & $4.1^{+2.0}_{-1.1}$ & $1.96^{+0.9}_{-0.52}$ \\
~~~~$R_P/R_*$\dotfill &Radius of planet in stellar radii \dotfill & $0.02485^{+0.00160}_{-0.00085}$ & $0.03143^{+0.00068}_{-0.00036}$ & $0.01518^{+0.001}_{-0.00084}$ & $0.02319^{+0.00077}_{-0.00063}$ & $0.01757^{+0.0011}_{-0.00073}$ \\
~~~~$a/R_*$\dotfill &Semi-major axis in stellar radii \dotfill & $14.76^{+0.71}_{-0.72}$ & $14.11^{+0.44}_{-0.46}$ & $23.16^{+0.76}_{-0.75}$ & $12.64^{+0.73}_{-0.63}$ & $12.03^{+0.5}_{-0.51}$ \\
~~~~$\delta$\dotfill &$\left(R_P/R_*\right)^2$ \dotfill & $0.000618^{+8.3e-05}_{-4.1e-05}$ & $0.000988^{+4.3e-05}_{-2.3e-05}$ & $0.00023^{+3.3e-05}_{-2.5e-05}$ & $0.000538^{+3.6e-05}_{-2.9e-05}$ & $0.000309^{+3.9e-05}_{-2.5e-05}$ \\
~~~~$Depth_{\rm K2}$\dotfill &Flux decrement at mid transit for \ktwo \dotfill & $0.000747^{+2.9e-05}_{-3.0-05}$ & $0.001308\pm 2.5e-05$ & $0.0003\pm 3e-05$ & $0.000645^{+3.2e-05}_{-3e-05}$ & $0.000365\pm 2.1e-05$ \\
~~~~$Depth_{\rm TESS}$\dotfill &Flux decrement at mid transit for \tess \dotfill & ---& $0.001226^{+2.3e-05}_{-2.4e-05}$ & ---& $0.000623\pm 2.9e-05$ & ---\\
~~~~$\tau$\dotfill &Ingress/egress transit duration (days)\dotfill & $0.00329^{+0.0025}_{-0.00077}$ & $0.0035^{+0.00071}_{-0.00025}$ & $0.00216^{+0.001}_{-0.00027}$ & $0.00462^{+0.0016}_{-0.00035}$ & $0.00183^{+0.0013}_{-0.00044}$ \\
~~~~$T_{14}$\dotfill &Total transit duration (days)\dotfill & $0.105^{+0.0028}_{-0.0026}$ & $0.10693^{+0.001}_{-0.00088}$ & $0.1293^{+0.0078}_{-0.008}$ & $0.1904^{+0.005}_{-0.0046}$ & $0.081^{+0.0036}_{-0.0035}$ \\
~~~~$b$\dotfill &Transit Impact parameter \dotfill & $0.48^{+0.25}_{-0.33}$ & $0.27^{+0.19}_{-0.18}$ & $0.33^{+0.29}_{-0.23}$ & $0.27^{+0.27}_{-0.19}$ & $0.49^{+0.24}_{-0.33}$ \\
~~~~$\rho_P$\dotfill &Density (cgs)\dotfill & $2.28^{+0.83}_{-0.54}$ & $5.7^{+1.6}_{-1.5}$ & $6.7^{+2.4}_{-1.3}$ & $1.98^{+0.75}_{-0.46}$ & $4.5^{+1.6}_{-1.1}$ \\
~~~~$logg_P$\dotfill &Surface gravity \dotfill & $3.06^{+0.13}_{-0.11}$ & $3.44^{+0.1}_{-0.13}$ & $3.187^{+0.12}_{-0.095}$ & $3.05^{+0.14}_{-0.11}$ & $3.13^{+0.13}_{-0.11}$ \\
~~~~$T_S$\dotfill &Time of eclipse (\bjdtdb)\dotfill & $2457140.4\pm 2.1$ & $2457148.3^{+0.22}_{-0.29}$ & $2457401.5^{+2.0}_{-2.1}$ & $2457400.0^{+1.5}_{-1.6}$ & $2457398.6^{+1.1}_{-1.2}$ \\
~~~~$T_{S,14}$\dotfill &Total eclipse duration (days)\dotfill & $0.109^{+0.058}_{-0.022}$ & $0.105^{+0.0072}_{-0.012}$ & $0.114^{+0.02}_{-0.041}$ & $0.144^{+0.032}_{-0.047}$ & $0.084^{+0.039}_{-0.017}$ \\
~~~~$e\cos{\omega_*}$\dotfill & \dotfill& $-0.0^{+0.50}_{-0.48}$ & $-0.003^{+0.074}_{-0.095}$ & $-0.0^{+0.32}_{-0.34}$ & $-0.0^{+0.33}_{-0.34}$ & $-0.0^{+0.43}_{-0.47}$ \\
~~~~$e\sin{\omega_*}$\dotfill & \dotfill& $0.08^{+0.2}_{-0.26}$ & $-0.009^{+0.035}_{-0.082}$ & $-0.073^{+0.097}_{-0.3}$ & $-0.16^{+0.12}_{-0.24}$ & $0.06^{+0.19}_{-0.25}$ \\
~~~~$M_P/M_*$\dotfill &Mass ratio \dotfill & $2.84e-05^{+1.1e-05}_{-7.1e-06}$ & $7.6e-05\pm 1.9e-05$ & $9.7e-06^{+3.5e-06}_{-2.1e-06}$ & $3.24e-05^{+1.3e-05}_{-8.1e-06}$ & $1.35e-05^{+5.1e-06}_{-3.3e-06}$ \\
~~~~$d/R_*$\dotfill &Separation at mid transit \dotfill & $10.3^{+3.9}_{-2.7}$ & $14.23^{+1.1}_{-0.91}$ & $22.7^{+3.0}_{-4.1}$ & $13.1\pm 2.2$ & $9.0^{+3.0}_{-2.3}$ \\
\enddata
\label{tab:med3}
 \begin{flushleft} 
  \footnotesize{
    \textbf{Notes.} See Table 3 in \citet{Eastman:2019} for a detailed description of all parameters. Gaussian and uniform priors are indicated as $\mathcal{G\mathrm{[median, width]}}$ and $\mathcal{U\mathrm{[lower~bound, upper~bound]}}$, respectively. Metallicity priors are adopted from TRES spectra unless otherwise indicated. $^\star$ Gaussian priors were placed on dilution in \tess only for systems with a contamination listed on EXOFOP.\\
    $^1$The metallicity of the star at birth. $^2$Corresponds to static points in a star's evolutionary history. See \S2 in \citet{Dotter:2016}. $^3$Time of conjunction is commonly reported as the "transit time". $^4$Time of minimum projected separation is a more correct "transit time". $^5$Optimal time of conjunction minimizes the covariance between $T_C$ and Period. $^6$Assumes no albedo and perfect redistribution.
    }
\end{flushleft}
\end{deluxetable*}
\providecommand{\bjdtdb}{\ensuremath{\rm {BJD_{TDB}}}}
\providecommand{\feh}{\ensuremath{\left[{\rm Fe}/{\rm H}\right]}}
\providecommand{\teff}{\ensuremath{T_{\rm eff}}}
\providecommand{\teq}{\ensuremath{T_{\rm eq}}}
\providecommand{\ecosw}{\ensuremath{e\cos{\omega_*}}}
\providecommand{\esinw}{\ensuremath{e\sin{\omega_*}}}
\providecommand{\msun}{\ensuremath{\,M_\Sun}}
\providecommand{\rsun}{\ensuremath{\,R_\Sun}}
\providecommand{\lsun}{\ensuremath{\,L_\Sun}}
\providecommand{\mj}{\ensuremath{\,M_{\rm J}}}
\providecommand{\rj}{\ensuremath{\,R_{\rm J}}}
\providecommand{\me}{\ensuremath{\,M_{\rm E}}}
\providecommand{\re}{\ensuremath{\,R_{\rm E}}}
\providecommand{\fave}{\langle F \rangle}
\providecommand{\fluxcgs}{10$^9$ erg s$^{-1}$ cm$^{-2}$}
\startlongtable
\begin{deluxetable*}{lcccccc}
\clearpage
\newpage
\centering
\tablecaption{Median values and 68\% confidence intervals.}
\tablehead{\omit}
\tabletypesize{\tiny}
\startdata
\multicolumn{2}{l}{Priors:}  & K2-211 & K2-225 & K2-226 & K2-237 & K2-250\\
\hline \\
$\pi$ &  Gaia Parallax (mas) \dotfill & $\mathcal{G}[3.60379, 0.05510]$ & $\mathcal{G}[2.79562, 0.04500]$ & $\mathcal{G}[4.80729, 0.07400]$ & $\mathcal{G}[3.29816, 0.07060]$ & $\mathcal{G}[2.47572, 0.03610]$  \\
$[{\rm Fe/H}]$  & Metallicity (dex) \dotfill  & $\mathcal{G}[0.115, 0.080]$ & $\mathcal{G}[0.471, 0.080]$  & $\mathcal{G}[-0.082, 0.080]$  & $\mathcal{G}[0.357, 0.080]$  & $\mathcal{G}[-0.227, 0.280]^\dagger$ \\
 $A_V$ & V-band extinction (mag) \dotfill & $\mathcal{U}[0, 0.08618]$ & $\mathcal{U}[0, 0.11997]$ & $\mathcal{U}[0, 0.21266]$ & $\mathcal{U}[0, 0.58342]$ & $\mathcal{U}[0, 0.11997]$ \\
$D_T^*$&  Dilution in \tess\  \dotfill & --- & ---  & ---  & $\mathcal{G}[0, 0.079682]$  & --- \\
$D_K$&  Dilution in \ktwo  \dotfill & --- & ---  & ---  & $\mathcal{G}[0, 0.050]$  & --- \\
\hline\hline
Parameter & Units & Values \\
Stellar Parameters: \\
~~~~$M_*$\dotfill &Mass (\msun)\dotfill & $0.851^{+0.039}_{-0.035}$ & $1.482^{+0.11}_{-0.096}$ & $0.856^{+0.042}_{-0.03}$ & $1.256^{+0.055}_{-0.062}$ & $0.809^{+0.044}_{-0.036}$ \\
~~~~$R_*$\dotfill &Radius (\rsun)\dotfill & $0.818^{+0.029}_{-0.027}$ & $1.700^{+0.094}_{-0.087}$ & $0.889^{+0.034}_{-0.032}$ & $1.236^{+0.043}_{-0.036}$ & $0.797^{+0.029}_{-0.027}$ \\
~~~~$L_*$\dotfill &Luminosity (\lsun)\dotfill & $0.414^{+0.017}_{-0.016}$ & $2.41\pm 0.11$ & $0.558\pm 0.029$ & $2.01^{+0.27}_{-0.21}$ & $0.359^{+0.017}_{-0.016}$ \\
~~~~$F_{Bol}$\dotfill &Bolometric Flux (cgs)\dotfill & $1.715e-10\pm 4.6e-12$ & $6.01e-10^{+1.8e-11}_{-2e-11}$ & $4.14e-10^{+1.6e-11}_{-1.9e-11}$ & $7e-10^{+9.1e-11}_{-7e-11}$ & $7.04e-11^{+2.6e-12}_{-2.4e-12}$ \\
~~~~$\rho_*$\dotfill &Density (cgs)\dotfill & $2.19^{+0.25}_{-0.23}$ & $0.425^{+0.047}_{-0.043}$ & $1.72^{+0.22}_{-0.19}$ & $0.937^{+0.073}_{-0.085}$ & $2.26^{+0.24}_{-0.22}$ \\
~~~~$\log{g}$\dotfill &Surface gravity (cgs)\dotfill & $4.543^{+0.033}_{-0.034}$ & $4.148\pm 0.025$ & $4.473^{+0.04}_{-0.034}$ & $4.353^{+0.022}_{-0.029}$ & $4.544^{+0.033}_{-0.032}$ \\
~~~~$T_{\rm eff}$\dotfill &Effective Temperature (K)\dotfill & $5117.0^{+85.0}_{-86.0}$ & $5520.0\pm 140.0$ & $5288.0\pm 99.0$ & $6180.0^{+160.0}_{-140.0}$ & $5003.0\pm 94.0$ \\
~~~~$[{\rm Fe/H}]$\dotfill &Metallicity (dex)\dotfill & $0.134^{+0.077}_{-0.078}$ & $0.486^{+0.061}_{-0.072}$ & $-0.05^{+0.068}_{-0.052}$ & $0.337\pm 0.076$ & $-0.02^{+0.24}_{-0.098}$ \\
~~~~$[{\rm Fe/H}]_{0}$\dotfill &Initial Metallicity$^{1}$ \dotfill & $0.13^{+0.078}_{-0.079}$ & $0.405^{+0.057}_{-0.067}$ & $-0.013^{+0.066}_{-0.06}$ & $0.309^{+0.069}_{-0.07}$ & $0.01^{+0.21}_{-0.1}$ \\
~~~~$Age$\dotfill &Age (Gyr)\dotfill & $6.5^{+4.8}_{-4.3}$ & $0.0121^{+0.0039}_{-0.0032}$ & $9.6^{+3.0}_{-4.7}$ & $1.09^{+1.5}_{-0.78}$ & $8.2^{+4.0}_{-5.0}$ \\
~~~~$EEP$\dotfill &Equal Evolutionary Phase$^{2}$ \dotfill & $344.0^{+24.0}_{-31.0}$ & $188.0^{+4.3}_{-4.8}$ & $373.0^{+20.0}_{-30.0}$ & $324.0^{+28.0}_{-40.0}$ & $347.0^{+19.0}_{-25.0}$ \\
~~~~$A_V$\dotfill &V-band extinction (mag)\dotfill & $0.05^{+0.026}_{-0.032}$ & $0.08^{+0.029}_{-0.046}$ & $0.164^{+0.037}_{-0.07}$ & $0.24^{+0.15}_{-0.14}$ & $0.071^{+0.035}_{-0.045}$ \\
~~~~$\sigma_{SED}$\dotfill & SED photometry error scaling\dotfill & $1.0^{+0.4}_{-0.25}$ & $0.78^{+0.32}_{-0.19}$ & $1.18^{+0.5}_{-0.29}$ & $1.81^{+0.73}_{-0.46}$ & $1.6^{+0.66}_{-0.4}$ \\
~~~~$\varpi$\dotfill &Parallax (mas)\dotfill & $3.599\pm 0.054$ & $2.792\pm 0.045$ & $4.812^{+0.073}_{-0.074}$ & $3.304^{+0.067}_{-0.068}$ & $2.478\pm 0.036$ \\
~~~~$d$\dotfill &Distance (pc)\dotfill & $277.9^{+4.2}_{-4.1}$ & $358.1^{+5.8}_{-5.7}$ & $207.8^{+3.3}_{-3.1}$ & $302.7^{+6.3}_{-6.0}$ & $403.6^{+6.0}_{-5.8}$ \\
Planetary Parameters:\\
~~~~$P$\dotfill &Period (days)\dotfill & $0.669561^{+3.1e-05}_{-3.2e-05}$ & $15.8723^{+0.0021}_{-0.0019}$ & $3.27109^{+0.00036}_{-0.00039}$ & $2.18053332\pm 5.4e-07$ & $4.01392\pm 0.00029$ \\
~~~~$R_P$\dotfill &Radius (\rj)\dotfill & $0.1188^{+0.0078}_{-0.0064}$ & $0.322^{+0.025}_{-0.020}$ & $0.1373^{+0.0092}_{-0.0077}$ & $1.433^{+0.056}_{-0.049}$ & $0.242^{+0.016}_{-0.012}$ \\
~~~~$M_P$\dotfill &Mass (\mj)\dotfill & $0.0087^{+0.003}_{-0.0017}$ & $0.046^{+0.017}_{-0.011}$ & $0.0108^{+0.004}_{-0.0025}$ & $1.366^{+0.11}_{-0.092}$ & $0.0282^{+0.011}_{-0.0068}$ \\
~~~~$T_C$\dotfill &Time of conjunction$^{3}$ (\bjdtdb)\dotfill & $2457393.8134\pm 0.0023$ & $2457587.3665^{+0.0049}_{-0.0055}$ & $2457584.0262^{+0.0049}_{-0.0046}$ & $2457656.463914\pm 3.3e-05$ & $2457584.1282\pm 0.0031$ \\
~~~~$T_T$\dotfill &Time of minimum projected separation$^{4}$ (\bjdtdb)\dotfill & $2457393.8134\pm 0.0021$ & $2457587.3665^{+0.0046}_{-0.0051}$ & $2457584.0262^{+0.0048}_{-0.0044}$ & $2457656.463912\pm 3.1e-05$ & $2457584.1281^{+0.003}_{-0.0029}$ \\
~~~~$T_0$\dotfill &Optimal conjunction Time$^{5}$ (\bjdtdb)\dotfill & $2457432.6479\pm 0.0013$ & $2457619.1111^{+0.0031}_{-0.0030}$ & $2457620.0082\pm 0.002$ & $2457706.61618\pm 3e-05$ & $2457620.2535\pm 0.0015$ \\
~~~~$a$\dotfill &Semi-major axis (AU)\dotfill & $0.01419^{+0.00021}_{-0.0002}$ & $0.1409^{+0.0033}_{-0.0031}$ & $0.04095^{+0.00066}_{-0.00049}$ & $0.03552^{+0.00051}_{-0.0006}$ & $0.04606^{+0.00082}_{-0.00068}$ \\
~~~~$i$\dotfill &Inclination (Degrees)\dotfill & $84.1^{+3.8}_{-3.0}$ & $88.64^{+0.85}_{-0.60}$ & $87.6^{+1.5}_{-1.1}$ & $88.37^{+1.0}_{-0.88}$ & $88.26^{+1.1}_{-0.94}$ \\
~~~~$e$\dotfill &Eccentricity \dotfill & $0.19^{+0.3}_{-0.14}$ & $0.23^{+0.39}_{-0.16}$ & $0.22^{+0.37}_{-0.16}$ & $0.03^{+0.033}_{-0.021}$ & $0.21^{+0.39}_{-0.16}$ \\
~~~~$\omega_*$\dotfill &Argument of Periastron (Degrees)\dotfill & $-150.0^{+130.0}_{-110.0}$ & $-160.0^{+130.0}_{-120.0}$ & $-33.0^{+99.0}_{-130.0}$ & $73.0^{+57.0}_{-58.0}$ & $-40.0^{+92.0}_{-120.0}$ \\
~~~~$T_{eq}$\dotfill &Equilibrium temperature$^{6}$ (K)\dotfill & $1873.0^{+22.0}_{-21.0}$ & $923.0^{+13.0}_{-14.0}$ & $1187.0^{+16.0}_{-17.0}$ & $1759.0^{+49.0}_{-42.0}$ & $1003.0\pm 14.0$ \\
~~~~$\tau_{\rm circ}$\dotfill &Tidal circularization timescale (Gyr)\dotfill & $0.107^{+0.087}_{-0.095}$ & $4000.0^{+4500.0}_{-3900.0}$ & $53.0^{+57.0}_{-51.0}$ & $0.0231^{+0.0044}_{-0.0043}$ & $19.0^{+20.0}_{-19.0}$ \\
~~~~$K$\dotfill &RV semi-amplitude (m/s)\dotfill & $2.38^{+0.89}_{-0.5}$ & $3.13^{+1.5}_{-0.83}$ & $1.79^{+0.81}_{-0.46}$ & $184.0^{+14.0}_{-11.0}$ & $4.5^{+2.2}_{-1.2}$ \\
~~~~$R_P/R_*$\dotfill &Radius of planet in stellar radii \dotfill & $0.01489^{+0.00079}_{-0.00059}$ & $0.01942^{+0.00091}_{-0.00064}$ & $0.01586^{+0.00082}_{-0.00068}$ & $0.119^{+0.0024}_{-0.0023}$ & $0.0311^{+0.0016}_{-0.001}$ \\
~~~~$a/R_*$\dotfill &Semi-major axis in stellar radii \dotfill & $3.73\pm 0.13$ & $17.82^{+0.64}_{-0.62}$ & $9.91^{+0.41}_{-0.37}$ & $6.18^{+0.16}_{-0.19}$ & $12.44^{+0.43}_{-0.42}$ \\
~~~~$\delta$\dotfill &$\left(R_P/R_*\right)^2$ \dotfill & $0.000222^{+2.4e-05}_{-1.7e-05}$ & $0.000377^{+3.6e-05}_{-2.4e-05}$ & $0.000252^{+2.7e-05}_{-2.1e-05}$ & $0.01416^{+0.00058}_{-0.00053}$ & $0.00097^{+0.0001}_{-6.2e-05}$ \\
~~~~$Depth_{\rm K2}$\dotfill &Flux decrement at mid transit for \ktwo \dotfill & $0.000287\pm 1.9e-05$ & $0.000465\pm2.6e-05$ & $0.000314\pm 2.3e-05$ & $0.0169^{+0.00072}_{-0.00067}$ & $0.001284^{+7e-05}_{-7.2e-05}$ \\
~~~~$Depth_{\rm TESS}$\dotfill &Flux decrement at mid transit for \tess \dotfill & ---& ---& ---& $0.01624^{+0.00072}_{-0.00068}$ & ---\\
~~~~$\tau$\dotfill &Ingress/egress transit duration (days)\dotfill & $0.00092^{+0.00037}_{-0.00012}$ & $0.00572^{+0.0026}_{-0.00082}$ & $0.00176^{+0.00076}_{-0.00025}$ & $0.01347^{+0.00054}_{-0.00041}$ & $0.00336^{+0.0014}_{-0.00041}$ \\
~~~~$T_{14}$\dotfill &Total transit duration (days)\dotfill & $0.0538^{+0.003}_{-0.0031}$ & $0.258^{+0.0073}_{-0.0064}$ & $0.0968^{+0.0046}_{-0.0047}$ & $0.12197^{+0.00038}_{-0.00034}$ & $0.0986\pm 0.0037$ \\
~~~~$b$\dotfill &Transit Impact parameter \dotfill & $0.36\pm 0.24$ & $0.38^{+0.24}_{-0.25}$ & $0.38^{+0.24}_{-0.25}$ & $0.171^{+0.091}_{-0.11}$ & $0.35^{+0.25}_{-0.23}$ \\
~~~~$\rho_P$\dotfill &Density (cgs)\dotfill & $6.4^{+2.2}_{-1.3}$ & $1.67^{+0.61}_{-0.4}$ & $5.1^{+1.9}_{-1.2}$ & $0.577^{+0.069}_{-0.064}$ & $2.41^{+0.88}_{-0.56}$ \\
~~~~$logg_P$\dotfill &Surface gravity \dotfill & $3.182^{+0.12}_{-0.09}$ & $3.03^{+0.13}_{-0.11}$ & $3.15^{+0.13}_{-0.1}$ & $3.218\pm 0.039$ & $3.07^{+0.13}_{-0.11}$ \\
~~~~$T_S$\dotfill &Time of eclipse (\bjdtdb)\dotfill & $2457393.48^{+0.11}_{-0.12}$ & $2457595.3^{3.4}_{3.5}$ & $2457582.39\pm 0.68$ & $2457655.38^{+0.031}_{-0.021}$ & $2457586.13\pm 0.81$ \\
~~~~$T_{S,14}$\dotfill &Total eclipse duration (days)\dotfill & $0.0517^{+0.0079}_{-0.011}$ & $0.249^{+0.046}_{-0.081}$ & $0.092^{+0.016}_{-0.027}$ & $0.1266^{+0.0087}_{-0.0049}$ & $0.092^{+0.014}_{-0.029}$ \\
~~~~$e\cos{\omega_*}$\dotfill & \dotfill& $-0.0^{+0.26}_{-0.27}$ & $0.0\pm ^{0.34}_{0.35}$ & $0.0^{+0.32}_{-0.33}$ & $0.005^{+0.023}_{-0.015}$ & $-0.0^{+0.31}_{-0.32}$ \\
~~~~$e\sin{\omega_*}$\dotfill & \dotfill& $-0.024^{+0.083}_{-0.18}$ & $-0.023^{+0.097}_{-0.28}$ & $-0.033^{+0.094}_{-0.25}$ & $0.019^{+0.035}_{-0.02}$ & $-0.038^{+0.083}_{-0.26}$ \\
~~~~$M_P/M_*$\dotfill &Mass ratio \dotfill & $9.8e-06^{+3.4e-06}_{-1.9e-06}$ & $2.94e-05^{+1.1e-05}_{-7e-06}$ & $1.2e-05^{+4.5e-06}_{-2.8e-06}$ & $0.001041^{+8.2e-05}_{-6.4e-05}$ & $3.32e-05^{+1.3e-05}_{-8e-06}$ \\
~~~~$d/R_*$\dotfill &Separation at mid transit \dotfill & $3.62^{+0.46}_{-0.52}$ & $16.6^{+2.5}_{-3.1}$ & $9.4^{+1.3}_{-1.7}$ & $6.06^{+0.27}_{-0.39}$ & $12.0^{+1.5}_{-2.0}$ \\
\label{tab:med4}
\enddata
 \begin{flushleft} 
  \footnotesize{
    \textbf{Notes.} See Table 3 in \citet{Eastman:2019} for a detailed description of all parameters. Gaussian and uniform priors are indicated as $\mathcal{G\mathrm{[median, width]}}$ and $\mathcal{U\mathrm{[lower~bound, upper~bound]}}$, respectively. Metallicity priors are adopted from TRES spectra unless otherwise indicated. $^\star$ Gaussian priors were placed on dilution in \tess only for systems with a contamination listed on EXOFOP. $^\dagger$ From \cite{Huber:2016}. \\
    $^1$The metallicity of the star at birth. $^2$Corresponds to static points in a star's evolutionary history. See \S2 in \citet{Dotter:2016}. $^3$Time of conjunction is commonly reported as the "transit time". $^4$Time of minimum projected separation is a more correct "transit time". $^5$Optimal time of conjunction minimizes the covariance between $T_C$ and Period. $^6$Assumes no albedo and perfect redistribution.
    }
\end{flushleft}
\end{deluxetable*}
\providecommand{\bjdtdb}{\ensuremath{\rm {BJD_{TDB}}}}
\providecommand{\feh}{\ensuremath{\left[{\rm Fe}/{\rm H}\right]}}
\providecommand{\teff}{\ensuremath{T_{\rm eff}}}
\providecommand{\teq}{\ensuremath{T_{\rm eq}}}
\providecommand{\ecosw}{\ensuremath{e\cos{\omega_*}}}
\providecommand{\esinw}{\ensuremath{e\sin{\omega_*}}}
\providecommand{\msun}{\ensuremath{\,M_\Sun}}
\providecommand{\rsun}{\ensuremath{\,R_\Sun}}
\providecommand{\lsun}{\ensuremath{\,L_\Sun}}
\providecommand{\mj}{\ensuremath{\,M_{\rm J}}}
\providecommand{\rj}{\ensuremath{\,R_{\rm J}}}
\providecommand{\me}{\ensuremath{\,M_{\rm E}}}
\providecommand{\re}{\ensuremath{\,R_{\rm E}}}
\providecommand{\fave}{\langle F \rangle}
\providecommand{\fluxcgs}{10$^9$ erg s$^{-1}$ cm$^{-2}$}
\startlongtable
\begin{deluxetable*}{lccccccc}
\clearpage
\newpage
\centerwidetable
\tablecaption{Median values and 68\% confidence intervals.}
\tablehead{\omit}
\tabletypesize{\tiny}
\startdata
\multicolumn{2}{l}{Priors:}  & K2-260 & \multicolumn{2}{c}{K2-261$^\dagger$} & K2-265 & K2-277 & K2-321$^\prime$\\
\hline \\
$\pi$ &  Gaia Parallax (mas) \dotfill & $\mathcal{G}[1.49761, 0.04250]$ & \multicolumn{2}{c}{$\mathcal{G}[4.68526, 0.04270]$} & $\mathcal{G}[7.18885, 0.05050]$ & $\mathcal{G}[8.84150, 0.06199]$ & --- \\
$[{\rm Fe/H}]$  & Metallicity (dex) \dotfill  & $\mathcal{G}[0.386, 0.080]$ & \multicolumn{2}{c}{$\mathcal{G}[0.382, 0.080]$}  & $\mathcal{G}[0.08, 0.08]$  & $\mathcal{G}[0.064, 0.080]$  & --- \\
 $A_V$ & V-band extinction (mag) \dotfill & $\mathcal{U}[0, 0.82243]$ & \multicolumn{2}{c}{$\mathcal{U}[0, 0.12679]$} & $\mathcal{U}[0, 0.11408]$ & $\mathcal{U}[0, 0.19933]$ & --- \\
$D_T^*$&  Dilution in \tess\  \dotfill & $\mathcal{G}[0, 0.036733]$  & \multicolumn{2}{c}{$\mathcal{G}[0, 0.0021478]$}  & ---  & $\mathcal{G}[0, 0.0020849]$  & $\mathcal{G}[0, 0.0035631]$ \\
\hline\hline
Parameter & Units & Values \\
Stellar Parameters: \\    
~~~~$M_*$\dotfill &Mass (\msun)\dotfill & $1.637^{+0.065}_{-0.069}$ & $1.107^{+0.042}_{-0.047}$ & $1.264\pm0.041$  & $0.901^{+0.051}_{-0.042}$ & $0.974^{+0.053}_{-0.058}$ & \textcolor{red}{$0.578\pm 0.03$} \\
~~~~$R_*$\dotfill &Radius (\rsun)\dotfill & $1.755^{+0.061}_{-0.078}$ & $1.663^{+0.071}_{-0.068}$& $1.609^{+0.061}_{-0.057}$  & $0.92^{+0.034}_{-0.032}$ & $0.973^{+0.038}_{-0.035}$ & \textcolor{red}{$0.595\pm 0.031$} \\
~~~~$L_*$\dotfill &Luminosity (\lsun)\dotfill & $6.14^{+0.48}_{-0.52}$ & $2.259^{+0.087}_{-0.083}$ & $2.275^{+0.085}_{-0.084}$  &  $0.657\pm 0.024$ & $0.89^{+0.048}_{-0.044}$ & \textcolor{red}{$0.063^{+0.042}_{-0.028}$} \\
~~~~$F_{Bol}$\dotfill &Bolometric Flux (cgs)\dotfill & $4.36e-10^{+1.9e-11}_{-3e-11}$ & $1.588e-09^{+5.5e-11}_{-5.1e-11}$ & $1.599e-09^{+5.2e-11}_{-5.3e-11}$ &   $1.087e-09\pm 3.7e-11$ & $2.22e-09^{+1.2e-10}_{-1.1e-10}$ & ---\\
~~~~$\rho_*$\dotfill &Density (cgs)\dotfill & $0.424^{+0.067}_{-0.039}$ & $0.339^{+0.046}_{-0.042}$ & $0.427^{+0.046}_{-0.041}$ &  $1.63^{+0.22}_{-0.19}$ & $1.5\pm 0.2$ & \textcolor{red}{$3.86^{+0.7}_{-0.57}$} \\
~~~~$\log{g}$\dotfill &Surface gravity (cgs)\dotfill & $4.161^{+0.044}_{-0.03}$ & $4.040^{+0.038}_{-0.041}$ & $4.126^{+0.030}_{-0.028}$ &  $4.465^{+0.042}_{-0.039}$ & $4.452^{+0.041}_{-0.049}$ & \textcolor{red}{$4.65^{+0.051}_{-0.049}$} \\
~~~~$T_{\rm eff}$\dotfill &Effective Temperature (K)\dotfill & $6860.0^{+150.0}_{-160.0}$ & $5490\pm110$ & $5587\pm100$ &  $5420.0\pm 100.0$ &   $5680.0\pm 120.0$ & \textcolor{red}{$3750.0^{+510.0}_{-520.0}$} \\
~~~~$[{\rm Fe/H}]$\dotfill &Metallicity (dex)\dotfill & $0.334^{+0.052}_{-0.061}$ & $0.372^{+0.069}_{-0.072}$ & $0.401^{+0.064}_{-0.070}$   &  $0.033^{+0.099}_{-0.1}$ & $0.078^{+0.072}_{-0.065}$ & \textcolor{red}{$0.148\pm 0.062$} \\
~~~~$[{\rm Fe/H}]_{0}$\dotfill &Initial Metallicity$^{1}$ \dotfill & $0.44^{+0.04}_{-0.056}$ & $0.368^{+0.064}_{-0.068}$ & $0.396^{+0.057}_{-0.064}$ &  $0.051^{+0.096}_{-0.091}$ & $0.084^{+0.072}_{-0.069}$ & ---\\
~~~~$Age$\dotfill &Age (Gyr)\dotfill & $0.65^{+0.43}_{-0.37}$ & $8.8^{+1.7}_{-1.2}$ &$4.78^{+0.72}_{-0.78}$  & $7.8^{+4.0}_{-4.6}$ & $4.9^{+4.8}_{-3.4}$ & ---\\
~~~~$EEP$\dotfill &Equal Evolutionary Phase$^{2}$ \dotfill & $327.0^{+13.0}_{-26.0}$ & $455.4^{+4.6}_{-5.7}$ &  $413.3^{+9.4}_{-15}$ &  $366.0^{+29.0}_{-32.0}$ & $350.0^{+45.0}_{-32.0}$ & ---\\
~~~~$A_V$\dotfill &V-band extinction (mag)\dotfill & $0.766^{+0.041}_{-0.079}$ & $0.060^{+0.045}_{-0.041}$ & $0.074^{+0.037}_{-0.047}$  & $0.071^{+0.031}_{-0.042}$ & $0.099^{+0.065}_{-0.066}$ & ---\\
~~~~$\sigma_{SED}$\dotfill & SED photometry error scaling\dotfill & $0.56^{+0.24}_{-0.14}$ & $0.81^{+0.32}_{-0.20}$ & $0.82^{+0.33}_{-0.20}$  &  $0.95^{+0.41}_{-0.26}$ & $0.75^{+0.27}_{-0.17}$ & ---\\
~~~~$\varpi$\dotfill &Parallax (mas)\dotfill & $1.486\pm 0.041$ & $4.688\pm0.043$ & $4.687\pm0.042$ &  $7.19^{+0.051}_{-0.05}$ & $8.838\pm 0.062$ & ---\\
~~~~$d$\dotfill &Distance (pc)\dotfill & $672.0^{+19.0}_{-18.0}$ & $213.3^{+2.0}_{-1.9}$ & $213.3\pm1.9$  & $139.08^{+0.98}_{-0.97}$ & $113.14^{+0.8}_{-0.78}$ & ---\\
Planetary Parameters:\\
~~~~$P$\dotfill &Period (days)\dotfill & $2.62669762\pm 6.6e-07$ & $11.6334681\pm 4.4e-06$ &  $11.6334681\pm4.4e-06$ &   $2.36902^{+5.8e-05}_{-5.9e-05}$ & $6.326768^{+1.5e-05}_{-1.2e-05}$ & $2.2979749^{+1.7e-06}_{-1.9e-06}$ \\
~~~~$R_P$\dotfill &Radius (\rj)\dotfill & $1.643^{+0.058}_{-0.073}$ & $0.856^{+0.038}_{-0.036}$ & $0.827^{+0.033}_{-0.030}$  &  $0.1524^{+0.01}_{-0.0072}$ & $0.195^{+0.018}_{-0.011}$ & $0.183^{+0.015}_{-0.012}$ \\
~~~~$M_P$\dotfill &Mass (\mj)\dotfill & $1.72^{+0.34}_{-0.35}$ & $0.194^{+0.024}_{-0.028}$ &  $0.217^{+0.025}_{-0.029}$ &   $0.0231^{+0.0045}_{-0.0044}$ & $0.0197^{+0.0076}_{-0.0048}$ & $0.0178^{+0.0076}_{-0.0046}$ \\
~~~~$T_C$\dotfill &Time of conjunction$^{3}$ (\bjdtdb)\dotfill & $2457820.737343\pm 6.3e-05$ & $2457906.84110^{+0.00028}_{-0.00033}$ & $2457906.84108^{+0.00026}_{-0.00029}$ &   $2456981.6455^{+0.0011}_{-0.001}$ & $2457221.2291\pm 0.0011$ & $2457909.17213^{+0.00069}_{-0.00071}$ \\
~~~~$T_T$\dotfill &Time of minimum projected separation$^{4}$ (\bjdtdb)\dotfill & $2457820.737341\pm 6e-05$ & $2457906.84130\pm0.00022$ & $2457906.84124\pm0.00022$ &   $2456981.64531^{+0.00099}_{-0.00098}$ & $2457221.22914^{+0.00064}_{-0.00063}$ & $2457909.17213\pm 0.00051$ \\
~~~~$T_0$\dotfill &Optimal conjunction Time$^{5}$ (\bjdtdb)\dotfill & $2457894.284876^{+6e-05}_{-5.9e-05}$ & $2458151.14394^{+0.00027}_{-0.00032}$ & $2458151.14392^{+0.00024}_{-0.00027}$ &   $2457017.18078^{+0.00055}_{-0.00054}$ & $2457303.4771\pm 0.001$ & $2458141.26759^{+0.00064}_{-0.00068}$ \\
~~~~$a$\dotfill &Semi-major axis (AU)\dotfill & $0.04392^{+0.00057}_{-0.00063}$ & $0.1040^{+0.0013}_{-0.0015}$ &  $0.1086\pm0.0012$ &  $0.03359^{+0.00062}_{-0.00053}$ & $0.0664^{+0.0012}_{-0.0013}$ & $0.02839^{+0.00048}_{-0.0005}$ \\
~~~~$i$\dotfill &Inclination (Degrees)\dotfill & $89.18^{+0.58}_{-0.83}$ & $88.24^{+1.0}_{-0.67}$ &  $88.58^{+0.85}_{-0.57}$ &  $87.01^{+1.8}_{-1.0}$ & $86.83^{+1.6}_{-0.96}$ & $86.1^{+2.5}_{-1.1}$ \\
~~~~$e$\dotfill &Eccentricity \dotfill & $0.053^{+0.051}_{-0.036}$ & $0.331^{+0.063}_{-0.066}$ & $0.274^{+0.065}_{-0.061}$ &   $0.16^{+0.16}_{-0.11}$ & $0.52^{+0.24}_{-0.26}$ & $0.44^{+0.26}_{-0.23}$ \\
~~~~$\omega_*$\dotfill &Argument of Periastron (Degrees)\dotfill & $-72.0^{+74.0}_{-79.0}$ & $137^{+13}_{-16}$ & $145^{+13}_{-15}$ &   $-57.0^{+52.0}_{-27.0}$ & $46.0^{+96.0}_{-120.0}$ & $45.0^{+96.0}_{-140.0}$ \\
~~~~$T_{eq}$\dotfill &Equilibrium temperature$^{6}$ (K)\dotfill & $2090.0^{+35.0}_{-39.0}$ & $1058\pm11$ & $1036\pm10.$  &  $1366.0\pm 17.0$ & $1049.0\pm 15.0$ & $830.0\pm 110.0$ \\
~~~~$\tau_{\rm circ}$\dotfill &Tidal circularization timescale (Gyr)\dotfill & $0.0381^{+0.011}_{-0.0099}$ & $15.8^{+12}_{-7.3}$ & $33^{+19}_{-14}$  &  $23.0^{+14.0}_{-15.0}$ & $31.0^{+210.0}_{-31.0}$ & $0.76^{+2.8}_{-0.74}$ \\
~~~~$K$\dotfill &RV semi-amplitude (m/s)\dotfill & $182.0^{+36.0}_{-37.0}$ & $17.3^{+2.1}_{-2.5}$ &  $17.4^{+1.9}_{-2.3}$ &   $3.87^{+0.77}_{-0.75}$ & $2.73^{+1.3}_{-0.75}$ & $4.6^{+2.3}_{-1.3}$ \\
~~~~$R_P/R_*$\dotfill &Radius of planet in stellar radii \dotfill & $0.09617^{+0.00024}_{-0.00019}$ & $0.05279^{+0.00079}_{-0.00054}$ & $0.05274^{+0.00076}_{-0.00049}$ &   $0.01695^{+0.00095}_{-0.00045}$ & $0.02045^{+0.0019}_{-0.00074}$ & $0.0315^{+0.0021}_{-0.0013}$ \\
~~~~$a/R_*$\dotfill &Semi-major axis in stellar radii \dotfill & $5.37^{+0.27}_{-0.17}$ & $13.43\pm0.58$ & $14.51^{+0.51}_{-0.47}$ &   $7.86^{+0.33}_{-0.31}$ & $14.68^{+0.64}_{-0.7}$ & $10.25^{+0.58}_{-0.53}$ \\
~~~~$\delta$\dotfill &$\left(R_P/R_*\right)^2$ \dotfill & $0.00925^{+4.5e-05}_{-3.7e-05}$ & $0.002787^{+8.5e-05}_{-5.6e-05}$ & $0.002781^{+8.1e-05}_{-5.2e-05}$ &   $0.000287^{+3.3e-05}_{-1.5e-05}$ & $0.000418^{+7.9e-05}_{-3e-05}$ & $0.000992^{+0.00014}_{-7.8e-05}$ \\
~~~~$Depth_{\rm K2}$\dotfill &Flux decrement at mid transit for \ktwo \dotfill & $0.01042\pm 0.00011$ & $0.003571^{+6.7e-05}_{-6.4e-05}$ & $0.003553^{+6.5e-05}_{-6.2e-05}$  &  $0.00035\pm 1.1e-05$ & $0.000482^{+1.7e-05}_{-1.9e-05}$ & $0.001167^{+8.6e-05}_{-6.5e-05}$ \\
~~~~$Depth_{\rm TESS}$\dotfill &Flux decrement at mid transit for \tess \dotfill & $0.01007\pm 0.00014$ & $0.003353\pm 5.2e-05$   & $0.003332\pm5.0e-05$ & ---& $0.000469^{+1.6e-05}_{-1.7e-05}$ & $0.00114^{+8.7e-05}_{-9.2e-05}$ \\
~~~~$\tau$\dotfill &Ingress/egress transit duration (days)\dotfill & $0.015594^{+0.00033}_{-9.5e-05}$ & $0.0117^{+0.0016}_{-0.0010}$ & $0.01163^{+0.0016}_{-0.00092}$  &   $0.00198^{+0.0013}_{-0.00037}$ & $0.00227^{+0.003}_{-0.00065}$ & $0.00183^{+0.0017}_{-0.00041}$ \\
~~~~$T_{14}$\dotfill &Total transit duration (days)\dotfill & $0.17496^{+0.00035}_{-0.00032}$ &$0.2138^{+0.0015}_{-0.0011}$ & $0.2137^{+0.0014}_{-0.0011}$  &  $0.0962^{+0.0018}_{-0.0016}$ & $0.0819^{+0.0028}_{-0.0023}$ & $0.0478\pm 0.0017$ \\
~~~~$b$\dotfill &Transit Impact parameter \dotfill & $0.08^{+0.08}_{-0.056}$ & $0.30^{+0.14}_{-0.18}$ & $0.29^{+0.14}_{-0.18}$ &   $0.44^{+0.26}_{-0.28}$ & $0.53^{+0.28}_{-0.36}$ & $0.47^{+0.29}_{-0.33}$ \\
~~~~$\rho_P$\dotfill &Density (cgs)\dotfill & $0.48^{+0.12}_{-0.11}$ & $0.381^{+0.078}_{-0.072}$& $0.473^{+0.082}_{-0.079}$  &  $7.9^{+2.1}_{-1.9}$ & $3.18^{+1.2}_{-0.76}$ & $3.55^{+1.4}_{-0.88}$ \\
~~~~$logg_P$\dotfill &Surface gravity \dotfill & $3.199^{+0.085}_{-0.1}$ & $2.815^{+0.066}_{-0.080}$ & $2.895^{+0.058}_{-0.070}$ &   $3.385^{+0.09}_{-0.1}$ & $3.1^{+0.13}_{-0.11}$ & $3.11^{+0.14}_{-0.12}$ \\
~~~~$T_S$\dotfill &Time of eclipse (\bjdtdb)\dotfill & $2457819.432^{+0.082}_{-0.063}$ & $2457910.91^{+0.61}_{-0.62}$ & $2457911.04^{+0.51}_{-0.55}$ &   $2456980.58\pm 0.1$ & $2457218.1\pm 2.2$ & $2457908.02^{+0.7}_{-0.72}$ \\
~~~~$T_{S,14}$\dotfill &Total eclipse duration (days)\dotfill & $0.167^{+0.011}_{-0.016}$ & $0.310^{+0.053}_{-0.043}$ & $0.281^{+0.035}_{-0.031}$  &  $0.082^{+0.013}_{-0.016}$ & $0.082^{+0.062}_{-0.042}$ & $0.05^{+0.034}_{-0.011}$ \\
~~~~$e\cos{\omega_*}$\dotfill & \dotfill& $0.005^{+0.049}_{-0.037}$ & $-0.233^{+0.082}_{-0.085}$ &  $-0.218^{+0.070}_{-0.077}$ &  $0.075^{+0.067}_{-0.065}$ & $0.0^{+0.56}_{-0.54}$ & $-0.0^{+0.49}_{-0.5}$ \\
~~~~$e\sin{\omega_*}$\dotfill & \dotfill& $-0.024^{+0.032}_{-0.051}$ & $0.220^{+0.059}_{-0.064}$ & $0.155^{+0.050}_{-0.057}$ &   $-0.093^{+0.093}_{-0.21}$ & $0.15^{+0.26}_{-0.33}$ & $0.14^{+0.2}_{-0.24}$ \\
~~~~$M_P/M_*$\dotfill &Mass ratio \dotfill & $0.001\pm 0.0002$ & $0.000168^{+2.0e-05}_{-2.4e-05}$ & $0.000164^{+1.8e-05}_{-2.2e-05}$ &   $2.44e-05\pm 4.7e-06$ & $1.94e-05^{+7.7e-06}_{-4.9e-06}$ & $2.94e-05^{+1.3e-05}_{-7.6e-06}$ \\
~~~~$d/R_*$\dotfill &Separation at mid transit \dotfill & $5.48^{+0.56}_{-0.33}$ & $9.8^{+1.2}_{-1.1}$ & $11.6^{+1.2}_{-1.0}$  &  $8.58^{+1.6}_{-0.9}$ & $8.4^{+5.4}_{-2.4}$ & $6.6^{+2.9}_{-1.8}$ \\
\enddata
\label{tab:med5}
 \begin{flushleft} 
  \footnotesize{
    \textbf{Notes.} See Table 3 in \citet{Eastman:2019} for a detailed description of all parameters. Gaussian and uniform priors are indicated as $\mathcal{G\mathrm{[median, width]}}$ and $\mathcal{U\mathrm{[lower~bound, upper~bound]}}$, respectively. Metallicity priors are adopted from TRES spectra unless otherwise indicated. $^\star$ Gaussian priors were placed on dilution in \tess only for systems with a contamination listed on EXOFOP. ~$^\prime$ The stellar parameters from the global fit are not considered reliable as the SED was not included within this fit. $^\dagger$The PDFs for K2-261 showed bimodality of the host star's mass and age (See \S \ref{sec:k2-261}). The two solutions are shown here, but we adopt the low mass solutions for figures as it has a higher probability. \\
     $^1$The metallicity of the star at birth. $^2$Corresponds to static points in a star's evolutionary history. See \S2 in \citet{Dotter:2016}. $^3$Time of conjunction is commonly reported as the "transit time". $^4$Time of minimum projected separation is a more correct "transit time". $^5$Optimal time of conjunction minimizes the covariance between $T_C$ and Period. $^6$Assumes no albedo and perfect redistribution.
    }
\end{flushleft}
\end{deluxetable*}

\begin{deluxetable*}{lccccc}
\centering
\tablecaption{Median values and 68\% confidence intervals for the global models for \ktwo fits only.}
\tablehead{\omit}
\tabletypesize{\footnotesize}
\startdata
System & \ktwo Campaign & \multicolumn{2}{c}{Wavelength Parameters} & \multicolumn{2}{c}{Transit Parameters} \\
 &  & \multicolumn{1}{c}{$u_1^\dagger$} & \multicolumn{1}{c}{$u_2^\ddagger$} & $\sigma^{2\star}$ & $F_0^*$ \\ \hline
K2-7 & C1  & $0.394\pm0.054$ & $0.261\pm0.051$ & $-0.0000000095^{+0.0000000051}_{-0.0000000045}$ & $0.999998^{+0.000018}_{-0.000019}$ \\
K2-54 & C3  & $0.44\pm0.17$ & $0.24^{+0.17}_{-0.18}$ & $0.0000000087^{+0.000000011}_{-0.0000000097}$ & $1.000000\pm0.000029$ \\
K2-57 & C3  & $0.677\pm0.054$ & $0.076^{+0.053}_{-0.054}$ & $0.000000001^{+0.000000013}_{-0.000000010}$ & $1.000021\pm0.000031$ \\
K2-147 & C7  & $0.28^{+0.17}_{-0.15}$ & $0.35\pm0.18$ & $0.0000000111^{+0.0000000033}_{-0.0000000030}$ & $1.0000077^{+0.000010}_{-0.0000100}$ \\
K2-181 & C5  & $0.480^{+0.042}_{-0.041}$ & $0.217\pm0.038$ & $0.0000000110^{+0.0000000027}_{-0.0000000024}$ & $0.999996^{+0.000012}_{-0.000011}$ \\
 & C18 & $0.480^{+0.042}_{-0.041}$ & $0.217\pm0.038$  & $0.0000000020^{+0.0000000025}_{-0.0000000022}$ & $1.000010\pm0.000012$ \\
K2-203 & C8  & $0.561\pm0.054$ & $0.158\pm0.052$ & $0.0000000038^{+0.0000000017}_{-0.0000000015}$ & $0.999997\pm0.000010$ \\
K2-204 & C8  & $0.372^{+0.055}_{-0.054}$ & $0.253^{+0.051}_{-0.052}$ & $0.0000000095^{+0.0000000033}_{-0.0000000030}$ &$0.999995\pm0.000012$\\
K2-208 & C8  & $0.447\pm0.055$ & $0.237^{+0.051}_{-0.053}$ & $0.00000000067^{+0.00000000100}_{-0.00000000091}$& $0.9999965^{+0.0000069}_{-0.0000070}$ \\
K2-211 & C8  & $0.548\pm0.055$ & $0.166\pm0.052$ & $0.0000000056\pm0.0000000012$& $1.0000020^{+0.0000057}_{-0.0000056}$ \\
K2-225 & C10  & $0.468\pm0.058$ & $0.210\pm0.054$ & $0.0000000029^{+0.0000000014}_{-0.0000000012}$& $1.000031\pm0.000011$\\
K2-226 & C10  & $0.492^{+0.056}_{-0.055}$ & $0.203\pm0.053$ & $0.0000000019^{+0.0000000013}_{-0.0000000012}$ & $1.0000034\pm0.0000081$ \\
K2-250 & C10  & $0.576\pm0.058$ & $0.157\pm0.054$ & $0.0000000115^{+0.0000000063}_{-0.0000000057}$ & $1.000006\pm0.000015$ \\
K2-265 & C3  & $0.466\pm0.054$ & $0.215\pm0.052$ & $-0.00000000033^{+0.00000000015}_{-0.00000000014}$& $0.9999979\pm0.0000025$  \\
\enddata
\label{tab:wavek2}
 \begin{flushleft} 
  \footnotesize{
    \textbf{Notes.}$^\dagger$Linear limb-darkening coefficient. $^\ddagger$Quadratic limb-darkening coefficient. $^\star$Added variance. $^*$ Baseline flux. 
    }
\end{flushleft}
\end{deluxetable*}

\begin{deluxetable*}{llccccc}
\centering
\tablecaption{Median values and 68\% confidence intervals for the global models for \ktwo and \tess fits.}
\tablehead{\omit}
\tabletypesize{\tiny}
\startdata
System  & Campaign/Sector & \multicolumn{2}{c}{Wavelength Parameters} & \multicolumn{2}{c}{Transit Parameters} & Dilution \\
 &  & $u_1^\dagger$ & $u_2^\ddagger$ & $\sigma^{2\star}$ & $F_0^*$ & $A_D$ \\ \hline
K2-77 & \ktwo C4 & $0.506\pm0.059$ & $0.151\pm0.055$ & $0.0000000036^{+0.0000000016}_{-0.0000000013}$ & $1.000009\pm0.000010$ &  \\
 & \tess S5 & $0.411^{+0.034}_{-0.035}$ & $0.218\pm0.029$ & $0.000000178^{+0.000000050}_{-0.000000049}$ & $0.999994\pm0.000016$ & $0.00000\pm0.00024$ \\
 & \tess S42 &  &  & $0.000000028^{+0.000000059}_{-0.000000058}$ & $1.000003\pm0.000019$ &  \\
 & \tess S43 &  &  & $0.000000118^{+0.000000054}_{-0.000000053}$ & $1.000007\pm0.000017$ &  \\
 & \tess S44 &  &  & $-0.000000085^{+0.000000050}_{-0.000000049}$ & $1.000012\pm0.000017$ &  \\\hline
K2-97 & \ktwo C5 & $0.652^{+0.040}_{-0.041}$ & $0.098\pm0.039$ & $0.0000000070^{+0.0000000031}_{-0.0000000028}$ & $0.999995\pm0.000011$ &  \\
 & \ktwo C18 &  &  & $0.0000000149^{+0.0000000048}_{-0.0000000043}$ & $0.999996\pm0.000014$ &  \\
 & \tess S7 & $0.488\pm0.029$ & $0.181\pm0.026$ & $0.000000159^{+0.000000070}_{-0.000000069}$ & $1.000000\pm0.000020$ & $-0.27\pm0.12$ \\
 & \tess S44 &  &  & $0.000000244^{+0.000000083}_{-0.000000082}$ & $1.000001\pm0.000022$ &  \\
 & \tess S45 &  &  & $-0.000000001^{+0.000000084}_{-0.000000083}$ & $0.999988\pm0.000022$ &  \\
 & \tess S46 &  &  & $0.000000148^{+0.000000072}_{-0.000000070}$ & $1.000008\pm0.000020$ &  \\\hline
K2-98 & \ktwo C5 & $0.325^{+0.037}_{-0.036}$ & $0.305\pm0.036$ & $0.0000000017^{+0.0000000012}_{-0.0000000010}$ & $1.0000098^{+0.0000075}_{-0.0000074}$ &  \\
 & \ktwo C18 &  &  & $0.0000000044^{+0.0000000020}_{-0.0000000018}$ & $0.999991\pm0.000010$ &  \\
 & \tess S7 & $0.235^{+0.028}_{-0.027}$ & $0.304\pm0.023$ & $-0.00000003^{+0.00000031}_{-0.00000029}$ & $1.000088\pm0.000085$ & $-0.0000\pm0.0019$ \\
 & \tess S34 &  &  & $0.00000057^{+0.00000045}_{-0.00000042}$ & $1.00015\pm0.00010$ &  \\
 & \tess S44 &  &  & $0.00000017^{+0.00000033}_{-0.00000031}$ & $0.999879^{+0.000088}_{-0.000087}$ &  \\
 & \tess S45 &  &  & $0.00000045^{+0.00000033}_{-0.00000031}$ & $0.999931^{+0.000086}_{-0.000085}$ &  \\
 & \tess S46 &  &  & $0.00000059^{+0.00000030}_{-0.00000028}$ & $1.000249\pm0.000079$ &  \\\hline
K2-114 & \ktwo C5 & $0.611^{+0.025}_{-0.026}$ & $0.106^{+0.036}_{-0.035}$ & $0.000000012^{+0.000000018}_{-0.000000016}$ & $1.000034\pm0.000032$ &  \\
 & \ktwo C18 &  &  & $0.00000082\pm0.00000013$ & $0.999745^{+0.000044}_{-0.000043}$ &  \\
 & \tess S7 & $0.473\pm0.026$ & $0.189^{+0.026}_{-0.025}$ & $-0.0000060^{+0.0000062}_{-0.0000058}$ & $0.99980^{+0.00042}_{-0.00043}$ & $-0.027^{+0.034}_{-0.035}$ \\
 & \tess S44 &  &  & $-0.0000000^{+0.0000076}_{-0.0000071}$ & $0.99925\pm0.00047$ &  \\
 & \tess S45 &  &  & $0.0000153^{+0.0000093}_{-0.0000088}$ & $0.99999\pm0.00051$ &  \\
 & \tess S46 &  &  & $-0.0000077^{+0.0000068}_{-0.0000064}$ & $0.99941\pm0.00044$ &  \\\hline
K2-115 & \ktwo C5 & $0.365\pm0.035$ & $0.278\pm0.036$ & $0.0000000117^{+0.0000000070}_{-0.0000000058}$ & $0.999993\pm0.000023$ &  \\
 & \ktwo C18 &  &  & $0.000000028^{+0.000000016}_{-0.000000013}$ & $0.999966\pm0.000037$ &  \\
 & \tess S7 & $0.274\pm0.032$ & $0.285\pm0.026$ & $0.0000013^{+0.0000015}_{-0.0000014}$ & $1.00008\pm0.00018$ & $-0.000000\pm0.000014$ \\
 & \tess S34 &  &  & $0.0000033^{+0.0000033}_{-0.0000030}$ & $0.99972\pm0.00034$ &  \\
 & \tess S45 &  &  & $-0.0000020^{+0.0000016}_{-0.0000015}$ & $0.99999^{+0.00020}_{-0.00021}$ &  \\
 & \tess S46 &  &  & $-0.0000024^{+0.0000020}_{-0.0000018}$ & $0.99930\pm0.00027$ &  \\\hline
K2-167 & \ktwo C3& $0.320\pm0.050$ & $0.314\pm0.050$ & $-0.00000000008^{+0.00000000022}_{-0.00000000018}$ & $0.9999986\pm0.0000041$ &  \\
 & \tess S2 & $0.224\pm0.031$ & $0.300\pm0.029$ & $0.0000000005^{+0.0000000047}_{-0.0000000044}$ & $0.999982\pm0.000011$ & $0.000000\pm0.000027$ \\
 & \tess S28 &  &  & $-0.0000015010^{+0.0000000039}_{-0.0000000019}$ & $1.000031^{+0.000038}_{-0.000032}$ &  \\
 & \tess S42 &  &  & $0.000000017^{+0.000000012}_{-0.000000010}$ & $1.000000\pm0.000021$ &  \\\hline
K2-180 & \ktwo C5 & $0.416^{+0.042}_{-0.041}$ & $0.253^{+0.038}_{-0.039}$ & $0.0000000016^{+0.0000000029}_{-0.0000000025}$ & $0.999998\pm0.000013$ &  \\
 & \ktwo C18 &  &  & $0.0000000196^{+0.0000000070}_{-0.0000000059}$ & $1.000017\pm0.000021$ &  \\
 & \tess S7 & $0.323^{+0.030}_{-0.029}$ & $0.268^{+0.24}_{-0.025}$ & $0.00000007^{+0.00000031}_{-0.00000022}$ & $0.99953\pm0.00019$ & $-0.012\pm0.039$ \\
 & \tess S34 &  &  & $0.0000011^{+0.0000011}_{-0.0000010}$ & $1.00028\pm0.00016$ &  \\
 & \tess S44 &  &  & $0.00000045^{+0.0000011}_{-0.00000097}$ & $0.99981\pm0.00016$ &  \\
 & \tess S45 &  &  & $0.00000168^{+0.00000081}_{-0.00000077}$ & $0.99983\pm0.00014$ &  \\
 & \tess S46 &  &  & $-0.00000158^{+0.00000069}_{-0.00000064}$ & $0.99985\pm0.00013$ &  \\\hline
K2-182 & \ktwo C5 & $0.528\pm0.035$ & $0.167\pm0.038$ & $0.00000000092^{+0.00000000059}_{-0.00000000053}$ & $0.9999964^{+0.0000051}_{-0.0000050}$ &  \\
 & \ktwo C18 &  &  & $0.00000000107^{+0.00000000095}_{-0.00000000085}$ & $1.0000071\pm0.0000067$ &  \\
 & \tess S7 & $0.416\pm0.027$ & $0.219\pm0.024$ & $-0.000000031^{+0.000000039}_{-0.000000031}$ & $1.000089\pm0.000052$ & $-0.00000\pm0.00029$ \\
 & \tess S34 &  &  & $0.00000022^{+0.00000019}_{-0.00000018}$ & $1.000116^{+0.000059}_{-0.000058}$ &  \\
 & \tess S44 &  &  & $0.00000002^{+0.00000019}_{-0.00000018}$ & $0.999984\pm0.000065$ &  \\
 & \tess S45 &  &  & $0.00000027^{+0.00000020}_{-0.00000019}$ & $1.000113\pm0.000067$ &  \\
 & \tess S46 &  &  & $0.00000009^{+0.00000018}_{-0.00000017}$ & $0.999995\pm0.000063$ &  \\\hline
K2-237 & \ktwo C11-1 & $0.337^{+0.014}_{-0.015}$ & $0.250\pm0.034$ & $0.0000000056^{+0.0000000021}_{-0.0000000018}$ & $1.0000000^{+0.0000096}_{-0.0000097}$ & $0.006^{+0.039}_{-0.040}$ \\
 & \ktwo C11-2 &  &  & $0.00000000385^{+0.0000000010}_{-0.00000000092}$ & $0.9999947\pm0.0000062$ &  \\
 & \tess S12 & $0.266\pm0.033$ & $0.305\pm0.035$ & $0.0000217\pm0.0000013$ & $1.00025\pm0.00015$ & $-0.050^{+0.042}_{-0.043}$ \\
 & \tess S39 &  &  & $0.0001052^{+0.0000031}_{-0.0000030}$ & $1.00057\pm0.00014$ &  \\\hline
K2-260 & \ktwo C13 & $0.228\pm0.017$ & $0.323\pm0.037$ & $0.0000000061^{+0.0000000012}_{-0.0000000011}$ & $0.9999861^{+0.0000058}_{-0.0000059}$ &  \\
 & \tess S5 & $0.164\pm0.026$ & $0.321\pm0.028$ & $-0.00000027^{+0.00000012}_{-0.00000011}$ & $1.000021^{+0.000072}_{-0.000073}$ & $0.027\pm0.011$ \\
 & \tess S32&  &  & $0.00000086^{+0.00000051}_{-0.00000049}$ & $1.000154^{+0.000082}_{-0.000083}$ &  \\
 & \tess S43 &  &  & $-0.00000027^{+0.00000036}_{-0.00000035}$ & $0.999843^{+0.000072}_{-0.000073}$ &  \\\hline
K2-261 & \ktwo C14 & $0.474\pm0.031$ & $0.208\pm0.046$ & $0.00000000157^{+0.00000000065}_{-0.00000000057}$ & $0.9999974\pm0.0000059$ &  \\
 & \tess S9 & $0.363\pm0.027$ & $0.257\pm0.027$ & $0.000000106^{+0.000000050}_{-0.000000047}$ & $1.000048\pm0.000034$ & $0.0002\pm0.0021$ \\
 & \tess S35 &  &  & $0.00000388^{+0.00000018}_{-0.00000017}$ & $1.000253\pm0.000041$ &  \\
 & \tess S45 &  &  & $0.000000017^{+0.000000054}_{-0.000000051}$ & $0.999950\pm0.000035$ &  \\
 & \tess S46 &  &  & $0.000000004^{+0.000000050}_{-0.000000046}$ & $1.000102\pm0.000034$ &  \\\hline
K2-277 & \ktwo C6 & $0.419^{+0.055}_{-0.054}$ & $0.250\pm0.052$ & $0.00000000110^{+0.00000000033}_{-0.00000000028}$ & $0.9999913^{+0.0000046}_{-0.0000045}$ &  \\
 & \tess S10 & $0.322\pm0.040$ & $0.270^{+0.036}_{-0.037}$ & $-0.000000010^{+0.000000044}_{-0.000000041}$ & $1.000020\pm0.000034$ & $-0.0000\pm0.0021$ \\
 & \tess S37 &  &  & $0.000000024^{+0.000000012}_{-0.000000011}$ & $1.0000220\pm0.0000080$ &  \\\hline
K2-321 & \ktwo C14 & $0.41^{+0.16}_{-0.17}$ & $0.33^{+0.17}_{-0.18}$ & $0.0000000045^{+0.0000000025}_{-0.0000000023}$ & $1.000003\pm0.000010$ &  \\
 & \tess S8 & $0.38\pm0.17$ & $0.30^{+0.17}_{-0.18}$ & $0.00000023\pm0.00000012$ & $1.000003\pm0.000028$ & $0.0001\pm0.0037$ \\
 & \tess S35 &  &  & $-0.000000009^{+0.000000044}_{-0.000000042}$ & $0.999998\pm0.000025$ &  \\
 & \tess S45 &  &  & $0.000000078^{+0.000000099}_{-0.000000096}$ & $1.000015\pm0.000024$ &  \\
 & \tess S46 &  &  & $-0.00000001^{+0.00000011}_{-0.00000010}$ & $1.000000^{+0.000024}_{-0.000025}$ & 
\enddata
\label{tab:wave_k2+tess}
 \begin{flushleft} 
  \footnotesize{
    \textbf{Notes.}$^\dagger$Linear limb-darkening coefficient. $^\ddagger$Quadratic limb-darkening coefficient. $^\star$Added variance. $^*$ Baseline flux. 
    }
\end{flushleft}
\end{deluxetable*}

\begin{deluxetable*}{lccccc}
\centering
\tablecaption{Median values and 68\% confidence intervals for the radial velocity parameters.}
\tablehead{\omit}
\tabletypesize{\scriptsize}
\startdata
K2-97\\
\multicolumn{2}{c}{Telescope Parameters:}&HIRES&LEVY\smallskip\\
~~~~$\gamma_{\rm rel}$\dotfill &Relative RV Offset (m/s)\dotfill &$-5.2\pm1.6$&$12^{+16}_{-19}$\\
~~~~$\sigma_J$\dotfill &RV Jitter (m/s)\dotfill &$5.8^{+1.6}_{-1.2}$&$26^{+35}_{-27}$\\
~~~~$\sigma_J^2$\dotfill &RV Jitter Variance \dotfill &$33^{+21}_{-12}$&$720^{+3100}_{-900}$\\\hline
K2-98\\
\multicolumn{2}{l}{Telescope Parameters:}&FIES&HARPS&HARPS-N\smallskip\\
~~~~$\gamma_{\rm rel}$\dotfill &Relative RV Offset (m/s)\dotfill &$76612.0^{+4.5}_{-4.3}$&$76747.7\pm3.3$&$76740.8\pm4.0$\\
~~~~$\sigma_J$\dotfill &RV Jitter (m/s)\dotfill &$7.03^{+0.94}_{-2.2}$&$1.55^{+0.12}_{-0.43}$&$0.00^{+1.6}_{-0.00}$\\
~~~~$\sigma_J^2$\dotfill &RV Jitter Variance \dotfill &$5^{+58}_{-41}$&$0.1^{+2.7}_{-2.8}$&$-0.1^{+2.8}_{-2.7}$\\\hline
K2-114\\
\multicolumn{2}{l}{Telescope Parameters:}&HIRES\smallskip\\
~~~~$\gamma_{\rm rel}$\dotfill &Relative RV Offset (m/s)\dotfill &$-40.8^{+6.7}_{-6.8}$\\
~~~~$\sigma_J$\dotfill &RV Jitter (m/s)\dotfill &$11.9^{+3.8}_{-4.9}$\\
~~~~$\sigma_J^2$\dotfill &RV Jitter Variance \dotfill &$141^{+100}_{-93}$\\\hline
K2-115\\
\multicolumn{2}{l}{Telescope Parameters:}&HIRES\smallskip\\
~~~~$\gamma_{\rm rel}$\dotfill &Relative RV Offset (m/s)\dotfill &$25^{+12}_{-13}$\\
~~~~$\sigma_J$\dotfill &RV Jitter (m/s)\dotfill &$26^{+21}_{-10.}$\\
~~~~$\sigma_J^2$\dotfill &RV Jitter Variance \dotfill &$710^{+1600}_{-440}$\\\hline
K2-180\\
\multicolumn{2}{l}{Telescope Parameters:}&HARPS-N\smallskip\\
~~~~$\gamma_{\rm rel}$\dotfill &Relative RV Offset (m/s)\dotfill &$-76614.40^{+0.58}_{-0.62}$\\
~~~~$\sigma_J$\dotfill &RV Jitter (m/s)\dotfill &$0.00$\\
~~~~$\sigma_J^2$\dotfill &RV Jitter Variance \dotfill &$-3.6^{+3.2}_{-1.0}$\\\hline
K2-182\\
\multicolumn{2}{l}{Telescope Parameters:}&HIRES\smallskip\\
~~~~$\gamma_{\rm rel}$\dotfill &Relative RV Offset$^{4}$ (m/s)\dotfill &$-1.8\pm1.6$\\
~~~~$\sigma_J$\dotfill &RV Jitter (m/s)\dotfill &$4.6^{+1.9}_{-1.2}$\\
~~~~$\sigma_J^2$\dotfill &RV Jitter Variance \dotfill &$21.6^{+21}_{-9.8}$\\\hline
K2-237\\
\multicolumn{2}{l}{Telescope Parameters:}&CORALIE&FIES&HARPS (Smith) &HARPS (Soto)\smallskip\\
~~~~$\gamma_{\rm rel}$\dotfill &Relative RV Offset (m/s)\dotfill &$-22252\pm40$&$-22507^{+15}_{-16}$&$-22325.7^{+9.0}_{-9.4}$&$-22252\pm14$\\
~~~~$\sigma_J$\dotfill &RV Jitter (m/s)\dotfill &$110^{+49}_{-31}$&$0.00^{+39}_{-0.00}$&$18^{+22}_{-16}$&$6.0^{+8.8}_{-6.0}$\\
~~~~$\sigma_J^2$\dotfill &RV Jitter Variance \dotfill &$12200^{+13000}_{-5800}$&$-70^{+1600}_{-520}$&$340^{+1300}_{-330}$&$40^{+180}_{-220}$\\\hline
K2-260\\
\multicolumn{2}{l}{Telescope Parameters:}&FIES\smallskip\\
~~~~$\gamma_{\rm rel}$\dotfill &Relative RV Offset (m/s)\dotfill &$29072^{+22}_{-24}$\\
~~~~$\sigma_J$\dotfill &RV Jitter (m/s)\dotfill &$0.00^{+60}_{-0.00}$\\
~~~~$\sigma_J^2$\dotfill &RV Jitter Variance \dotfill &$-600^{+4200}_{-2000}$\\\hline
K2-261\\
\multicolumn{2}{l}{Telescope Parameters:}&FIES&HARPS&HARPS-N\smallskip\\
~~~~$\gamma_{\rm rel}$\dotfill &Relative RV Offset (m/s)\dotfill &$-13.5^{+2.7}_{-2.8}$&$3340.2^{+2.3}_{-2.4}$&$3335.3^{+2.1}_{-2.6}$\\
~~~~$\sigma_J$\dotfill &RV Jitter (m/s)\dotfill &$4.6^{+4.0}_{-4.6}$&$6.6^{+2.7}_{-1.8}$&$5.6^{+4.2}_{-2.4}$\\
~~~~$\sigma_J^2$\dotfill &RV Jitter Variance \dotfill &$20^{+53}_{-25}$&$43^{+43}_{-20}$&$31^{+64}_{-21}$\\\hline
K2-265\\
\multicolumn{2}{l}{Telescope Parameters:}&HARPS\smallskip\\
~~~~$\gamma_{\rm rel}$\dotfill &Relative RV Offset (m/s)\dotfill &$-18185.56^{+0.52}_{-0.53}$\\
~~~~$\sigma_J$\dotfill &RV Jitter (m/s)\dotfill &$5.71^{+0.43}_{-0.39}$\\
~~~~$\sigma_J^2$\dotfill &RV Jitter Variance \dotfill &$32.6^{+5.1}_{-4.3}$\\
\enddata
\label{tab:rv_params}
 \begin{flushleft} 
  \footnotesize{
    \textbf{Notes.} $\sigma_J^2$ was bound to $\pm300$ m/s for K2-114 and the \cite{Soto:2018} RVs for K2-237. For K2-98 $\sigma_J^2$ was bound to $\pm100$ m/s for the FIES RVs and $\pm4$ m/s for HARPS and HARPS-N. See \S \ref{sec:RVs} for discussion.
    }
\end{flushleft}
\end{deluxetable*}

\end{document}